\documentclass[
aps,prd,
nofootinbib,
superscriptaddress,
showpacs,
tightenlines,
]{revtex4}
\usepackage{amsmath}
\usepackage{amssymb}
\usepackage{bm}
\usepackage{color,graphicx}

\begin{document}

\title{Flexible Parametrization of Generalized Parton Distributions:
The Chiral-Odd Sector}

\author{Gary R.~Goldstein} 
\email{gary.goldstein@tufts.edu}
\affiliation{Department of Physics and Astronomy, Tufts University, Medford, MA 02155 USA.}

\author{J. Osvaldo Gonzalez Hernandez} 
\email{jog4m@virginia.edu}
\affiliation{Department of Physics, University of Virginia, Charlottesville, VA 22904, USA.}

\author{Simonetta Liuti} 
\email{sl4y@virginia.edu}
\affiliation{Department of Physics, University of Virginia, Charlottesville, VA 22904, USA.}
\affiliation{Laboratori Nazionali di Frascati, INFN, Frascati, Italy}



\begin{abstract}
We present a physically motivated parameterization of the chiral-odd
generalized parton distributions. 
The parametrization is an extension of our previous one in the chiral-even 
sector  which was based on the reggeized diquark model. While for chiral even generalized distributions a quantitative fit
with uncertainty estimation can be performed using deep inelastic scattering data, nucleon electromagnetic, axial and pseudoscalar form factors measurements, and 
all available deeply virtual Compton scattering data, 
the chiral-odd sector is far less constrained. 
While awaiting the analysis of measurements on pseudoscalar mesons exclusive electroproduction which are key for the extraction of chiral odd GPDs, 
we worked out a 
connection between the chiral-even  and chiral-odd reduced helicity amplitudes using Parity transformations. The connection works for a class of models
including two-components models. 
This relation allows us to estimate the size of the various chiral odd contributions and it opens the way for future quantitative fits. 
\end{abstract}

\maketitle

\baselineskip 3.0ex
\section{Introduction}
\label{sec:1}
The proton's transversity structure functions, $h_1$, or the probability of finding a  transversely polarized quark inside a transversely polarized proton  has notoriously been an elusive quantity to extract from experiment. Being chirally odd, it can be observed in either Semi Inclusive Deep Inelastic Scattering (SIDIS) or in the Drell Yan process in conjunction with another chiral odd partner. $h_1$'s flavor dependence and its behavior in $x_{Bj}$, and in the four-momentum transfer, $Q^2$, were obtained only relatively recently from model dependent analyses of SIDIS experiments in a limited kinematical range. Similarly, the various related chiral odd Transverse Momentum Distributions  (TMDs)  which are necessary to give a complete description of the proton's transverse structure \cite{Mulders,BogMul}, are hard to extract from experiment (see  \cite{BarBradMar} and references therein). In Ref.\cite{AGL} a new avenue to access transversity was suggested. It was shown that a class of experiments including Deeply Virtual $\pi^o$ Production (DV$\pi^o$P), and more generally deeply virtual neutral pseudo-scalar meson production \cite{GL_charm}, are directly sensitive to the chiral-odd GPDs, $h_1$, and the moments of the chiral odd TMDs representing their forward  limits.

\begin{figure}
\includegraphics[width=9.cm]{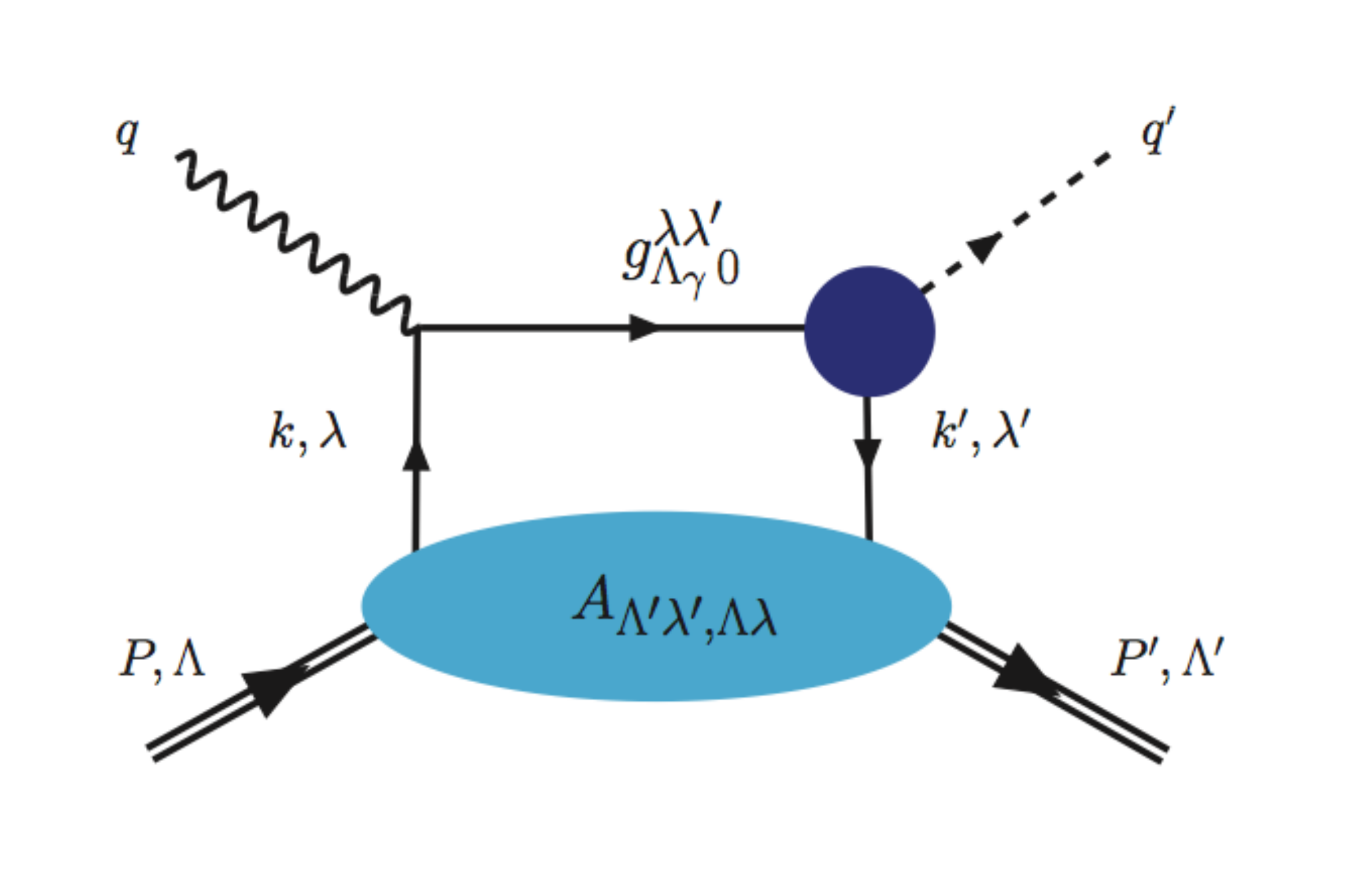}
\caption{Leading order amplitude for the DVCS/DVMP processes described in the text. Crossed diagrams are not shown in the figure.}
\label{fig1}
\end{figure}

Deeply Virtual Compton Scattering (DVCS)  and Deeply Virtual Meson Production (DVMP)  can be described within QCD factorization,
through the convolution of specific Generalized Parton Distributions (GPDs) and hard scattering amplitudes.  
In DVCS and DVMP processes where no net helicity transfer occurs, one identifies four chiral-even GPDs, $H, E, \widetilde{H}, \widetilde{E}$ \cite{Ji_even}. 
Four additional chiral-odd GPDs are known to exist by considering twist-two quark operators that flip the net helicity by one unit, 
$H_T, E_T, \widetilde{H}_T, \widetilde{E}_T$ \cite{Ji_odd,Diehl_odd}.  
All GPDs depend on two additional kinematical invariants besides the parton's  Light Cone (LC) momentum fraction, $x$, and the DVCS process' four-momentum transfer, $Q^2$,
namely $t=\Delta^2$ where $\Delta=P-P'$ is the momentum transfer between the initial and final protons, and $\xi$, or the fraction of LC momentum transfer, $\xi=\Delta^+/(P^+ + P'^+)$ 
(Figure \ref{fig1}). 
The observables containing the various GPDs are the so-called Compton Form Factors (CFFs), which are convolutions over $x$ of GPDs with the struck parton propagator. 
The CFFs are complex quantities  that depend on ($\xi \approx x_{Bj}/(2-x_{Bj})$, $t$, and $Q^2$).    
In the forward limit defined as: $t \rightarrow 0$, $\xi \rightarrow 0$,  the spin conserving GPDs, $H(x,0,0; Q^2)$, $\widetilde{H}(x,0,0; Q^2)$, and $H_T(x,0,0; Q^2)$ become the PDFs,
$f_1(x,Q^2)$, $g_1(x,Q^2)$, and $h_1(x,Q^2)$, respectively.     

In  Ref.\cite{AGL},  after showing how  DV$\pi^o$P can be described in terms of chiral-odd GPDs, we estimated all of their contributions to the various observables with particular attention to the ones which were sensitive to the values of the tensor charge. A sound, fully quantitative model/parametrization for chiral odd GPDs was however missing. 
In this paper we present such a  model. We consider an extension of the reggeized diquark model which was already discussed in detail in the chiral even sector in Refs.\cite{GGL,newFF}. 
Differently from the chiral even case where the GPDs integrate to the nucleon  form factors,  and $H$ and $\widetilde{H}$ have the PDFs $f_1$ and $g_1$ as their forward limits, very little can be surmised on the size/normalization, and on the 
$t$ and $x$ dependences of the chiral odd GPDs. Few constraints  from phenomenology exist, namely $H_T$ becomes the transversity structure function, $h_1$, in the forward limit, and it integrates to the still unknown tensor charge; the first moment of $2\widetilde{H}_T+E_T$ can be interpreted as  the proton's transverse anomalous magnetic moment \cite{Bur2},  and $\widetilde{E}_T$'s first moment is null \cite{Diehl_odd,DieHag_odd}. 

Our approach allows us to overcome this problem and to estimate more precisely the size of all the chiral odd GPDs since, owing to the Parity and Charge Conjugation symmetries obeyed by the various helicity structures in the reggeized diquark model,  we can write approximate relations between the  chiral-even and  
chiral-odd GPDs. 

Although the ultimate goal is to determine the chiral odd GPDs from a global analysis on its own merit using all of the pseudo-scalar meson production data, a necessary intermediate step is to gauge the various contributions to the cross sections and asymmetries. This paper aims to be a step in this direction.

A confirmation of the validity of our approach can be seen in that using our method we obtain as a side result an estimate of $h_1$ for the $u$ and $d$ quarks which is in line with current experimental extractions.

A debatable question, and one that perhaps spurred an additional analysis in Ref.\cite{GolKro,Kro_new},  is how to treat the $\pi^o$ production vertex. 
This is an important issue since the $Q^2$ dependence of DV$\pi^o$P largely depends on the description of the process $\gamma^* (q \bar{q}) \rightarrow \pi^o$, and experimental evidence
to date shows disagreement with theoretical predictions put forth prior to Ref.\cite{AGL}.     
Within a standard collinear factorization scheme it was initially proposed that: {\it i)} factorization in DVMP works rigorously for longitudinal virtual photon polarization \cite{ColFraStr}, the transverse polarization case being yet unproven; {\it ii)}  the only coupling that survives at the pion vertex in the large $Q^2$ limit is of the type $\gamma_\mu \gamma_5$, the other possible term $\propto \gamma_5 P$, being suppressed. The resulting amplitudes were written in terms of the chiral even GPDs, $\widetilde{H}$ and $\widetilde{E}$. 

In Ref.\cite{AGL} we took a different approach. We first of all assumed a form of factorization working for both longitudinal and transverse virtual photons. 
Factorization for transverse polarization has in fact not been disproven although a dedicated proof is missing. 
\footnote{Leading order factorization in the transverse channel is also supported by a duality argument for the GPDs Regge term proposed in Ref.\cite{newFF}.}
 We then proposed an alternative model to the standard
one gluon exchange model first adopted in DV$\pi^o$P in Ref.\cite{ManPil,MankWeigl,VdH}, in connection with collinear factorization for longitudinal virtual photons.       
In our model outlined in Ref.\cite{AGL}, the  form factor  of the out going pseudoscalar meson depends 
on the $J^{PC}$ quantum numbers in the $t$-channel. These were first introduced in the description of deeply virtual exclusive processes in Refs.\cite{ChenJi}  and \cite{Hagler} for the chiral even and chiral odd cases,
respectively (for a detailed discussion of $J^{PC}$  in deeply virtual exclusive processes see {\it e.g.}  \cite{Hagler}). $J^{PC}$ quantum numbers provide a way of counting the number of generalized form factors contributing to the hadronic tensor.
In Ref.\cite{AGL} we noticed that for pseudoscalar electroproduction one has at leading order $J^{PC} \equiv 1^{--}, 1^{+-}$, corresponding to either vector (V) or axial-vector  (A) fermion anti-fermion pairs. This, in turn, corresponds
to $^{2J+1}L_S \equiv ^3S_1,^1P_0$.
The transition from $\gamma^* (q\bar{q})$ into $\pi^o$ ($J^{PC} \equiv 0^{+-}$), therefore corresponds to a change of Orbital Angular Momentum (OAM), $\Delta L = 0$ for the vector case,  and  $\Delta L = 1$ for the axial-vector.
Our idea is to introduce orbital angular momentum in the calculation of the one gluon exchange mechanism for the transition form factor by using a technique similar to the one first introduced in \cite{BelJiYuan_F2} (see also \cite{Rad_new}).  By doing so we describe the pion vertex with two form factors,  an axial vector type, $F_A(Q^2)$,  suppressed by ${\cal O}(1/Q^2)$ with respect to the vector one, $F_V(Q^2)$. The two form factors enter the helicity amplitudes for the various processes in different combinations. This gives rise to a more articulated form of the $Q^2$ dependence, which is more flexible and apt to describe the features of the data than the standard one. In particular we can now understand and reproduce the persistence of a large transverse component in the multi-GeV region.      

Our paper is organized as follows:
in Section \ref{sec:2} we present our formalism and we outline the derivation of the helicity amplitudes entering the cross section 
for DV$\pi^o$P, including both chiral even and chiral odd contributions;  in Section \ref{sec:model} we present our model relating the chiral even to chiral odd GPDs;   in Section \ref{sec:xsec} we present our results for the various observables; finally in Section \ref{sec:conclusions} we draw our conclusions.

\section{Formalism}
\label{sec:2}
We start by defining GPDs at twist-two as the matrix elements of  the following projection of the unintegrated quark-quark proton correlator (see Ref.\cite{Metz} for a detailed overview),
\footnote{In what follows we can omit the Wilson gauge link without loss of generality \cite{Ji1}.}  
\begin{eqnarray}
W_{\Lambda', \Lambda}^\Gamma(x,\xi,t) & = & \frac{1}{2} \int \frac{d z^- }{2 \pi} e^{ix\overline{P}^+ z^-} \left. \langle P', \Lambda' \mid \overline{\psi}\left(-\frac{z}{2}\right) \Gamma \, \psi\left(\frac{z}{2}\right)\mid P, \Lambda \rangle \right|_{z^+=0,{\bf z}_T=0},
\label{matrix}
\end{eqnarray}
where $\Gamma=\gamma^+, \gamma^+\gamma_5, i\sigma^{i+}\gamma_5 (i=1,2)$, and the target's spins are $\Lambda, \Lambda^\prime$. For the chiral odd case, $\Gamma= i\sigma^{i+}\gamma_5$, $W_{\Lambda', \Lambda}^\Gamma$ was parametrized as \cite{Diehl_odd},
\begin{eqnarray}
\label{correlator}
W_{\Lambda', \Lambda}^{[i\sigma^{i+}\gamma_5]}(x,\xi,t)  & = &  \frac{1}{2\overline{P}^+} \overline{U}(P',\Lambda') \left( i \sigma^{+i} H_T(x,\xi,t) +
 \frac{\gamma^+ \Delta^i - \Delta^+ \gamma^i}{2M} E_T(x,\xi,t)   \right. \nonumber \\
& + & \left.  \frac{P^+ \Delta^i - \Delta^+ P^i}{M^2}  \widetilde{H}_T(x,\xi,t)  +
\frac{\gamma^+ P^i - P^+ \gamma^i}{2M} \widetilde{E}_T(x,\xi,t) \right) U(P,\Lambda)
\end{eqnarray}
As we will show below, the spin structures of GPDs that are directly related to spin dependent observables are most effectively expressed in term of helicity amplitudes, developed extensively for the covariant description of two body scattering processes 
(for a detailed description of the helicity amplitudes formalism in deeply virtual scattering processes see also Ref.\cite{Diehl_hab}). Before proceeding with the helicity amplitudes we introduce the kinematics.
\subsection{Kinematics}
\label{sec:basic}
The correlator in Eqs.(\ref{matrix},\ref{correlator}) is expressed in terms of kinematical variables defined in the ``symmetric frame", where we define: $\overline{P}=(P+P')/2$, the average proton momentum, and $\Delta = P-P'$. $\overline{P}$  is along the $z$-axis with momentum, $\overline{P}_3 \approx \overline{P}^+$.  
The four-momenta LC components ($v \equiv (v^+,v^-,\vec{v}_T)$, where $v^\pm=1/\sqrt{2}(v_o \pm v_3)$) are:

\noindent \underline{\em Symmetric}
\begin{subequations}
\label{kin:sym}
\begin{eqnarray}
\overline{P} & \equiv & \left( \overline{P}^+, \frac{M^2}{\overline{P}^+}, 0 \right) \nonumber  \\
\Delta & \equiv &  \left( \xi \, (2 \overline{P}^+), \frac{ t+ {\bf \Delta}_T^2}{2 \xi \overline{P}^+},  {\bf \Delta}_T  \right) \\
P & \equiv &   \left((1+\xi) \overline{P}^+,  \frac{M^2+  {\bf \Delta}_T^2/4}{(1+\xi)\overline{P}^+},   {\bf \Delta}_T/2 \right) \nonumber  \\
P' & \equiv &  \left( (1-\xi) \overline{P}^+,  \frac{M^2+  {\bf \Delta}_T^2/4}{(1-\xi)\overline{P}^+},- {\bf \Delta}_T/2  \right) \nonumber  
\label{coord_asym}
\end{eqnarray}
\end{subequations}
The coordinates of the off-shell struck parton are,
\begin{subequations}
\begin{eqnarray}
k & \equiv &  \left( (x+\xi)\overline{P}^+,  k^-, {\bf k}_T +  {\bf \Delta}_T /2 \right),  \nonumber \\
k' & \equiv &  \left( (x-\xi)\overline{P}^+,  k'^-,{\bf k}_T - {\bf \Delta}_T/2   \right)  
\end{eqnarray}
\end{subequations}
Another choice of frame is the ``asymmetric frame", where $\overline{P}$ is longitudinal:

\noindent \underline{\em Asymmetric}
\begin{subequations}
\label{kin:asym}
\begin{eqnarray}
P & \equiv & \left(P^+, \frac{M^2}{P^+},0 \right) \nonumber  \\
P' & \equiv &  \left( (1-\zeta)P^+, \frac{M^2+ \vec{\Delta}_T^2}{(1-\zeta)P^+},- \vec{\Delta}_T \right) \nonumber  \\
\overline{P} & \equiv & \left( \left(1-\zeta/2 \right) P^+, \frac{\left(1- \zeta/2 \right) M^2+ {\bf \Delta}_T^2/2}{(1-\zeta)P^+}, - {\bf \Delta}_T  \right) \nonumber  \\
\Delta & \equiv &  \left( \zeta P^+, \frac{\left(1- \zeta/2 \right) M^2+ {\bf \Delta}_T^2/2}{(1-\zeta)P^+},  {\bf \Delta}_T  \right) 
\label{coord_asym}
\end{eqnarray}
\end{subequations}
and, 
\begin{subequations}
\begin{eqnarray}
k & \equiv & \left(XP^+,  k^-, {\bf k}_T \right),  \nonumber \\
k' & \equiv &  \left( (X-\zeta)P^+,  k'^-,{\bf k}_T - {\bf \Delta}_T \right)  
\end{eqnarray}
\end{subequations}
$\displaystyle t=  t_o - \Delta_\perp^2/(1-\zeta)$, $t_o= - \zeta^2 M^2/(1-\zeta)$.  

The two frames are entirely equivalent and one can connect from one another with simple transformations. We find the asymmetric frame more useful when referring to the partonic picture,  while the symmetric frame is more convenient for symmetry transformations and sum rules derivations. In this paper we will use either notation according to these criteria. 

Other useful variables are: 
\[ \hat{s} = (k+q)^2 \approx - Q^2(X-\zeta)/\zeta , \;\;\;\;  \hat{u} = (k^\prime -q)^2 \approx Q^2 X/\zeta, \;\;\;\; q^- \approx(Pq)/P^+ = Q^2/(2\zeta P^+). \]
The loop diagram in Fig.\ref{fig1} integrated over the struck quark's momentum is performed using the variables: $d^4 k \equiv d k^+ d k^- d^2 k_\perp \equiv P^+ dX dk^- d^2 k_\perp$.  
\subsection{Helicity Amplitudes}
\label{sec:helamp}
The connection of the correlator, Eq.(\ref{correlator}), with the helicity amplitudes proceeds by introducing \cite{AGL,GGL}, 
\begin{eqnarray}
f_{\Lambda_\gamma 0}^{\Lambda \Lambda^\prime} (\zeta,t)& = & \sum_{\lambda,\lambda^\prime} 
g_{\Lambda_\gamma 0}^{\lambda \lambda^\prime} (X,\zeta,t,Q^2)  \otimes
A_{\Lambda^\prime \lambda^\prime, \Lambda \lambda}(X,\zeta,t), 
\label{facto}
\end{eqnarray}
where the helicities of the virtual photon and the initial proton are, $\Lambda_\gamma$, $\Lambda$, 
and the helicities of  the produced pion and final proton are $0$, and $\Lambda^\prime$, respectively.
Factorization theorems at large $Q^2$ have been proven strictly for the process $\gamma_L^* p \rightarrow  M p$. 
Large transverse photon polarization contributions have been however observed in the experimental data. In a previous publication 
a possible scenario beyond collinear factorization was introduced for $\gamma_T^* p \rightarrow M p$ which involves directly the chiral odd GPDs. 
In Eq.(\ref{facto})  we describe the factorization into a ``hard part'', 
$g_{\Lambda_\gamma 0}^{\lambda  \lambda^\prime}$ for the partonic subprocess 
$\gamma^* + q \rightarrow \pi^0 + q$, 
and a ``soft part'' given by the quark-proton helicity amplitudes, $A_{\Lambda^\prime,\lambda^\prime;\Lambda,\lambda}$ 
that contain the GPDs.

\subsubsection{Chiral Odd Quark-Proton Helicity Amplitudes, $A_{\Lambda^\prime \lambda^\prime, \Lambda \lambda}$}
The amplitudes $A_{\Lambda^\prime \lambda^\prime, \Lambda \lambda}$ implicitly contain an integration over the unobserved quark's transerve momentum, $k_T$,
and are functions of 
$x_{Bj} =Q^2/2M\nu \approx \zeta, t$ and $Q^2$. The convolution integral in  Eq.(\ref{facto}) 
 is given by $\otimes \rightarrow \int_{-\zeta+1}^1 d X$, as we explain in detail later on.

The connection with the correlator is carried out by considering,
\begin{eqnarray}
A_{\Lambda' \lambda', \Lambda \lambda} = \int \frac{d z^-}{2 \pi} e^{ix\overline{P}^+ z^-} \left. \langle P', \Lambda' \mid {\cal O}_{\lambda' \lambda}(z) \mid P, \Lambda \rangle \right|_{z^+=0, {\bf z}_T=0}, 
\end{eqnarray} 
where, 
\begin{eqnarray}
{\cal O}_{-+}(z) & = & -i \bar{\psi}\left(-\frac{z}{2}\right) (\sigma^{+1} - i \sigma^{+2})  \psi\left(\frac{z}{2}\right) \\
{\cal O}_{+ -}(z) & = & i \bar{\psi}\left(-\frac{z}{2}\right) (\sigma^{+1} + i \sigma^{+2}) \psi\left(\frac{z}{2}\right), 
\end{eqnarray}
By taking this into account in Eq.(\ref{correlator}), and by adding and subtracting the expressions corresponding to $i=1,2$, respectively, one obtains the expressions for  the chiral odd helicity amplitudes in terms of 
GPDs   \cite{Diehl_odd,Diehl_hab},
\begin{widetext}
\begin{subequations}
\label{GPDodd}
\begin{eqnarray}
A_{++,--} & = &  \sqrt{1-\xi^2}  \left[ { H}_ T + \frac{t_0-t}{4M^2} \widetilde{ H}_T      
- \frac{\xi^2}{1-\xi^2}  { E}_T  + \frac{\xi}{1-\xi^2} \widetilde{ E}_T \right]  \nonumber \\ 
& = & 
\frac{\sqrt{1-\zeta}}{1-\zeta/2} \left[ H_ T + \frac{t_0-t}{4M^2} \widetilde{H}_T  - \frac{\zeta^2/4}{1-\zeta} E_T  + 
\frac{\zeta/2}{1-\zeta} \widetilde{E}_T \right]  \\
A_{+-,-+} & = &  - e^{i2 \varphi}  \sqrt{1-\xi^2}  \,  \frac{t_0-t}{4M^2} \, \widetilde{ H}_T 
 = 
- e^{i2 \varphi} \frac{\sqrt{1-\zeta}}{1-\zeta/2} \, \, \frac{t_0-t}{4M^2} \, \widetilde{H}_T   \\
A_{++,+-} & = & e^{i \varphi} \frac{\sqrt{t_0-t}}{4M} \left[ 2\widetilde{ H}_T  + (1-\xi)  \left({ E}_T + \widetilde{ E}_T \right) \right] \nonumber \\ 
& = & 
 e^{i \varphi} \frac{\sqrt{t_0-t}}{2M} \,    \left[ \widetilde{H}_T + \frac{1-\zeta}{1-\zeta/2}\left(E_T  + \widetilde{E}_T \right) \right],  \\
A_{-+,--} & = & e^{i \varphi} \frac{\sqrt{t_0-t}}{4M}  \,  \left[  2\widetilde{ H}_ T + (1+\xi) \left( { E}_T - \widetilde{ E}_T \right) \right] \nonumber \\ & = &
e^{i \varphi}  \frac{\sqrt{t_0-t}}{2M} \,   \left[ \widetilde{H}_ T +  \frac{1}{1-\zeta/2} \left(E_T - \widetilde{E}_T \right) \right] .   
\end{eqnarray}
\end{subequations}
where the first (second) line in each equation uses the symmetric (asymmetric) notation, $\phi$ is a phase given by the azymuthal angle of the vector {\bf D} with length $\mid {\bf D} \mid =\sqrt{t_o-t}/(1-\xi^2)$, \cite{Diehl_hab} and we have used the relations, 
\begin{subequations}
\begin{eqnarray}
\frac{1}{2 \overline{P}^+} \overline{U}(P^\prime,\Lambda^\prime) i \sigma^{+i}  U(P,\Lambda) 
& = &  f  \;    (\Lambda  \delta_{i1} + i \delta_{i2} ) \delta_{\Lambda,-\Lambda'}  \\ 
 \overline{U}(P^\prime,\Lambda^\prime) \frac{\overline{P}^+ \Delta^i - \Delta^+ \overline{P}^i}{M^2}   U(P,\Lambda)   
& = &  
\frac{1}{\sqrt{1-\zeta}} \, \left[   (1-\zeta/2) \, \frac{\Delta_i}{M} \delta_{\Lambda, \Lambda'}  + (\Lambda \Delta_1+ i  \Delta_2) \frac{\Delta_i}{2M^2} \delta_{\Lambda,- \Lambda'}  \right]  \\
   \overline{U}(P^\prime,\Lambda^\prime)   \frac{\gamma^+ \Delta^i - \Delta^+ \gamma^i}{2M}  U(P,\Lambda)
  &=&
f \left[\left( \frac{\Delta_i}{2M} + i \Lambda \,  \frac{\zeta/2}{1-\zeta/2} \, \epsilon^{03ji} \frac{\Delta_j}{2M} \right)  \delta_{\Lambda \Lambda'} \right. 
\left. +  
\frac{\zeta^2}{4} (\Lambda \delta_{i1} + i \delta_{i2}) \delta_{\Lambda,- \Lambda'}  \right ] 
 \\
 \overline{U}(P^\prime,\Lambda^\prime)   \frac{\gamma^+ \overline{P}^i - \overline{P}^+ \gamma^i}{M}  U(P,\Lambda)    
& =  &
 f \left[ - \left( \frac{\zeta/2}{1-\zeta/2} \frac{\Delta_i}{2M} + i \, \Lambda \, \epsilon^{03ji} \frac{\Delta_j}{2M} \right) \delta_{\Lambda \Lambda'}  \right.  
 \left. +
\frac{\zeta}{2} \left(1-\frac{\zeta}{2}\right)( \Lambda \delta_{i1} + i \delta_{i2} ) \delta_{\Lambda,- \Lambda'}  \right ]  \nonumber \\
\end{eqnarray}
\end{subequations}
\end{widetext}
with,
$f =  \sqrt{1-\zeta}/(1-\zeta/2) $.

\subsubsection{Hard Process Helicity Amplitudes, $g_{\Lambda_\gamma 0}^{\lambda \lambda^\prime}$}
\label{sec:g}
\begin{figure}
\includegraphics[width=9.cm]{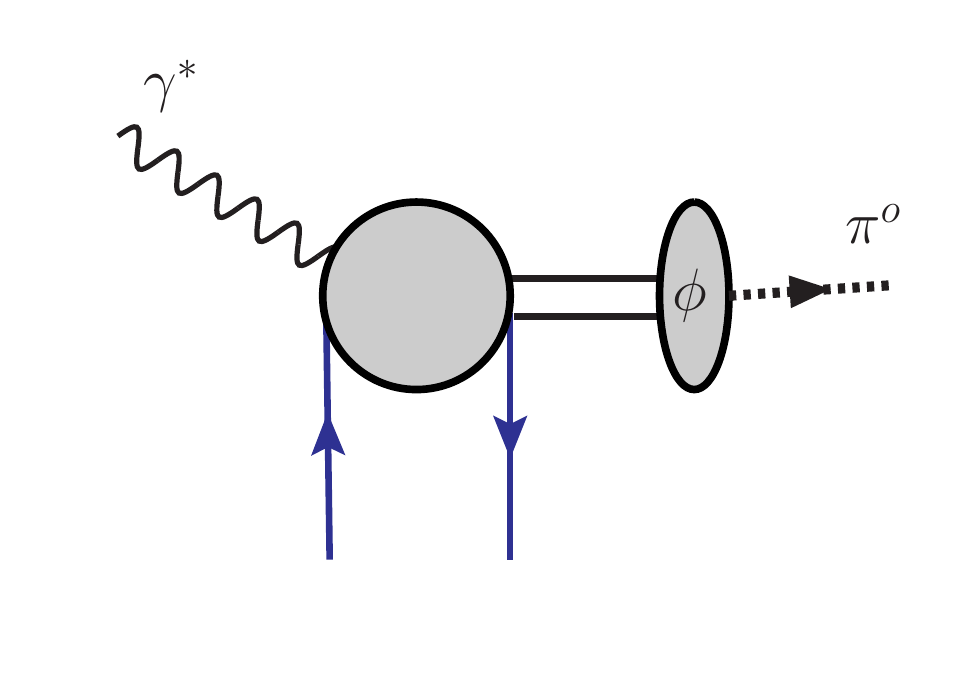}
\caption{Hard scattering contribution, $\gamma^* q \rightarrow \pi^o q'$; $\phi$ is the outgoing pseudoscalar meson distribution amplitude.}
\label{fig:g}
\end{figure}

The subprocess $\gamma^* q \rightarrow \pi^o q'$ is shown in Figure \ref{fig:g} For chiral-odd coupling at the pion vertex it is given by,
\begin{eqnarray} 
g_{\Lambda_\gamma 0}^{\lambda  \lambda^\prime \; (odd)} & =  & g_\pi^{V(A), odd}(Q^2)   \,  q^- 
\left[ \bar{u}(k^\prime,\lambda^\prime) \gamma^\mu \gamma^+  \gamma_5 u(k,\lambda) \right]  
 \epsilon_\mu^{\Lambda_\gamma}   \left( \frac{1}{\hat{s} - i \epsilon } - \frac{1}{\hat{u} - i \epsilon} \right).
\label{g_odd}
\end{eqnarray}
where the specific $Q^2$ dependence of the form factor $g_\pi^{V(A), odd}(Q^2)$ is discussed in Ref.\cite{AGL} and in Appendix \ref{appa}.

\begin{figure}
\includegraphics[width=9.cm]{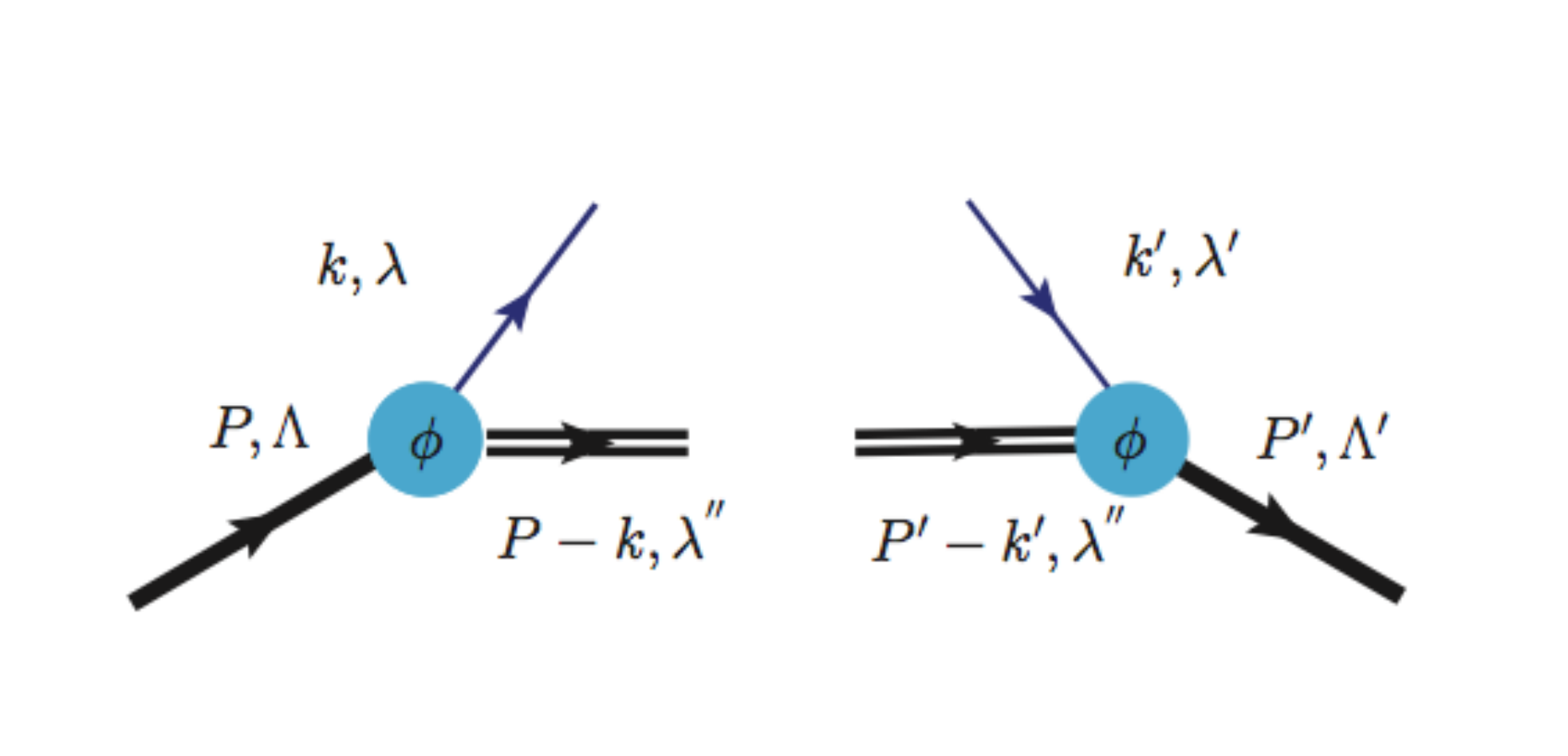}
\caption{Vertex structures defining the spectator model tree level diagrams.  
}
\label{fig_LCWF}
\end{figure}

By using the relation
\begin{eqnarray}
\bar{u}(k^\prime,\lambda^\prime) \gamma^\mu \gamma^+  \gamma_5 u(k,\lambda)    =  
 N N^\prime \,  {\rm Tr} 
\left\{ (\!\not{k}  +m) \, \hat{\mathcal O}_{\lambda,\lambda^\prime} (\!\not{k}^\prime + m) \gamma^\mu \gamma_5 \gamma^+  \right\} 
\end{eqnarray}
where $N = 1/\sqrt{P^+(X-\zeta)}$ and $N^\prime=1/\sqrt{XP^+}$ are the quark spinors normalizations, and,
\begin{subequations}
\begin{eqnarray}
\hat{\mathcal O}_{\pm \pm } = \frac{1}{4}(1+\gamma^o) (1\pm \gamma_5 \gamma_3) && \\ 
 \hat{\mathcal O}_{\pm \mp}  = -\frac{1}{4}(1+\gamma^o) \gamma_5(\gamma_1 \mp i\gamma_2) && 
 \end{eqnarray}
\end{subequations}
we see that for transverse photon polarization, in the $Q^2 \rightarrow \infty$ limit, the only contributing amplitude is 
$g_{10}^{+-}$,
\begin{eqnarray}
\label{ampg2}
g_{10}^{+-}  & \approx & -\frac{K}{\sqrt{k^{\prime \, +} k^+}} \, \left[ k^o p^{\prime \, +} -(k k^\prime) + k^+ k^{\prime \, o} \right] (\epsilon_1^{+1} - i \epsilon_2^{+1})  
\end{eqnarray}
By evaluating $K=q^-(1/\hat{s}-1/\hat{u}) =  (Q^2/2\zeta P^+) \, (\zeta/Q^2 C^+) \equiv 1/(2P^+) C^+$, where
\begin{equation}
C^+(X,\zeta) =   \frac{1}{X- \zeta + i \epsilon }  + \frac{1}{X - i \epsilon }
\end{equation}
 one has,
\begin{equation}
g_{10}^{+-} = \sqrt{X (X-\zeta)}  \: C^+
\end{equation} 

For longitudinal photon polarization,
\begin{subequations}
\begin{eqnarray}
g_{00}^{+-}    
& \approx & \frac{K}{\sqrt{k^{\prime \, o} k^o}} \, \left[ (k^o k^{\prime \, 1} -  k^1 k^{\prime \, o} )\epsilon_+^0 + 
(k^1 k^{\prime \, +} + k^1 k^{\prime \, +} ) \epsilon_o^0  \right. 
 -  \left.  i ( 1 \leftrightarrow 2) \right]
\end{eqnarray}
\label{ampg5}
\end{subequations}
Similarly to Eq.(\ref{ampg2}) the last line was obtained in the $Q^2 \rightarrow \infty$ limit, considering the dominant LC components. 

By inserting all kinematical components, we have
\begin{eqnarray}
g_{00}^{+-} & = &   \frac{(k^\prime_1 - i k^\prime_2) X P^+}{\sqrt{k^o k^{\prime \, o} }}  \frac{2\nu}{\sqrt{Q^2}}  \, \frac{1}{2P^+} C^+  
\label{g5final1}
\end{eqnarray}
where we defined $P^+ = (Pq)/q^- = Q^2/2M\zeta^2$, and $\nu=Q^2/2M\zeta$. 
Notice that the energy term from the spinor normalization, $k^{\prime \, o}$ needs to be evaluated using its full form, $k^{\prime \, o} =  \sqrt{k^{\prime \, 2}_\perp +  k^{\prime \, 2}_3}= \sqrt{k^{\prime \, 2}_\perp +  (X-\zeta)^2P^{+ \, 2}}$, in order to avoid a singularity for $X=\zeta$ (see Appendix \ref{appb}). 
As a result, 
%
\begin{eqnarray}
g_{00}^{+-} = 
\displaystyle \sqrt{\frac{t_o-t}{Q^2}}  \frac{\zeta \sqrt{X}}{\sqrt{\left((X-\zeta)^2 + \langle k_\perp^{\prime 2} \rangle/(Q^2/2 M \zeta^2)\right)^{1/2}} } \; C^+ 
\label{g5final2}
\end{eqnarray}
\subsubsection{Convolution}

The convolution in Eq.(\ref{facto}) yields the following decomposition of the various helicity amplitudes,
\footnote{Notice the alternative use of notation in Ref. \cite{AGL} and references therein, namely, 
$f_1 = f_{1 0}^{++}, \, f_2 = f_{1 0}^{+-}, \,f_3 = f_{1 0}^{-+},$
$f_4 = f_{1 0}^{- -}, \,f_5 = f_{0 0}^{+-}, \, f_6 = f_{0 0}^{++}$.}
\begin{subequations}
\begin{eqnarray}
f_{10}^{++} &  = & g_{10}^{+-} \otimes A_{+-,++}  \\
f_{10}^{+-} &  = &  g_{10}^{+-} \otimes  A_{--,++}     \\
f_{10}^{-+} & = &  g_{10}^{+-} \otimes  A_{+-,-+}    \\  
f_{10}^{--} & = &  g_{10}^{+-}  \otimes  A_{++,+-}  \\ 
f_{00}^{+-} &  = & g_{00}^{+-}   \otimes  (A_{--,++}  - A_{+-,-+})    \\
f_{00}^{++} &  = & g_{00}^{+-}  \otimes ( A_{++,+-} -A_{+-,++}) , 
\end{eqnarray}
\label{chiral_odd}
\end{subequations}
where the four chiral odd  quark-proton helicity amplitudes, $A_{\Lambda^\prime,\, -\lambda;\Lambda,\lambda}$, enter. Notice that $A_{+-,++}$, $A_{++,+-}$ change sign under Parity while $A_{--,++} $, $A_{+-,-+} $, do not change sign; since $g_{10}^{+-}$ also changes sign, then $f_{10}^{++}$,  $f_{10}^{--}$ will not change sign under Parity, while   $f_{10}^{+-} $, and$ f_{10}^{+-} $ will change sign.
%

For a transverse photon, inserting the expressions for $g_{10}^{+-}$ and the $A$'s into Eqs.(\ref{chiral_odd}) we obtain,
\begin{subequations}
\label{helamps_gpd2}
\begin{eqnarray}
f_{10}^{++} &= &  g_\pi^{V, odd}(Q)  e^{ i\varphi}  \frac{\sqrt{t_0-t}}{4M} \left[ 2\widetilde{ \cal H}_T  + (1+\xi)  \left({\cal  E}_T - \widetilde{\cal  E}_T \right) \right] \nonumber \\
& = & g_\pi^{V, odd}(Q) e^{ i\varphi} \frac{\sqrt{t_0-t}}{2M} \,    \left[ \widetilde{\cal H}_T + \frac{1}{1-\zeta/2}\left({\cal E}_T  - \widetilde{\cal E}_T \right) \right], 
\\
f_{10}^{+-} & = &   \frac{g_\pi^{V, odd}(Q)+ g_\pi^{A, odd}(Q)}{2} \, \sqrt{1-\xi^2}  \left[ { \cal H}_ T + \frac{t_0-t}{4M^2} \widetilde{ \cal H}_T      
- \frac{\xi^2}{1-\xi^2}  {\cal  E}_T  + \frac{\xi}{1-\xi^2} \widetilde{\cal  E}_T \right]  \nonumber \\ 
& = &  \frac{g_\pi^{V, odd}(Q)+ g_\pi^{A, odd}(Q)}{2}  \, \frac{\sqrt{1-\zeta}}{1-\zeta/2} \left[ {\cal H}_ T + \frac{t_0-t}{4M^2} \widetilde{\cal H}_T  -  \frac{\zeta^2/4}{1-\zeta} {\cal E}_T  + 
\frac{\zeta/2}{1-\zeta} \widetilde{ \cal E}_T \right]   \\
 f_{10}^{-+} & = &  -   \frac{g_\pi^{A, odd}(Q)- g_\pi^{V, odd}(Q)}{2}  \, e^{ i 2 \varphi}  \sqrt{1-\xi^2}  \,  \frac{t_0-t}{4M^2} \, \widetilde{\cal  H}_T \nonumber \\
&   = & -  \frac{g_\pi^{A, odd}(Q)- g_\pi^{V, odd}(Q)}{2} \,   e^{ i 2 \varphi}  \frac{\sqrt{1-\zeta}}{1-\zeta/2} \, \, \frac{t_0-t}{4M^2} \, \widetilde{\cal H}_T  
 \\
f_{10}^{--} & = & g_\pi^{V, odd}(Q) e^{ i\varphi} \frac{\sqrt{t_0-t}}{4M}  \,  \left[  2\widetilde{\cal H}_ T + (1-\xi) \left( { \cal E}_T + \widetilde{\cal E}_T \right) \right]  \nonumber \\ 
& = & g_\pi^{V, odd}(Q) \,  e^{ i\varphi} \frac{\sqrt{t_0-t}}{2M} \,   \left[ \widetilde{\cal H}_ T +  \frac{1-\zeta}{1-\zeta/2}\left( {\cal E}_T + \widetilde{\cal E}_T \right) \right]
\end{eqnarray} 
\end{subequations}
where ${\cal H}_T$, etc., are the convolutions of the GPDs with $C^+(X,\zeta)$, or the Compton form factors which at leading order in PQCD are given by,  
\begin{equation}
\label{CFF_def}
{\cal F}_T(\zeta,t,Q^2) = \int_{-1+\zeta}^1 dX \; C^+ \, F_T(X,\zeta ,t,Q^2) 
\end{equation}
$F_T \equiv {\cal H}_T, {\cal E}_T, \widetilde{\cal H}_ T, \widetilde{\cal E}_ T$.  
$g_\pi^{V, odd}(Q)$, and  $g_\pi^{A, odd}(Q)$ are given in Appendix \ref{appd}. 
These two distinct contributions arise owing to the fact that, as observed in Ref.~\cite{AGL},  the chiral odd coupling $\propto \gamma^5$  contributes to $\pi^o$ electroproduction provided one goes beyond a simple one gluon exchange description of this vertex.    
In the $t$-channel picture, which has its roots in a Regge analysis of this process \cite{GolOwe}, one separates 
the $J^{PC}=1^{- -}$ and $J^{PC}=1^{+ -}$
contributions to the amplitudes for transverse and longitudinal virtual photons, respectively, thus generating two different types of $Q^2$ dependence at the $\pi^o$ 
vertex.
The result of our analysis is that, differently from other treatments of pion electroproduction~\cite{ManPil,MankWeigl,VdH}, relying solely on chiral even GPDs, $\widetilde{H}, \, \, \widetilde{E}$, the chiral-odd sector provides the dominant contribution. 

For a longitudinal photon one has the convolution of $g_{00}^{+-}$ with the $A$ helicity amplitudes, 
\begin{subequations}
\label{helamps_00}
\begin{eqnarray}
f_{00}^{+-}  & = &    g_\pi^{A, odd}(Q)  \,  \sqrt{1-\xi^2}   \left[ {\cal H}_ T    
 + \frac{\xi^2}{1-\xi^2}  {\cal E}_T  + \frac{\xi}{1-\xi^2} \widetilde{\cal E}_T \right]   \\ 
f_{00}^{++}  & = &  - g_\pi^{A, odd}(Q) \,  \frac{\sqrt{t_0-t}}{2M}  \,  \left[  \xi  {\cal E}_T +
 \widetilde{\cal E}_T  \right].
\end{eqnarray} 
\end{subequations}
(expressions are simpler in this case in symmetric notation).

\subsection{Chiral Even Sector}
\label{even:sec}
For completeness we also show results in the the chiral even sector. The hard scattering amplitude reads,
\begin{eqnarray}
\label{g_even}
g_{\Lambda_\gamma 0}^{\lambda \lambda \; (even)} &  =  & g_\pi^{even}(Q^2) e_q \, q^- 
\left[ \bar{u}(k^\prime,\lambda) \gamma^\mu \gamma^+ \gamma^\nu \gamma_5 u(k,\lambda) \right]  \nonumber \\ 
& \times &  \epsilon_\mu^{\Lambda_\gamma}  \, q^\prime_\nu \left( \frac{1}{\hat{s} - i \epsilon } +\frac{1} {\hat{u} -i \epsilon} \right) \nonumber \\
& \approx & g_\pi^{even}(Q^2) e_q \, \sqrt{X(X-\zeta)}  \: C^-. 
\end{eqnarray}
while the quark-proton helicity amplitudes are given by \cite{Diehl_hab},
\vspace{1cm}
\begin{subequations}
\label{GPDeven}
\begin{eqnarray}
A_{++,++}  & = & \frac{\sqrt{1-\zeta}}{1-\zeta/2} \left( \frac{H + \widetilde{H}}{2} - \frac{\zeta^2/4}{1-\zeta}  \frac{E + \widetilde{E}}{2} \right)  
\\
A_{+-,+-}  & = & \frac{\sqrt{1-\zeta}}{1-\zeta/2} \left( \frac{H - \widetilde{H}}{2} - \frac{\zeta^2/4}{1-\zeta} \frac{E - \widetilde{E}}{2} \right) 
\\
A_{++,-+} & = & - e^{- i \varphi} \frac{\sqrt{t_o-t}}{2M} \, \left(E - \frac{\zeta/2}{1-\zeta/2} \widetilde{E} \right) \\
A_{-+,++} & = &  e^{ i \varphi} \frac{\sqrt{t_o-t}}{2M} \, \left(E + \frac{\zeta/2}{1-\zeta/2} \widetilde{E} \right) 
\end{eqnarray}
\end{subequations}
Following from the chiral even case~\cite{GGL}, in $\pi^o$ electroproduction one obtains longitudinal photon amplitudes~\cite{VGG}
\begin{subequations}
\label{f_even_1}
\begin{eqnarray}
f_{00}^{+-, \, even} & = &  \frac{\zeta}{\sqrt{1-\zeta}}  \frac{1}{1-\zeta/2} \frac{\sqrt{t_o-t}}{2M} \: \widetilde{\cal E}, \\
f_{00}^{++, \, even} & =  &  \frac{\sqrt{1-\zeta}}{1-\zeta/2}  \widetilde{\cal H}  +  \frac{-\zeta^2/4}{\left(1-\zeta/2\right)\sqrt{1-\zeta}} \: \widetilde{\cal E},
\end{eqnarray}
\end{subequations}
where $\widetilde{{\cal H}}, \, \, \widetilde{{\cal E}}$ are the corresponding Compton form factors.

\section{Evaluation of Chiral Odd GPDs in Reggeized Diquark Model}
\label{sec:model}
We now present our model for evaluating the chiral odd GPDs. We extend our reggeized diquark model, which was already configured for chiral even GPDs, to the chiral-odd sector.  
Model GPDs must satisfy the so-called polynomiality property which is a consequence of Lorentz invariance of the nucleon matrix elements, and the positivity bound  with respect to the forward limit (see Ref.\cite{Diehl_hab} for a review of these properties). 
While Double Distributions based models satisfy polynomiality by definition \cite{Radyush_pos}, in our model this has to be imposed as a constraint.
As shown in Ref.\cite{GGL,newFF}, our model was therefore constructed in such a way as to satisfy this constraint. As shown in Ref.\cite{GGL} (Figure 8) we verified that the first few moments of the 
chiral even GPDs are polynomials in $\xi^2$. 
We defined our approach as a  ``flexible parametrization" in that, mostly owing to its recursive feature, the different components can be efficiently fitted  separately as new data come in.  
The parameters were initially fixed by a fit applied recursively first to PDFs, and to the nucleon form factors. 
The model was shown to reproduce data on different observables in DVCS (charge \cite{HERMES1,HERMES2}, longitudinal  \cite{HallB} and transverse \cite{HERMES1,HERMES2} single spin asymmetries). Below we summarize the main steps in the model's construction, and we discuss its parameters. Recently \cite{newFF}, we presented a new fit that uses the form factor flavor separated data from Ref.\cite{Cates}. We give the new parameters' values in Appendix \ref{appc}. A more detailed description of the chiral even sector and of the reggeization procedure can be also found in Ref.\cite{newFF}.  

\subsection{Light Cone Wave Functions Definitions}
\label{sec:LCWF}
The basic structures  in our model are the quark-proton scattering amplitudes at leading order with proton-quark-diquark vertices given in Fig.\ref{fig_LCWF}.
The quark parton helicity amplitudes introduced in the previous Section describe a two body process, $q^\prime(k^\prime) P \rightarrow q(k) P^\prime$,
where $q(k)$ corresponds to the ``struck quark" in Fig.\ref{fig1}. We adopt the asymmetric system kinematics, Eqs.(\ref{kin:asym}).
The intermediate diquark system, $X$, can have $J^P=0^+$ (scalar), or $J^P=1^+$ (axial vector). Its invariant mass, $M_X$ varies in our model according to a spectral function, thus generating Regge behavior at large $M_X^2$ \cite{BroCloGun}. We start from the region $X\geq \zeta$. At fixed $M_X$, the amplitudes read:

\vspace{0.3cm}
\noindent\underline{\em Scalar}
\begin{equation} 
\label{AmpS0}
A^{(0)}_{\Lambda^\prime \lambda^\prime, \Lambda \lambda}
  =  \int d^2k_\perp\phi^*_{\Lambda^\prime \lambda^\prime}(k^\prime,P^\prime) \phi_{\Lambda \lambda}(k,P),
\end{equation} 
with vertex structures
\begin{eqnarray} 
\phi_{\Lambda,\lambda}(k,P) & = & \Gamma(k) \frac{\bar{u}(k,\lambda) U(P,\Lambda)}{k^2-m^2} \\
\phi^*_{\Lambda^\prime \lambda^\prime}(k^\prime,P^\prime) & = & \Gamma(k^\prime) \frac{\overline{U}(P^\prime,\Lambda^\prime) u(k^\prime,\lambda^\prime)}{k^{\prime \,2}-m^2}. 
\end{eqnarray}
where we defined the proton-quark-diquark coupling as,
\begin{equation}
\label{coupling}
\Gamma = g_s \frac{k^2-m_q^2}{(k^2- M_\Lambda^{q \, 2})^2}.
\end{equation}
This form is consistentl with predictions from Dyson--Schwinger on the proton quark-diquark vertex (\cite{Roberts2} and references therein).

\noindent\underline{\em Axial vector}
\begin{equation} 
\label{AmpS1}
A^{(1)}_{\Lambda^\prime \lambda^\prime, \Lambda \lambda}
  =  \int d^2k_\perp\phi^{*\mu}_{\Lambda^\prime \lambda^\prime} (k^\prime,P^\prime) \sum_{\lambda^{\prime \prime}}  \epsilon_\mu^{* \, \lambda^{\prime \prime}}  \epsilon_\nu^{\lambda^{\prime \prime}}  \phi_{\Lambda,\lambda}^\nu (k,P),
\end{equation} 
where $\lambda^{\prime \prime}$ is the diquark's helicity, which in our model is taken as transverse only, and  
\begin{eqnarray}
\phi_{\Lambda \lambda}^\nu (k,P) & = & \Gamma(k) \frac{\bar{u}(k,\lambda) \gamma^5 \gamma^\mu U(P,\Lambda)}{k^2-m^2} \\
\phi^{*\nu}_{\Lambda^\prime \lambda^\prime}(k^\prime,P^\prime) & = & \Gamma(k^\prime) \frac{\overline{U}(P^\prime,\Lambda^\prime) \gamma^5 \gamma^\mu u(k^\prime,\lambda^\prime)}{k^{\prime \,2}-m^2}.  
\end{eqnarray}
Notice that the amplitudes, $A_{\Lambda^\prime \lambda^\prime, \Lambda \lambda}$ are composed by the   
following  ``building blocks", or vertex structures connecting the incoming and outgoing protons and quarks, respectively (Fig.\ref{fig_LCWF}),
\[
\begin{array}{|c|c|}
\hline
S=0 & S=1 \\
 \hline 
 & \\
\phi^*_{\Lambda^\prime \lambda^\prime} \phi_{\Lambda \lambda}  & 
 \phi^\mu_{\Lambda^\prime \lambda^\prime} \left( \sum_{\lambda^{\prime \prime}}  \epsilon_\mu^{* \, \lambda^{\prime \prime}}  \epsilon_\nu^{\lambda^{\prime \prime}} \right)
\phi^\nu_{\Lambda \lambda}  \\ 
& \\
\hline
\end{array} 
\]
We obtain for $S=0$,  
\begin{subequations}
\begin{eqnarray}
& \phi^*_{++}(k,P) & =  {\cal A}  \left(m+M X \right),           \\
& \phi^*_{+-}(k,P) & =  {\cal A}  ({k}_1 + i {k}_2),        \\   
& \phi_{--}(k,P)   &  =      \phi_{++}(k,P)    \\
&  \phi_{-+}(k,P)&  =  -\phi^*_{+-}(k,P).
\end{eqnarray}
\end{subequations}
For $S=1$ the factorization of the vertices breaks: there is angular momentum exchange between the LHS and RHS. We find,
\begin{subequations}
 \begin{eqnarray}
& \phi_{++}^+(k,P) = &    {\cal A} \,  \frac{k_1 - i k_2}{1-X}    \\
& \phi_{++}^-(k,P) = &   -{\cal A} \,  \frac{(k_1 + i k_2)X}{1-X}  \\
& \phi_{+-}^+(k,P) = &   0   \\
& \phi_{+-}^-(k,P) = &  -{\cal A} \left(m+M X\right)  \\
& \phi_{-+}^+(k,P) = &  -{\cal A} \left(m+M X\right)  \\
& \phi_{-+}^-(k,P) = & 0 
\end{eqnarray}
\end{subequations}
where, 
\[{\cal A}= \frac{1}{\sqrt{X}} \frac{\Gamma(k)}{k^2-m^2}, \;\;\; {\rm and}  \;\;\; k^2-m^2  =  X M^2 - \frac{X}{1-X} M_X^2 - m^2  - \frac{k_\perp^2}{1-X} .\]
For $(k,P) \rightarrow (k',P')$, $X \rightarrow X^\prime = (X-\zeta)/(1-\zeta)$ and $k_i \rightarrow \tilde{k}_i = k_i - (1-X)/(1-\zeta) \Delta_i$, $(i=1,2)$. Details of the calculation are presented in Appendix \ref{appd}.

\subsection{Chiral Odd GPDs from Helicity Amplitudes}
\label{sec:DGLAP}
The chiral odd GPDs are obtained by inverting Eqs.(\ref{GPDodd}). 
For simplicity we show results for $\zeta=0$ (numerical calculations in the rest of this paper were conducted using the full $\zeta$ dependent expressions). One has,
\begin{subequations}
\label{gpdsodd:eq}
\begin{eqnarray}
\label{AHTbar}  
 \tau   \left[ 2 \widetilde{H}_T(X, 0, t)  + E_T(X,0,t) \right]  & = &  A_{++,+-}   + A_{-+,--} \nonumber \\
& = &  A^{T_Y}_{++,++} - A^{T_Y}_{+-,+-} + A^{T_Y}_{-+,-+} - A^{T_Y}_{--,--}    \\
 \label{AHTX} 
  H_T(X, 0, t) & = & A_{++,--}  + A_{-+,+-} \nonumber \\ 
& = &   A^{T_X}_{++,++} - A^{T_X}_{+-,+-} - A^{T_X}_{-+,-+} + A^{T_X}_{--,--}   \\
\label{AHTtildeX}
  \tau^2 \widetilde{H}_T(X, 0, t)   & = &  - A_{-+,+-} \nonumber \\  
& = &   A^{T_Y}_{++,++}-A^{T_Y}_{+-,+-}-A^{T_X}_{-+,-+}+A^{T_X}_{--,--}   \\
 \widetilde{E}_T(X, 0, t) & = &  A_{++,+-}   - A_{-+,--}   = 0
\end{eqnarray}
\end{subequations}
where $\tau=\sqrt{t_o-t}/2M$. 
Although we calculate the GPDs using the helicity amplitudes, we illustrate the physical meaning of each GPD (or combination of GPDs) in the second lines where each equation represents the helicity amplitudes in the two possible choices for transversity bases, $T_Y$, where the transverse spin is orthogonal to ${\bf \Delta}_T$ (which without loss of generality we can assume along the $x$ axis), and $T_X$, where the transverse spin is along $x$. By inspecting the spin content of the various equations we see that Eq.(\ref{AHTbar}) has the same spin content  of, and therefore reduces in the forward limit to the Boer Mulders function $h_1^\perp$ \cite{BoerMul};  Eq.(\ref{AHTX}) gives transversity, $h_1$; Eq.(\ref{AHTtildeX})  gives  the first $k_T$ moment of $h_{1T}^\perp$; while $\widetilde{E}_T(X, 0, t)$ decouples from the TMDs.    

In Eqs.(\ref{AmpS0},\ref{AmpS1}) we evaluated the quark-proton helicity amplitudes $A^{(0,1)}_{\Lambda^\prime \lambda^\prime, \Lambda \lambda}$ corresponding to the diquark spin components, $S=0,1$. We then obtained the chiral odd GPDs from the amplitudes in Eqs.(\ref{gpdsodd:eq}). The GPDs for each quark flavor are obtained from these equations, in turn, by using the SU(4) symmetry of the proton wave function,
\begin{eqnarray}
\label{FTu}
F_T^u & = &  \frac{3}{2} F_T^{(0)} - \frac{1}{6} F_T^{(1)}    \\
\label{FTd}
F_T^d & = & - \frac{1}{3} F_T^{(1)}, 
\end{eqnarray}
where $F_T^{q} \equiv \{ H_T^q, E_T^q, \widetilde{H}_T^q, \widetilde{E}_T^q \}$, $q=u,d$. 

\begin{figure}
\includegraphics[width=8.cm]{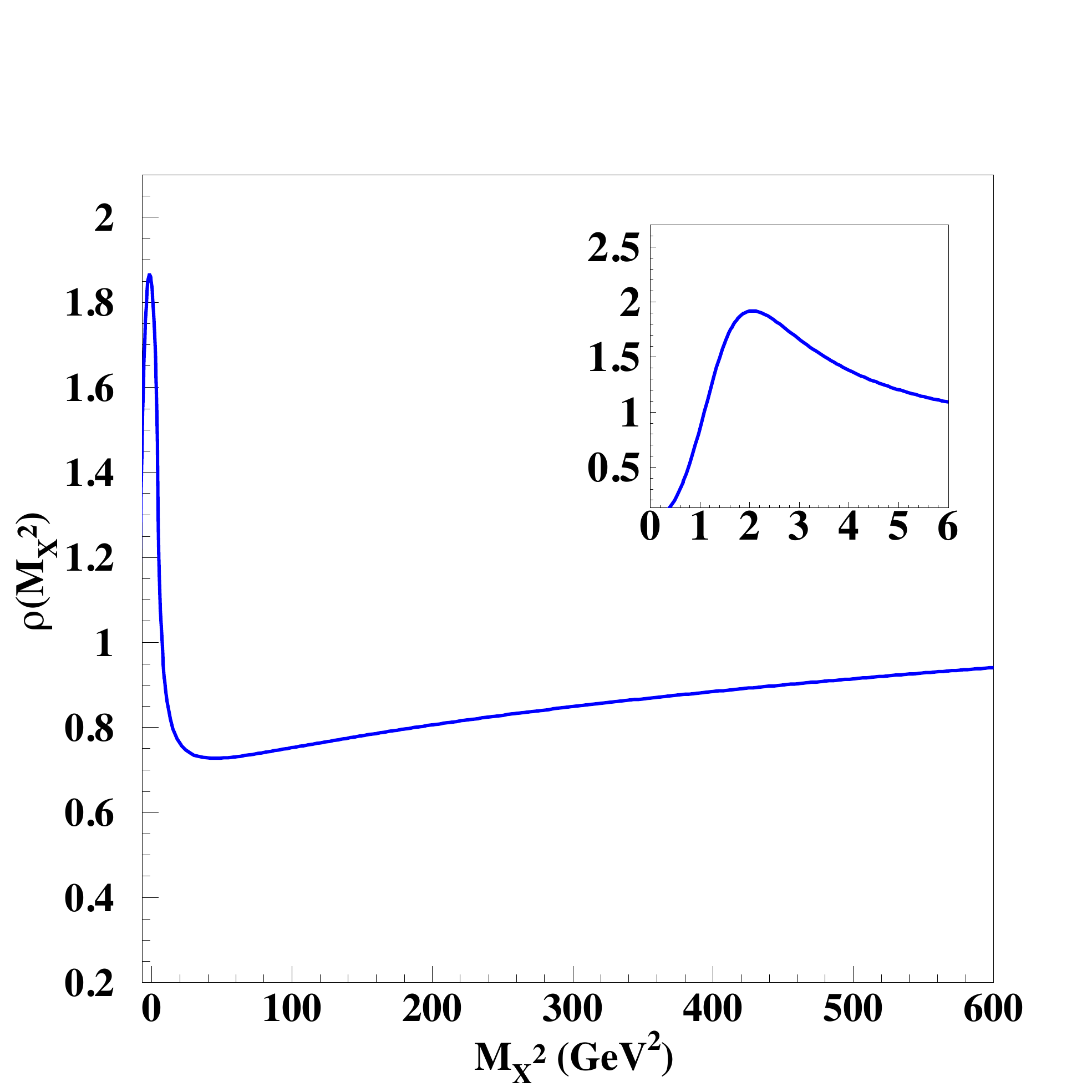}
\caption{The spectral function,  $\rho(M_X^2)$, described in Eqs.(\ref{reggeization}) and (\ref{rho_approx}). The inset shows the details of the behavior at values of the spectator mass in the multi-GeV region.}
\label{fig:regge}
\end{figure}
Next  we consider reggeization (see \cite{Forshaw}, Ch.3 and references therein), that is we extend the diquark model formalism to low $X$ by allowing the spectator system's mass 
to vary up to very large values.
This is accomplished by convoluting the GPD structures obtained in Eqs.(\ref{gpdsodd:eq}) with a spectral function, $\rho(M_X^2)$, where $M_X^2$ is the spectator's mass,  
\begin{eqnarray}
\label{reggeization}
F_T^q(X,\zeta,t) & = & {\cal N}_q  \int_0^\infty  d M_X^2  \rho(M_X^2)  F_{T}^{(m_q, M_\Lambda^q)}(X,\zeta,t; M_X). 
\end{eqnarray}
The spectral function was constructed in Refs.\cite{newFF,GGL} so that it approximately behaves as (Fig.\ref{fig:regge}), 
\begin{eqnarray}
\label{rho_approx}
 \rho(M_X^2) \approx  \left\{  
 \begin{array}{crrl}  (M_X^2)^{\alpha} & & M_X^2 &  \rightarrow \infty \\
 & \\ 
 \delta(M_X^2-\overline {M}_X^2) & & M_X^2 &  {\rm few \; GeV^2}
 \end{array}  \right.
\end{eqnarray}
where $0 < \alpha <1$, and $\overline {M}_X$ is in the GeV range. Upon integration over the mass in Eq.(\ref{reggeization}) one obtains the desired $X^{-\alpha}$ behavior for small $X$, while for intermediate and large $X$ the integral is dominated by the $\delta$ function, yielding a result consistent with the diquark model (more details are given in Ref.\cite{newFF}). 

Inserting $\rho(M_X^2)$ in Eq.(\ref{reggeization}) one obtains an expression that we parametrized in a practical form as,
\begin{equation}
F_T^q(X,\zeta,t)   \approx    {\cal N}_q X^{-\alpha_q +\alpha'_q(X) t}   \,  F_{T}^{(m_q, M_\Lambda^q)}(X,\zeta,t; \overline{M}_X) = R^{\alpha_q,\alpha^\prime_q}_{p_q}(X,\zeta,t) \,  
 G_{M_X,m}^{M_\Lambda}(X,\zeta,t)
%
\label{fit_form}
\end{equation}
where $\alpha_q^\prime(X) =  \alpha_q^\prime  (1-X)^{p_q}$.
The  functions $G_{M_X,m_q}^{M_\Lambda^q}$ and $R^{\alpha_q,\alpha^\prime_q}_{p_q}$ are the quark-diquark and Regge contributions, respectively.
 
Summarizing, the dominant components of the reggeized diquark model are quark-diquark correlations where the diquark system now has both a finite radius and a variable mass, $M_X$, differently from constituent type models. 
At low mass values one recovers compact diquark systems with spin $J=0^+,1^+$. Using the SU(4) symmetry the spin 0 and 1 components translate  into different values for the $u$ and $d$ quark distributions. 
More complex correlations ensue at large mass values which are regulated by the Regge behavior of the quark-proton amplitude, $\propto \hat{u}^{\alpha(t)} = (M_X^2)^{\alpha(t)}$.  
\footnote{The need for introducing a Regge term within the diquark model for GPDs 
was realized in previous phenomenological studies \cite{AHLT}. 
It stems from the observation that standard diquark models cannot reproduce the behavior of the structure functions at low $x$. While this might be of minor importance in kinematical regions centered at relatively large $x$ where most data in the multi-GeV region are, it is, however, a necessary contribution to obtain the normalization of the GPDs to the proton form factors  correctly. The Regge term is therefore an essential ingredient in model building (see also discussions in Refs.\cite{newFF,Rad11,Szcz11}). }

\subsection{$0 < X < \zeta $ $(-\xi <x <\xi)$}
\label{sec:ERBL}
In the ERBL region  ($0 < X<\zeta, -\xi <x <\xi)$ we adopt a functional form  that preserves the GPDs properties of
continuity at the crossover points ($X=0$ and $X=\zeta$), and polynomiality, {\it i.e.} the fundamental consequence of Lorentz covariance by which the $x^n F_q$ integrals (or the Mellin moments) are required to be be polynomials in $\xi$ of order $n+1$ ($n$ even) \cite{Diehl_hab}. The following symmetry relations for $x \rightarrow -x$ are also imposed,
\begin{eqnarray}
F_q^-(x,\xi,t)  & = & F_q^-(-x,\xi,t)  \\
F_q^+(x,\xi,t)  & = & - F_q^+(-x,\xi,t),
\end{eqnarray} 
where,
\begin{eqnarray}
F_q^- & = & F_q(x,\xi) - F_{\bar{q}}(x,\xi) \\
F_q^+ & = & F_q(x,\xi) +  F_{\bar{q}}(x,\xi), 
\end{eqnarray} 
with,
\begin{eqnarray}
F_{\bar{q}}(x,\xi) & = & -  F_q(x,\xi)    \; \; \;  x<0.                            
\end{eqnarray}
$F_q^-$ corresponds to the flavor non singlet  distribution, and $\sum_{q} F_q^+$ to the flavor singlet. 
In the asymmetric system the axis of symmetry is shifted to $\zeta/2$ and one has, over the $[-1+\zeta,1]$ region, 
\begin{eqnarray}
\label{eq:sym}
F_q^-(X,\zeta,t)  & = & F_q^-(\zeta - X,\zeta,t)  \\
\label{eq:asym}
F_q^+(X,\zeta,t)  & = & - F_q^+(\zeta-X,\zeta,t). 
\end{eqnarray} 
Similarly to the light cone wave functions overlap representation \cite{BroDieHwa,BroHwa} (see Ref.\cite{Diehl_hab} for a detailed discussion), in our approach 
polynomiality and continuity at $x= \pm \xi$ are not conditions built in the model. These conditions need to be satisfied by working out a mechanism that produces a covariant result (and consequently polynomiality \cite{Diehl_hab}) from the behavior of the wave functions in both the DGLAP and ERBL regions. As shown in  \cite{BroDieHwa,BroHwa} the ERBL region admits a wave function representation given by  the overlap of wave functions with different particle content. 
While general relations between light cone wave functions in the two regions are not known, we observe, on one side that these can be explored by implementing the equations of motion
(see Refs.\cite{Diehl_hab,GolLiu_z_diag} and in preparation), while
on the other side, one can perform practical calculations limited to the lowest Fock states. In Ref.\cite{BroEst} for instance, a calculation of the valence component was performed which displays polynomiality within a simplified realization of the overlap formulae  (disregarding, however, the full spin content of the partons).
A version of this model including spin was instead adopted in \cite{GolLiu_disp}.   

Here, with the aim of providing a parametrization for the DVCS type experimental data,  we match our  covariant representation in the DGLAP region (Section \ref{sec:DGLAP}),
with the following parametric form in the ERBL region,
\begin{eqnarray}
\label{h+}
F^-_{X<\zeta}(X,\zeta,t) & = & a^-(\zeta,t) X^2 -a^-(\zeta,t) \zeta X + F(\zeta,\zeta,t) \\
\label{h-}
F^+_{X<\zeta}(X,\zeta,t) & = & a^+(\zeta,t) X^3 -a^+(\zeta,t) \zeta X^2 + c(\zeta,t) X+ d(\zeta,t)   
\end{eqnarray}
where we distinguish the two separate functional forms, $F^\pm_{X<\zeta}$, respectively satisfying the symmetric, Eq.(\ref{eq:sym})  and anti-symmetric, Eq.(\ref{eq:asym}),  conditions (see also Eqs.(45,46), Ref.\cite{GGL,newFF}).  
The coefficients, $a^\pm(\zeta,t)$, $c(\zeta,t)$, and $d(\zeta,t)$ are determined so that $F^\pm_{X<\zeta}$ satisfy both the polynomiality property, and continuity at the point $X=\zeta$ (more details can be found in \cite{GGL_param}).

The valence and sea quarks contributions are then obtained as,
\begin{subequations}
\label{ERBL}
\begin{eqnarray}
\label{NS}
F^{NS}_{X<\zeta}(X,\zeta,t) & = &  F^+_{X<\zeta}(X,\zeta,t) + F^-_{X<\zeta}(X,\zeta,t)   \\
\label{S}
F^{S}_{X<\zeta}(X,\zeta,t) & = &    F^+_{X<\zeta}(X,\zeta,t) - F^-_{X<\zeta}(X,\zeta,t)
\end{eqnarray}
\end{subequations}

\subsection{Parametric Form: discussion of parameters}
\label{sec:param}
By putting together the results from Sections \ref{sec:DGLAP} and \ref{sec:ERBL} we obtain our final parametrization,
\begin{equation}
\label{final_form}
F_T^q(X,\zeta,t)  =  \left\{ \begin{array}{ll}
F_{T, \, X>\zeta}^{q (NS, S)}(X,\zeta,t),  \;\;\; {\rm Eq.(\ref{fit_form})}  & \mbox{if $\zeta \leq X \leq1$}  \\ 
 & \\
 F_{T, \, X<\zeta}^{q (NS, S)}(X,\zeta,t)  \;\;\; {\rm Eq.(\ref{NS},\ref{S})} & \mbox{if $1 - \zeta \leq X  < \zeta $}. 
\end{array} \right.
\end{equation}
$F_T^q$ satisfies all fundamental requirements, such as polynomiality, positivity, symmetry at ($x=0$, $X=\zeta/2$), hermiticity, and time reversal invariance (a similar expression was first derived for chiral even GPDs where the helicity amplitudes were evaluated according to  Eqs.(\ref{GPDeven})). 

To fit the chiral odd GPDs one faces, however, an important practical problem: while various models, {\it e.g.} our reggeized diquark model, can be extended to the chiral odd sector as we explained in Sections \ref{sec:LCWF}, \ref{sec:DGLAP} (Eqs.(\ref{fit_form},\ref{ERBL},\ref{final_form})), and the ERBL region can be treated as outlined in Section \ref{sec:ERBL},  both the forward limit and the integral of the GPDs (providing the normalization to the form factors in the chiral even case, and the lowest order polynomiality relation), are largely unknown. The only constraints which can be obtained from independent measurements are given by the transversity structure function,
\begin{eqnarray}
H_T(X,0,0) & = & h_1(X), 
\end{eqnarray}     
by the model dependent relations with the Boer-Mulders, $h_{1}^\perp(X)$, and $h_{1T}^\perp(X)$ functions, respectively given by \cite{Metz},
\footnote{These are valid in spectator models such as the one presented here.}
\begin{eqnarray}
\underset{t \rightarrow 0}{\rm lim} \, \frac{t}{4M^2} \widetilde{H}_T(X,0,t)  &  \leftrightarrow & h_{1T}^\perp(X) = \int d^2 k_T \, h_{1T}^\perp(X,k_T)  \\
2 \widetilde{H}_T(X,0,t) + E_T(X,0,t)  & \leftrightarrow &   h_{1}^\perp(X) = \int d^2 k_T \, h_{1}^\perp(X,k_T),
\end{eqnarray}     
and by the integration to form factors at $t=0$, giving the tensor charge,
\begin{eqnarray}
\delta_q = \int_0^1 dx H_T(X,0,0) 
\end{eqnarray}
Since $h_1$ and the related chiral odd functions are exactly the observables that we want to determine using GPDs, these constraints are not really useful quantitatively. They can just be used as indications. 

 It is, however, important to be able to determine the size of the various chiral odd GPDs especially in the first steps of  the analysis of available $\pi^o$ electroproduction data. The data are in fact not sufficient at this stage, to fully discern among the various possible contributions. 

An alternative way to gauge the contribution of the various chiral odd GPDs in deeply virtual exclusive experiments consists in exploiting Parity relations among the proton-quark-diquark vertex functions.  As we explain in the next Section, these Parity relations will allow us to establish relations between the chiral odd and chiral even GPDs, which are valid within the class of models including the spectator/diquark models.  By constraining the chiral odd GPDs using their chiral even counterparts we will be able to gauge the size of the GPDs contributions to pseudo-scalar electroproduction data. Before moving on to the discussion of the Parity relations, we end this Section by outlining the way our parametrization was worked out in the chiral even sector.  
 
Our parametrization is written in the form given in Eq.(\ref{final_form}) for both the chiral even and chiral odd sectors. 
The model's parameters can be divided into three sets: 
\begin{eqnarray}
& (1) &\;\;\;\;\;\;     \alpha_q, M^q_\Lambda, m_q, M_X^q, {\cal N}_q, \nonumber \\
& (2) & \;\;\;\;\;\;   \alpha'_q, p_q,  \nonumber \\
& (3)  & \;\;\;\;\;\;   a_q^\pm(\zeta,t), c(\zeta,t), d(\zeta,t) \nonumber
\end{eqnarray}
where $q=u,d$.  
Set $(1)$ contributes to PDFs ({\it i.e.} they determine the part of the GPD that survives when $t=\zeta=0$ {\it e.g.} in Eq.(\ref{fit_form})); set $(2)$  determines the $t$ dependence in Eq.(\ref{fit_form}), and set $(3)$ determines the $\zeta$ dependence (Section \ref{sec:ERBL}). 

A fit using these sets of parameters in the chiral even sector was performed in Ref.\cite{GGL}. 

Owing to the fact that we use a diverse body of data which are divided into proton and deuteron DIS data (set (1)), proton and neutron electromagnetic and electroweak form factor data (set (2)), and DVCS data (set(3)), we  used a {\em recursive} fitting procedure.  

In the recursive fit we first obtained the parameters in set (1) from DIS data. Keeping these parameters fixed, we then obtained the parameters of  set $(2)$ by fitting the electromagnetic, axial and pseudoscalar nucleon form factors given by the integrals over each GPD. The final step in the recursive fit was done by fitting the $\zeta$ dependent parameters from set $(3)$ to DVCS data from Hall B \cite{HallB}.

In the fitting procedure perturbative QCD evolution was taken into account  as follows:
in step (1) we used the PDFs, obtained as the limiting GPDs model's $X=0, t=0$ case, as inputs at the model's initial scale, $Q_o^2$. These PDF forms were evolved to $Q^2$ values that are typical for DIS fits by directly solving the DGLAP equations at NLO in $x$-space. The numerical value of the initial scale is an additional parameter. We found that a good fit of PDFs at $Q^2$ in the multi-GeV region could be obtained by choosing, $Q_o^2 = 0.1$ GeV$^2$, as determined in a previous study \cite{AHLT}.  This value is expectedly low, as featured in many versions of the diquark model (see {\it e.g.} Ref.\cite{BacConRad}).
Once the parameters were fixed, the parameters obtained in step(2) did not involve the $Q^2$ dependence.  The parameters in step (3) were instead determined by evolving 
the GPD forms with their full $X, \zeta$  and $t$ dependences to the $Q^2$ of interest. Evolution was performed at $X>\zeta$ again by solving the DGLAP equations at NLO in $x$-space, with appropriately modified kernels to describe GPDs evolution according to Refs.\cite{MusRad,GolMar}. In this last step the values of the parameters, $a^\pm(\zeta,t)$, $c(\zeta,t)$, and $d(\zeta,t)$. One does not have to solve evolution equations explicitly  in the $X<\zeta$ sector, however, the parameters $a^\pm(\zeta,t)$, $c(\zeta,t)$, and $d(\zeta,t)$ are effectively $Q^2$ dependent, producing different curves in the ERBL region for different $Q^2$ values. 

As new DVCS and meson electroproduction data become available, it will be possible to perform a global fit using simultaneously all sets of data.  At the present stage our approach provides a controlled procedure where the various kinematical dependences can be more readily tested. 

\subsection{Parity relations among amplitudes in the diquark model}
The Parity relations for the vertices in Fig.\ref{fig1} read,
 \begin{equation}
 \phi_{-\Lambda -\lambda} = (-1)^{\Lambda-\lambda} \phi^*_{\Lambda \lambda}. 
 \label{Parity0}
 \end{equation}
Since for $S=0$ the helicity structure of Fig.\ref{fig1} corresponds to 
a factorized form -- the product of two independently varying $\phi$ functions -- and, as shown in Eq.(\ref{Parity0}), these two components transform under Parity  
independently from one another.
The  following relations hold between the chiral odd amplitudes and the chiral even ones for $S=0$,
\begin{subequations}
\label{Amp0}
\begin{eqnarray}
A_{++,--}^{(0)}   & =  & A_{++,++}^{(0)}   \\
A_{++,+-} ^{(0)}    &  = &  - A_{++,-+}^{(0)*}  \\ 
A_{+-,++}^{(0)}   &  = &  - A_{-+,++}^{(0)*}  , 
\end{eqnarray}
\end{subequations}
Notice that  these relations are valid only if one of the two $\phi$ functions is real. By using Parity symmetry one cannot connect directly the chiral odd amplitude $A_{+-,-+}$, with its chiral even counterpart $A_{+-,+-}$ since both involve complex $\phi$ functions. Physically this corresponds to the fact that $A_{+-,-+}$
involves a double spin flip, and it must therefore be proportional to $\Delta_\perp^2 = (t_0-t)(1-\zeta)$, while $A_{+-,+-}$ is non-flip. 
By evaluating $A_{+-,-+}$ directly one has,
\begin{eqnarray}
A_{+-,-+}^{(0)}   & = &  \int d^2 k_\perp \phi^*_{+-}(k^\prime,P^\prime) \phi_{-+}(k,P)   \nonumber \\
& = & \mathcal{N}  \displaystyle\frac{\sqrt{1-\zeta}}{1-X}   \int d^2 k_\perp \frac{k_1\tilde{k}_1 + \tilde{k}_2(-k_2)}{(k^2-M_\Lambda^2)^2(k^{\prime \, 2}-M_\Lambda^2)^2} \nonumber \\
& = &  \mathcal{N}  \displaystyle\frac{\sqrt{1-\zeta}}{1-X}  \int d k_\perp k_\perp \nonumber \\
& \times & \int d \phi \frac{k_\perp^2 - \widetilde{X}\Delta_\perp k_\perp \cos \phi - 2 k_\perp^2 \sin^2 \phi}{(k^2-M_\Lambda^2)^2(k^{\prime \, 2}-M_\Lambda^2)^2} \nonumber \\
& \approx & \mathcal{N}  \displaystyle\frac{\sqrt{1-\zeta}}{1-X} \int d k_\perp k_\perp \nonumber \\
& \times & \int d \phi \frac{- \widetilde{X}\Delta_\perp k_\perp \cos \phi}{(k^2-M_\Lambda^2)^2(k^{\prime \, 2}-M_\Lambda^2)^2}, 
\end{eqnarray}
where $k_\perp^2 = k_1^2 + k_2^2$,  $\Delta_\perp^2 = \Delta_1^2 + \Delta_2^2$, and $\widetilde{X} = (1-X)/(1-\zeta)$. One can in fact demonstrate that an almost exact cancellation occurs 
between the terms $\propto$ $k_\perp^2$ and $k_\perp^2 \, 2 \sin^2 \phi$. From the explicit expressions for the integrals given in Appendix A of Ref.\cite{GGL}
we can see that the angular integration gives rise to an extra factor of $\Delta_\perp$.
A connection between the GPD $\widetilde{H}_T$ and the chiral even ones can be found by using the results from Ref.~\cite{GGL}, 
expressing the quark proton helicity amplitudes in terms of chiral even GPDs, we therefore obtain
\begin{equation}
A_{+-,-+}^{(0)}  = \frac{t_0-t}{4M} \frac{1}{\sqrt{1-\zeta}} \frac{1}{(1-\zeta/2)} \frac{\tilde{X}}{m+MX^\prime}  \left[ E - (\zeta/2) \widetilde{E} \right]
\end{equation}

For $S=1$ we obtain,
\begin{subequations}
\label{Amp1}
\begin{eqnarray}
A_{++,--}^{(1)}  & = &  - \displaystyle\frac{X+X^\prime}{1+ X X^\prime} \; A_{++,++} ^{(1)} \\
A_{+-,-+}^{(1)}  & = & 0 \\
A_{++,+-}^{(1)}  & = & - \sqrt{ \frac{\langle \tilde{k}_\perp^2\rangle /P^{+ \, 2} }{X^{\prime \, 2} + \langle \tilde{k}_\perp^2 \rangle /P^{+ \, 2} }  } \; A_{++,-+} ^{(1)*} \\
A_{+-,++}^{(1)}  & = & - \sqrt{ \frac{\langle k_\perp^2  \rangle /P^{+ \, 2} }{X^2 + \langle k_\perp^2 \rangle/P^{+ \, 2} }  } \; A_{-+,++}^{(1)*},
\end{eqnarray}
\end{subequations}
Notice that for $X=\zeta$, {\it i.e.} for the calculation of the imaginary parts of the CFFs,
\begin{subequations}
\begin{eqnarray}
A_{++,--}^{(1)} & = &  - A_{++,++}^{(1)} \\
A_{++,+-}^{(1)} & = & - A_{++,-+}^{(1)*} \\
A_{+-,++}^{(1)} & = & - A_{-+,++}^{(1)*},
\end{eqnarray}
\end{subequations}
we obtain 
the same relations as for the scalar diquark case.
Furthermore, the double helicity flip amplitude $A_{+-, -+}$ vanishes to order $\Delta_\perp^2$. This can be understood when considering the near collinear circumstance. For nucleon to quark flipping helicity from $+ \frac{1}{2}$ to $-\frac{1}{2}$ the diquark must carry $+1$ near forward. That is rejoined by a near forward quark of helicity $+\frac{1}{2}$ to form a $+\frac{3}{2}$ system in the near forward case. That cannot happen unless there is highly non-forward kinematics, because the helicity carried by the diquark cannot be compensated. 
These simple relations are a general feature of the axial diquark coupling, given that we have omitted the helicity 0 component. 
The chiral even and chiral odd helicity amplitudes in terms of GPDs were given above.

Using the Parity relations between amplitudes in Eqs.(\ref{Amp0}, \ref{Amp1}), 
we now give the functions $F_T^{(0),(1)}$, evaluated by 
inverting the expressions for the helicity amplitudes in Eqs.(\ref{GPDodd}) (with full account of the $\zeta$ dependence), for $S=0$,

\begin{subequations}
\label{S0}
\begin{eqnarray}
\widetilde{H}_T^{(0)} & = &   -\frac{1}{F} \left(E^{(0)}  -  \frac{\zeta}{2} \widetilde{E}^{(0)}  \right) \\ 
E_T^{(0)} & = &  \frac{2}{1-\xi^2} \left[E^{(0)} -  \widetilde{H}_T^{(0)} - \xi^2 \widetilde{E}^{(0)} \right] = 
2 \, \frac{(1-\zeta/2)^2}{1-\zeta} \left[E^{(0)} - \widetilde{H}_T^{(0)} - \left(\frac{\zeta/2}{1-\zeta/2}\right)^2 \widetilde{E}^{(0)}  \right]  \\
\widetilde{E}_T^{(0)} & = & \frac{2}{1-\xi^2} \left[ \xi^2 E^{(0)} -  \xi^2 \widetilde{H}_T^{(0)} - \widetilde{E}^{(0)} \right] = 
2 \, \frac{(1-\zeta/2)^2}{1-\zeta} \left[ \left(\frac{\zeta/2}{1-\zeta/2}\right)^2 E^{(0)} -  \left(\frac{\zeta/2}{1-\zeta/2}\right)^2 \widetilde{H}_T^{(0)} -  \widetilde{E}^{(0)}  \right] \\
H_T^{(0)}  & = &  \frac{H^{(0)}  + \widetilde{H}^{(0)} }{2} - \frac{\zeta^2/4}{1-\zeta}\frac{E^{(0)} + \widetilde{E}^{(0)} }{2} -  \frac{\zeta^2/4}{(1-\zeta/2)(1-\zeta)} E_T^{(0)}  + 
\frac{\zeta/4 (1-\zeta/2)}{1-\zeta} \widetilde{E}^{(0)} _T -
\frac{t_0-t}{4M^2} \frac{1}{F} \left(E^{(0)}  -  \frac{\zeta}{2} \widetilde{E}^{(0)} \right) \nonumber \\
\end{eqnarray}
\end{subequations}
and for $S=1$, 
\begin{widetext}
\begin{subequations}
\label{S1}
\begin{eqnarray}
\widetilde{H}_T^{(1)} & = &  0   \\
E_T^{(1)} & = &    \frac{1-\zeta/2}{1-\zeta} \left[ \tilde{a} \left(E^{(1)} -  \frac{\zeta/2}{1-\zeta/2} \widetilde{E}^{(1)} \right)   + a \left(E^{(1)} + \frac{\zeta/2}{1-\zeta/2} \widetilde{E}^{(1)} \right) \right] \\
\widetilde{E}_T^1 & = &   \frac{1-\zeta/2}{1-\zeta} \left[ \tilde{a} \left(E^{(1)} -  \frac{\zeta/2}{1-\zeta/2} \widetilde{E}^{(1)} \right)  -  a \left(E{(1)} + \frac{\zeta/2}{1-\zeta/2} \widetilde{E}^{(1)} \right) \right]  \\
H_T^{(1)} & = &  G \left[  \frac{H^{(1)} + \widetilde{H}^{(1)}}{2} - \frac{\zeta^2/4}{1-\zeta}\frac{E^{(1)} + \widetilde{E}^{(1)}}{2} \right] -  \frac{\zeta^2/4}{1-\zeta} E_T^{(1)} + 
\frac{\zeta/4}{1-\zeta} \widetilde{E}_T^{(1)}  
\end{eqnarray}
\end{subequations}
\end{widetext}
where the various kinematical factors are, 
\[ F= \left(\frac{m_q}{M} + X^\prime\right) \, \frac{1-\zeta}{\widetilde{X} },  \;\;\;\;    G= \displaystyle\frac{X+X^\prime}{1+ X X^\prime}, \;\;\;\; \widetilde{X} = \frac{1-X}{1-\zeta} \]
and
\[a =  \displaystyle\sqrt{ \frac{\langle k_\perp^2 \rangle}{X^2 + \langle k_\perp^2 \rangle \left(\frac{2M\zeta^2}{Q^2}\right)^2} }, \;\;\;\;
\tilde{a} =   \displaystyle\sqrt{ \frac{\langle \tilde{k}_\perp^2 \rangle}{(X-\zeta)^2 + \langle \tilde{k}_\perp^2 \rangle \left(\frac{2M\zeta^2}{Q^2}\right)^2} } \]
The GPDs, $F_T^u$, and $F_T^d$ are then calculated from $F_T^0$ (\ref{S0}), and $F_T^1$ (\ref{S1}), using the SU(4) relations in Eqs.(\ref{FTu}) and (\ref{FTd}), respectively.

\begin{figure}
\includegraphics[width=16.cm]{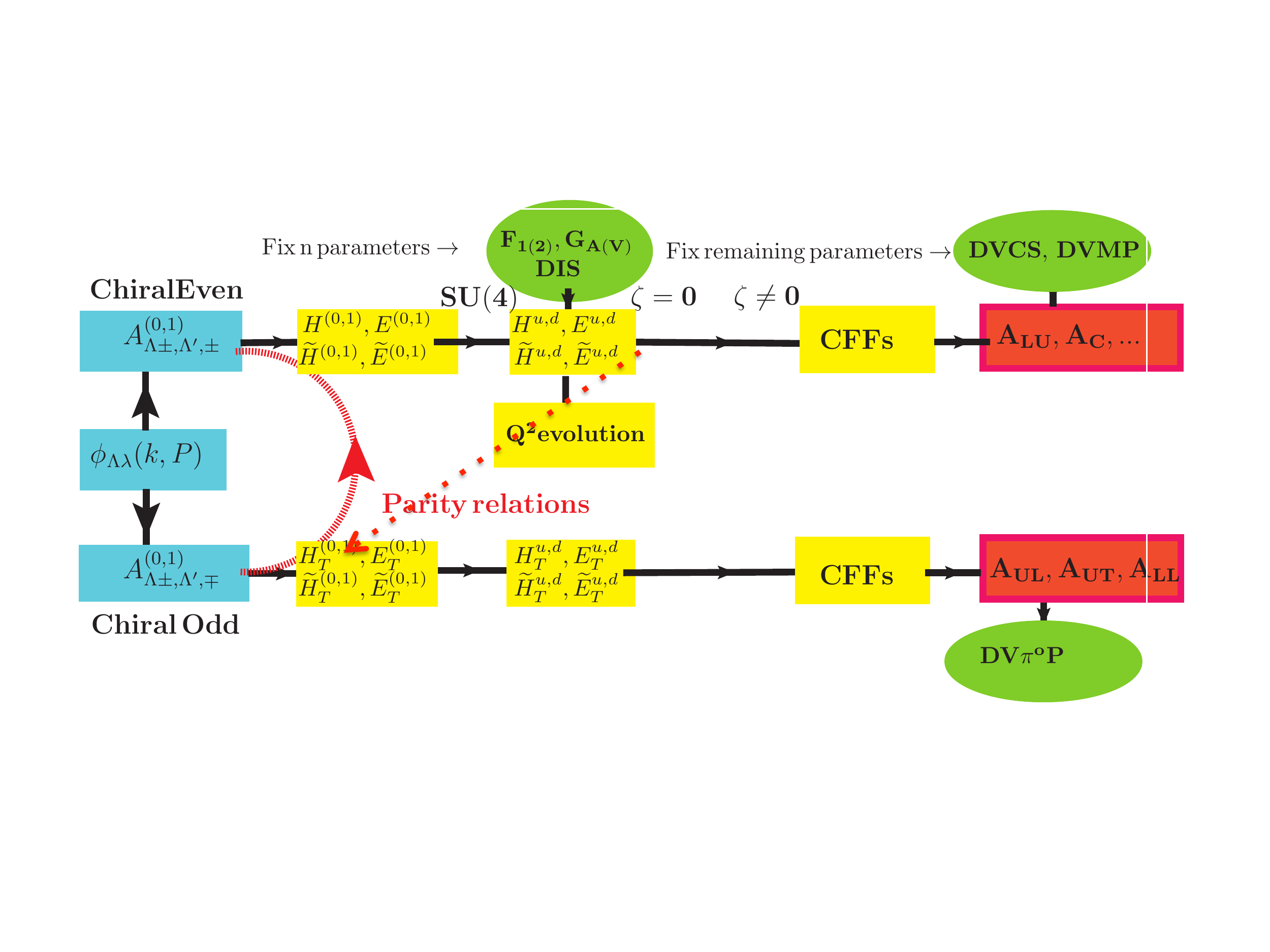}
\caption{(Color online) Flowchart for the GPD fitting procedure described in the text.}
\label{fig:flow}
\end{figure}

The fitting procedure of GPDs is quite complicated owing to its many different steps. In Fig.\ref{fig:flow} we summarize with a flowchart  the various steps described so far, {\it i.e.}, proceeding from left to right: 1) the construction of chiral odd helicity amplitudes; 2) the connection of these amplitudes to the chiral even ones using Parity relations within spectator models (curved upwards arrow); 3) the fixing of chiral even parameters at an initial scale, $Q_o^2$, using the nucleon form factors and PQCD evolution to match DIS data; 4) the determination of chiral odd GPDs (dotted arrow in the figure); 5) the construction of the corresponding Compton form factors, and of the pseudoscalar meson electroproduction observables. 

\begin{figure}
\includegraphics[width=8.cm]{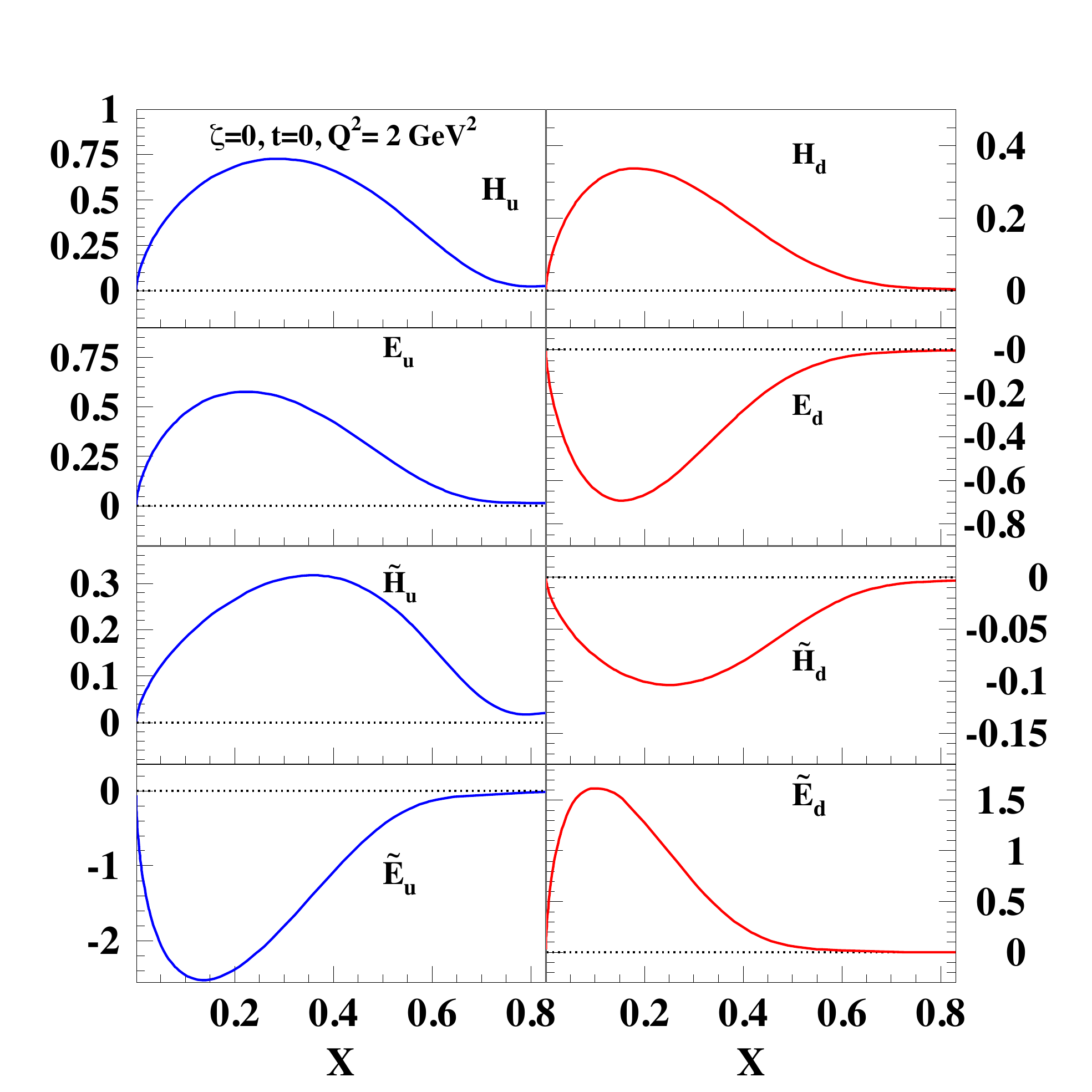}
\hspace{0.3cm}
\includegraphics[width=8.cm]{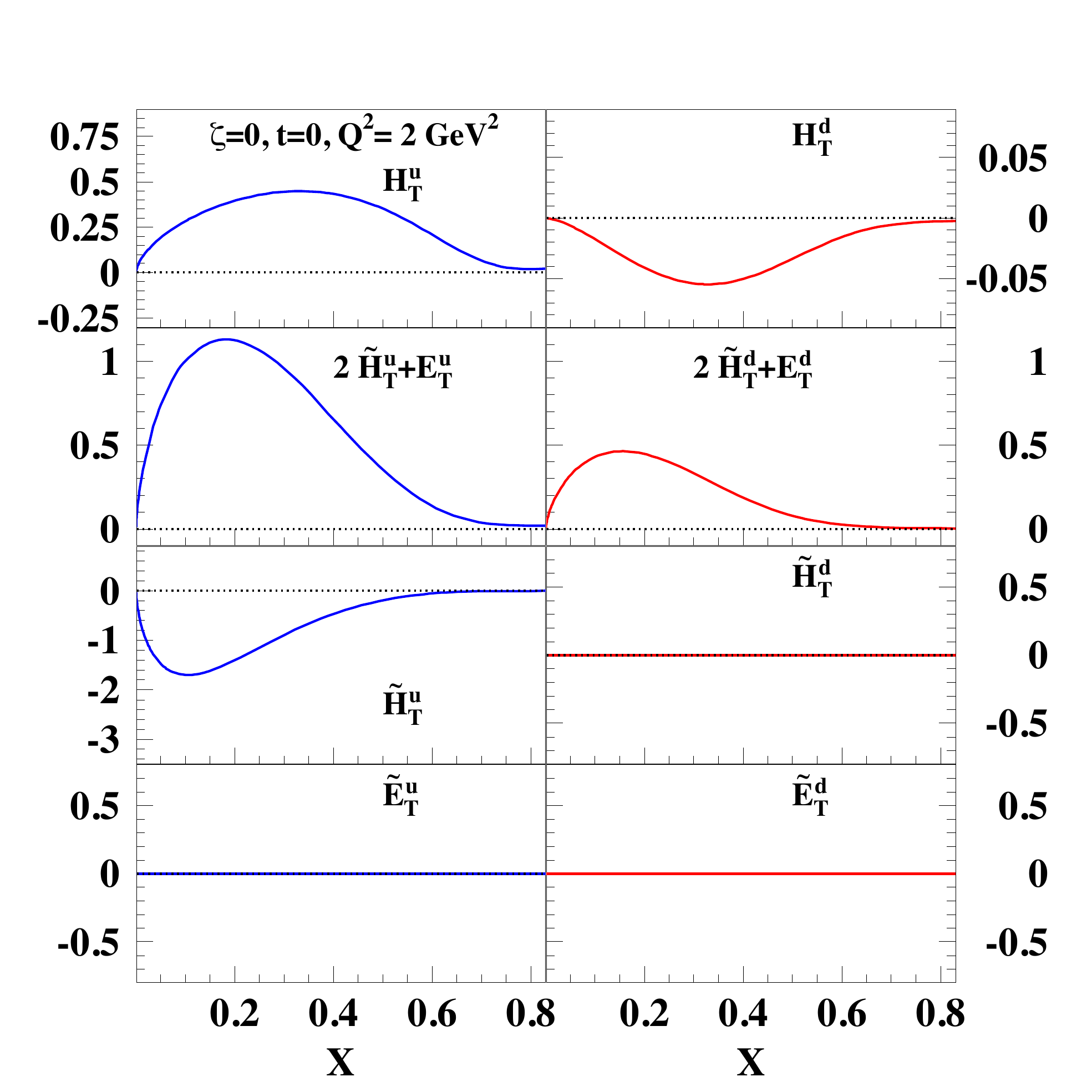}
\caption{(Color online) The chiral even (left panel) and chiral odd GPDs (right panel) evaluated using the model described in the text at $\zeta=0$, $t=0$, plotted vs. $X$ at fixed $Q^2=  2$ GeV$^2$.}
\label{fig:gpds}
\end{figure}
\begin{figure}
\includegraphics[width=8.cm]{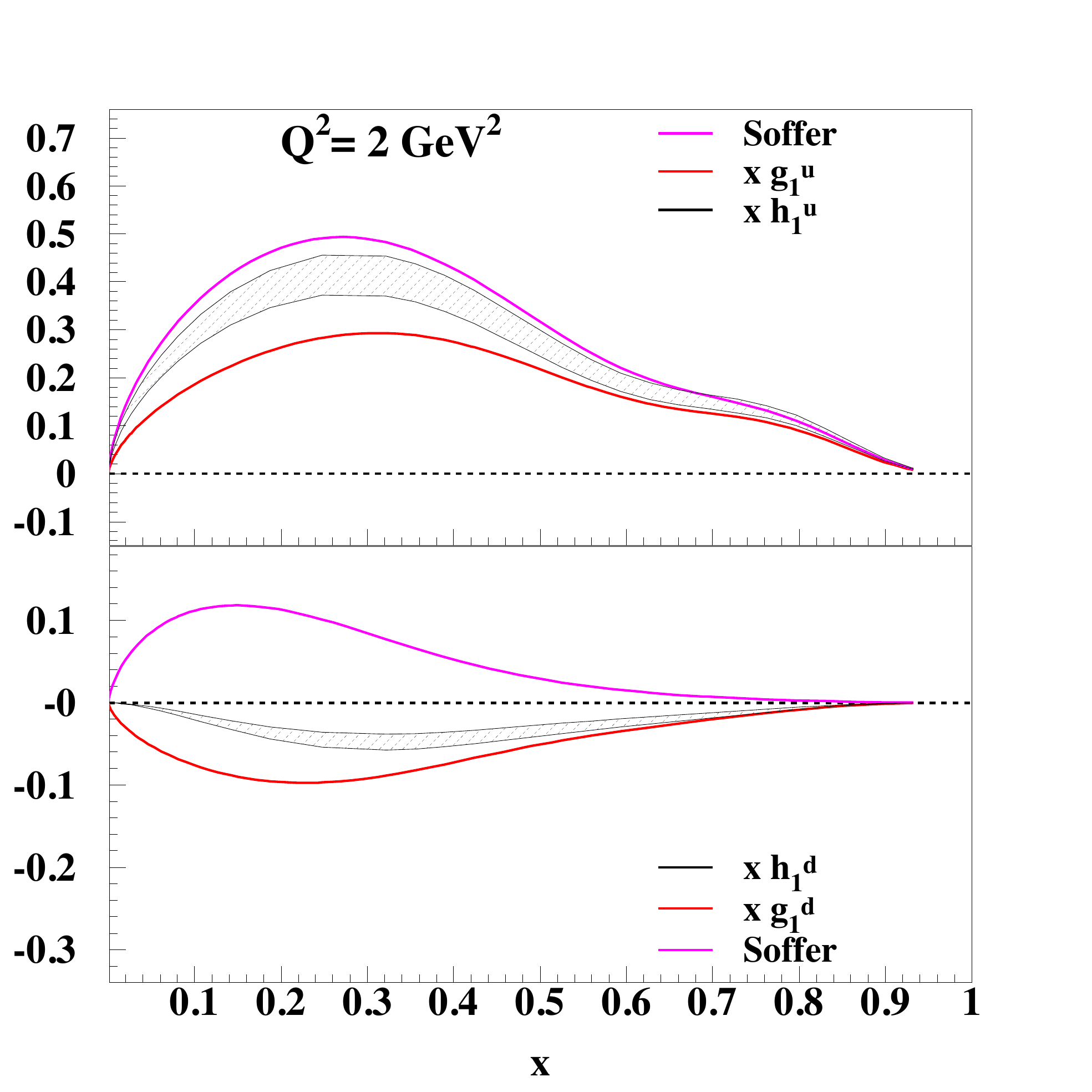}
\caption{(Color online) The transversity function, $h_1(x,Q^2) \equiv \widetilde{H}_T(X,0,0, Q^2)$ plotted along with theoretical errors (hashed area) for the up (top panel) and down (bottom panel) quarks. The theoretical uncertainties are propagated from the parameters errors from PDF fits in the chiral even sector. The other curves in the figure represent the Soffer bound \cite{Soffer} on the magnitude of $h_1$, and the values of $g_1^{u,d}$, respectively (adapted from Ref.\cite{Losini}).}
\label{fig:transversity}
\end{figure}

\begin{figure*}
\includegraphics[width=8.cm]{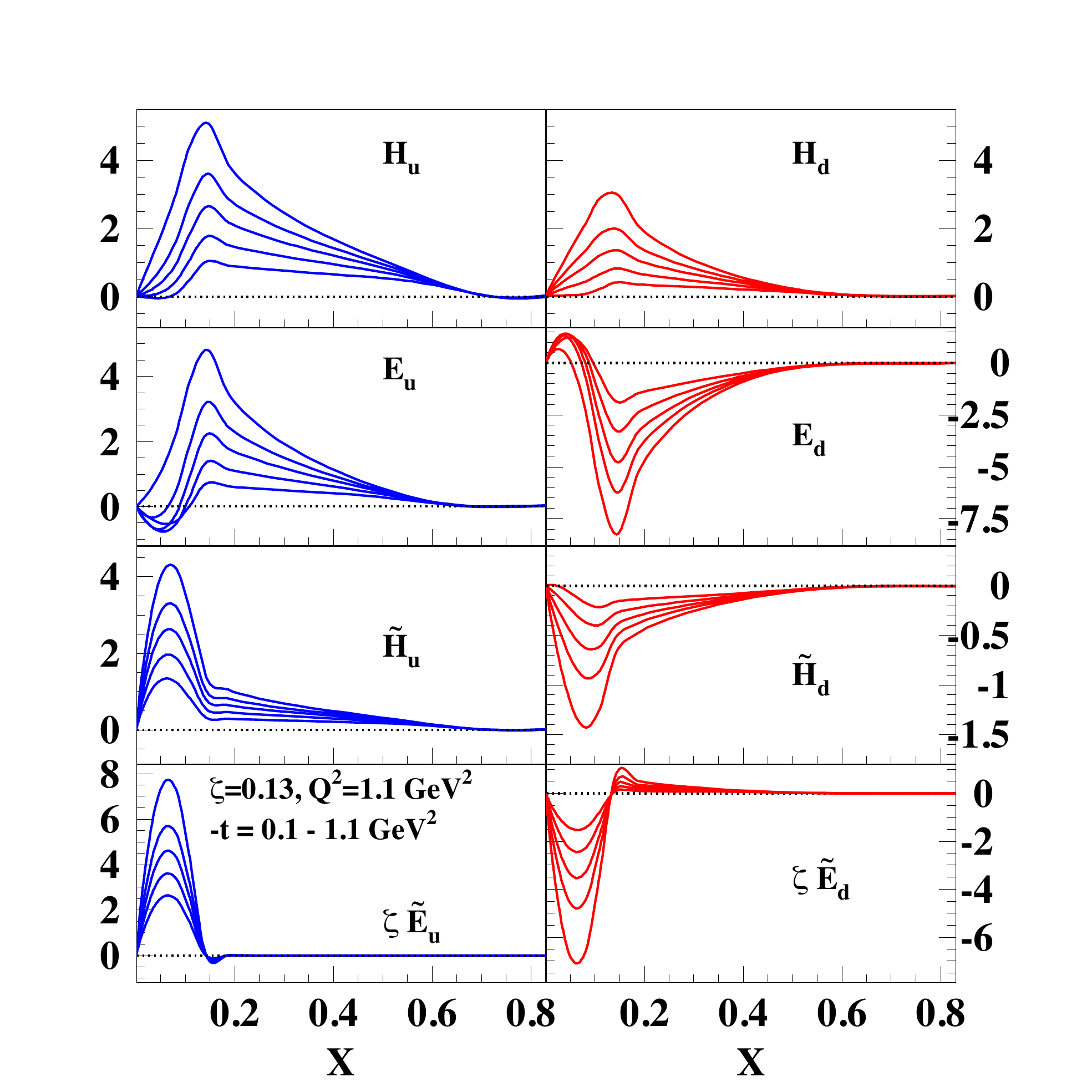}
\hspace{0.3cm}
\includegraphics[width=8.cm]{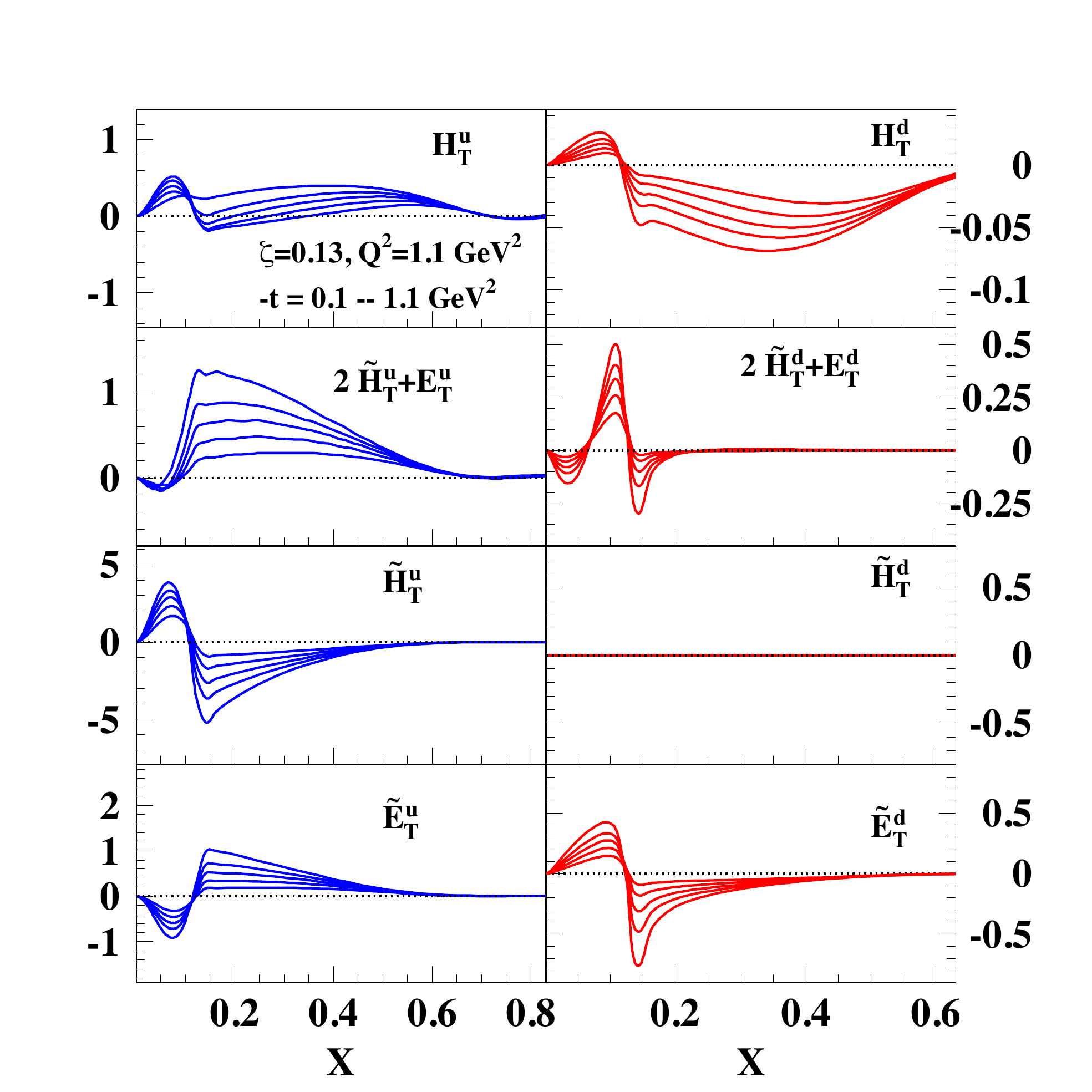}
\caption{(Color online) The chiral even (left panel) and chiral odd GPDs (right panel) evaluated using the model described in the text plotted vs. X at $x_{Bj}=\zeta=0.13$, $Q^2=  2$ GeV$^2$. The range in $-t$ is: $0.1  \leq -t \leq 1.1 \, {\rm GeV}^2$. Curves with the largest absolute values correspond to the lowest $t$.}
\label{fig:gpds_t}
\end{figure*}
In Fig.\ref{fig:gpds} we show both the chiral even GPDs (left panel) and the chiral odd GPDs (right panel) evaluated using the model described in this paper at $\zeta=0$, $t=0$, plotted vs. $X$ at fixed $Q^2=  2$ GeV$^2$. The chiral even GPDs were already evaluated in Ref.\cite{GGL} by using the recursive fitting procedure described above. Notice that as a byproduct of our analysis we obtain an independent extraction of, $H_T^q(X,0,0; Q^2) \equiv h_1^q(X,Q^2)$ (upper panels). 
In Fig.\ref{fig:transversity} we show transversity in more detail, compared with $g_1^q(X)$, and the Soffer bound, $f_1(X)+ g_1(X)$. It is interesting to notice how from exclusive pseudoscalar electroproduction data we obtain an independent extraction of this quantity. 
A more detailed description of the other transversity functions including the first moment of $h_1^\perp \equiv 2\widetilde{H}_T^q + E_T^q$, whose integral over $X$ gives the transverse anomalous magnetic moments
\cite{Bur3}, will be given in \cite{GGL_param}.

In Fig.\ref{fig:gpds_t} we show the $t$-dependent GPDs that enter the helicity amplitudes evaluated in Section II in a kinematical bin ($x_{Bj}=0.13, Q^2=1.1$ GeV$^2$) consistent with the Jefferson Lab kinematical coverage. 
The chiral even GPDs are shown in the left panel, and the chiral odd GPDs in the right panel.  

\begin{figure}
\includegraphics[width=9.cm]{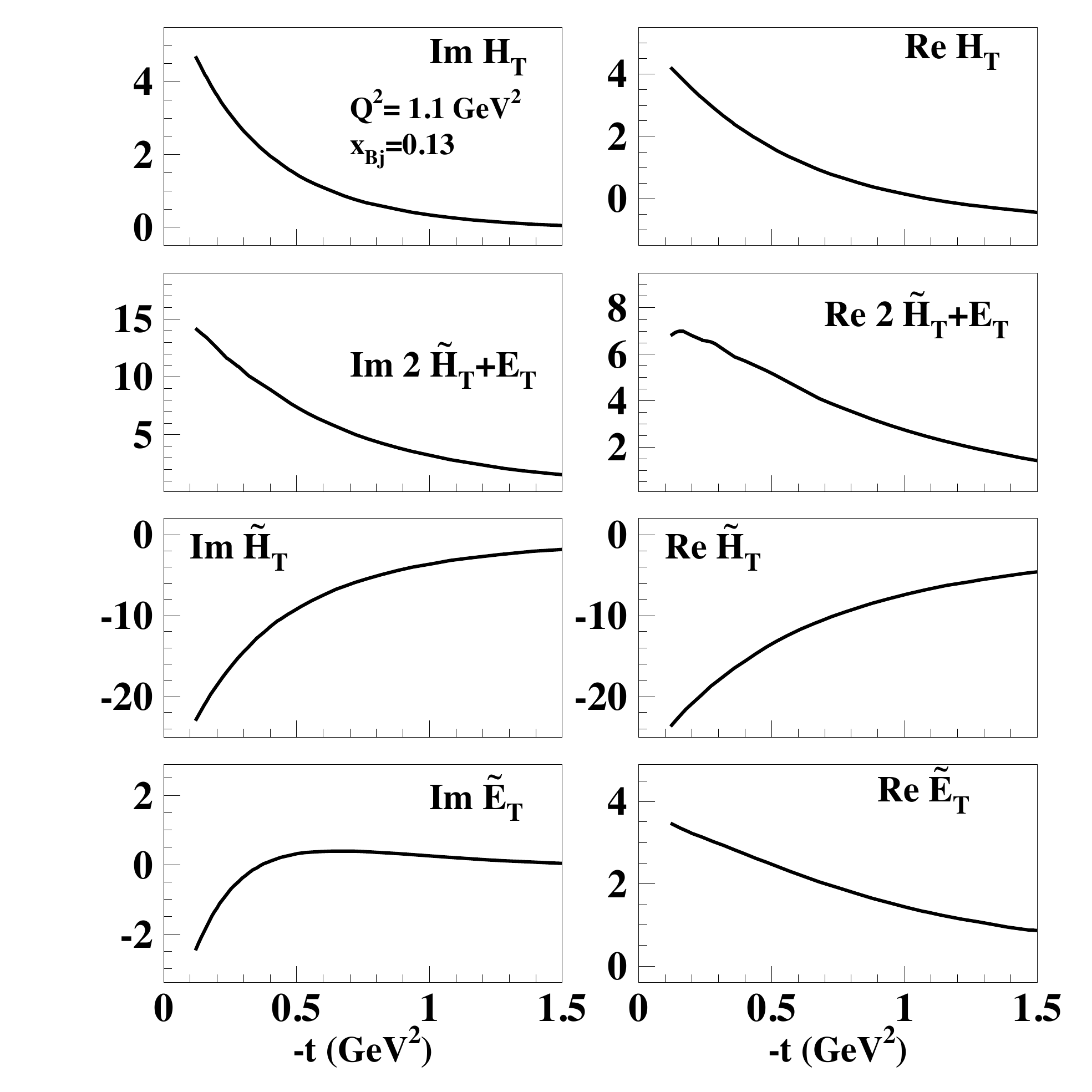}
\caption{Chiral Odd CFFs, Eqs.(\ref{CFF_def}), entering the process $\gamma^* p \rightarrow \pi^o p'$. From top to bottom  $\Im m{\cal H}_T$ (left), $\Re e{\cal H}_T$ (right);  $\Im m[2 \widetilde{\cal H}_T + {\cal E}_T]$ (left),  $\Re e[2 \widetilde{\cal H}_T + {\cal E}_T]$ (right); $\Im m \widetilde{\cal H}_T$ (left), $\Re e \widetilde{\cal H}_T$ (right); $\Im m \widetilde{\cal E}_T$ (left), $\Re e \widetilde{\cal E}_T$ (right). The various CFFs are  plotted vs. $-t$ for the kinematic bin $x_{Bj}=0.13$, $Q^2=1.1$ GeV$^2$.}
\label{fig:cff_odd}
\end{figure}

In Fig.\ref{fig:cff_odd} we show the proton CFFs, Eq.(\ref{CFF_def}), which enter the $\gamma^* p \rightarrow \pi^o p'$ reaction. 
The flavor content of both the chiral even and chiral odd GPDs
follows from the  SU(3) flavor symmetry for the pseudo-scalar meson octet. In particular, for the chiral odd sector we have,
\begin{eqnarray}
\label{octet}
{\cal F}_T^{\pi^o}  & =  & \frac{1}{\sqrt{2}} (e_u {\cal F}_T^u - e_d {\cal F}_T^d) \\
{\cal F}_T^{\eta}  & =  &  \frac{1}{\sqrt{6}} (e_u {\cal F}_T^u + e_d {\cal F}_T^d - 2 e_s {\cal F}_T^s)\\
\end{eqnarray}
where $F_T^{q} = H_T^{q}, E_T^{q}, \widetilde{H}_T^{q}, \widetilde{E}_T^{q}$, and $e_q$, $q=u,d,s$, is the quark's charge. Notice, however, that in our calculations we have set the $s$ quark GPDs to zero.

%

\subsection{The t-channel Analysis}
\label{sec2d} 
It is of interest to separate the role of chiral odd GPDs from chiral even GPDs in $\pi^0$ electroproduction. It is widely stated that the chiral even $\widetilde{H}$ and $\widetilde{E}$ are the sole contributions to the longitudinal photon cross section. Yet the t-channel decompositions for these suggest otherwise. Let us examine this proposition. In order to match the definite negative C-parity of the $\gamma^* \pi^0$,  the crossing and spin symmetry behavior of the GPDs must be selected. Furthermore the Dirac structure of the quark correlator must have the negative parity structure guaranteed by a factor of $\gamma^5$, so only  $\widetilde{H}$ and $\widetilde{E}$ contribute. The antisymmetric combinations under $x\rightarrow - x$ have negative C-parity~\cite{ChenJi,Hagler,GGL_short}. Further analysis in Ref.~\cite{GGL_short} shows that the $\widetilde{H}$ contains the series $2^{- -}, 4^{- -}, . . . $, while $\widetilde{E}$ contains $1^{+-}, 2^{- -}, 3^{+-}, 4^{- -} . . . $. 

Consider $\widetilde{E}$ first. Now an important observation is that because of C-parity there is no $0^{-+} \pi$ pole contribution to $\widetilde{E}$. It is expected from various model calculations that the remaining contributions to $\widetilde{E}$ will be appreciably smaller~\cite{Goeke,GolKro}, although we see that for the antisymmetric crossing combination there can be the quantum numbers of the $b_1, h_1$ axial vector mesons. While the first moment of the symmetric $\widetilde{E}(x,\xi,t)$ is the pseudoscalar form factor, for the first moment of the antisymmetric $\widetilde{E}(x,\xi,t)$ there is no simple phenomenological connection, except to lattice calculations of the generalized form factors. Nevertheless, we will keep the possibility of an axial vector pole contribution in mind below. 

The contribution of $\widetilde{H}$ to the $\pi^0$ involves a series that begins with $2^{- - }$. There are no known particle candidates for that state, either isoscalar or isovector. If we consider the Regge pole contributions to the antisymmetric $\widetilde{H}$, then, the trajectory would have to be lower than the well established trajectories and the first physical pole, far from the scattering region. Note that the absence of a $0^{- -}$ would require a factor of $\alpha(t)$ in the overall Regge residue (nonsense zero) for this case. Similar to $\widetilde{E}$, the first moment of the symmetric $\widetilde{H}(x,\xi,t)+\widetilde{H}(-x,\xi,t)$ is the axial vector form factor corresponding to the $a_1$ quantum numbers. But the antisymmetric case does not have such an interpretation. The $\widetilde{H}^q(x,0,0) = g_1^q(x)$, so it is known that $\widetilde{H}^q$ cannot be small at the boundary. For the process here, however, the $J^{PC}=2^{--}$ will suppress the non-singlet contribution for small $x$ and $\mid t \mid$. The considerably smaller value of the longitudinal cross section for $\pi^0$ corroborates this conclusion. 

\section{Cross Sections and Asymmetries}
\label{sec:xsec}
The various GPDs calculated in Section \ref{sec:model} enter the cross section terms for $\pi^o$ electroproduction, which, using the notation of Ref.\cite{Bacchetta_review} (based on \cite{DieSap}),  can be defined as, 
\begin{widetext}
\begin{eqnarray} 
\label{xs}
\frac{d^4\sigma}{dx_{Bj} dy d\phi dt} & = & \Gamma \left\{ \left[ F_{UU,T} + \epsilon F_{UU,L}+ \epsilon \cos 2\phi F_{UU}^{\cos 2 \phi} 
+ \sqrt{\epsilon(\epsilon+1)} \cos \phi F_{UU}^{\cos \phi}   + 
h  \, \sqrt{\epsilon(1-\epsilon)} \, \sin \phi F_{LU}^{\sin \phi} \right] \right. \nonumber \\
& + & S_{||} \left[   \sqrt{\epsilon(\epsilon+1)} \sin \phi F_{UL}^{\sin \phi}  + \epsilon \sin 2 \phi F_{UL}^{\sin 2 \phi}  + h \, 
\left( \sqrt{1 - \epsilon^2} \, F_{LL} +    \sqrt{\epsilon(1-\epsilon)} \, \cos \phi \, F_{LL}^{\cos \phi}  \right) \right]   \nonumber \\
%
& - &  S_\perp \left[ \sin(\phi-\phi_S) \left(F_{UT,T}^{\sin(\phi-\phi_S)} + \epsilon  F_{UT,L}^{\sin(\phi-\phi_S)}  \right) + 
\frac{\epsilon}{2} \left( \sin(\phi+\phi_S) F_{UT}^{\sin(\phi+\phi_S)}  +  \sin(3\phi-\phi_S) F_{UT}^{\sin(3\phi-\phi_S)} \right)  \right. \nonumber \\
& + & \left. \sqrt{\epsilon(1+\epsilon)} \left( \sin\phi_S F_{UT}^{\sin \phi_S} + \sin(2\phi-\phi_S) F_{UT}^{\sin(2\phi-\phi_S)} \right) \right]   \nonumber \\
& + & \left. S_\perp h \left[ \sqrt{1-\epsilon^2} \cos(\phi-\phi_S) F_{LT}^{\cos(\phi-\phi_S)} +
\sqrt{\epsilon(1-\epsilon)} \left(\cos \phi_S F_{LT}^{\cos\phi_S} + \cos(2\phi-\phi_S) F_{LT}^{ \cos(2\phi-\phi_S)} \right) \right] \right\} \nonumber \\
\end{eqnarray}
\end{widetext}
where 
$S_{||}$ and ${\bf S}_\perp$ refer to lab frame target polarization parallel and perpendicular to the virtual photon direction, $h$ is the lepton beam helicity, $\phi$ is the azimuthal angle between the lepton plane and the hadron scattering plane, $\phi_S$ is the azimuthal angle of the transverse spin vector ${\bf S}_\perp$ and $t$ is the square of the invariant momentum transfer between the initial and final nucleon.
%

The photon polarization parameter $\epsilon$, the ratio of longitudinal photon and transverse photon flux, can be written in terms of invariants as,
\begin{equation}
\epsilon^{-1} = 1 + 2\left( 1+\frac{\nu^2}{Q^2} \right)\left(4 \dfrac{\nu^2}{Q^2} \dfrac{1-y}{y^2}-1\right)^{-1}. 
\end{equation}
\noindent 
$\Gamma$ is given by,
\begin{equation}
\label{Gamma}
\Gamma = \frac{\alpha^2 \, y^2 (1-x_{Bj})}{2\pi x_{Bj}(1-\epsilon)Q^2}.
\end{equation}
%
%
%

We also list for completeness the alternative notations that are frequently used in the literature,
\begin{eqnarray}
F_{UU,T} & = & \frac{d\sigma_T}{dt} = \sigma_T,  \;\;\; F_{UU,L}=\frac{d\sigma_L}{dt} = \sigma_L, \;\;\;     F_{UU}^{\cos \phi} =\frac{d\sigma_{LT}}{dt} = \sigma_{LT}, \nonumber \\ 
F_{UU}^{\cos 2\phi} & = &  \frac{d\sigma_{TT}}{dt} = \sigma_{TT},  \;\;\; F_{LU}^{\sin \phi}  =   \frac{d\sigma_{LT'}}{dt} = \sigma_{LT'}, \;\;\;  {\rm etc..}  
\label{xsecs}
\end{eqnarray}
\begin{figure}
\includegraphics[width=8.cm]{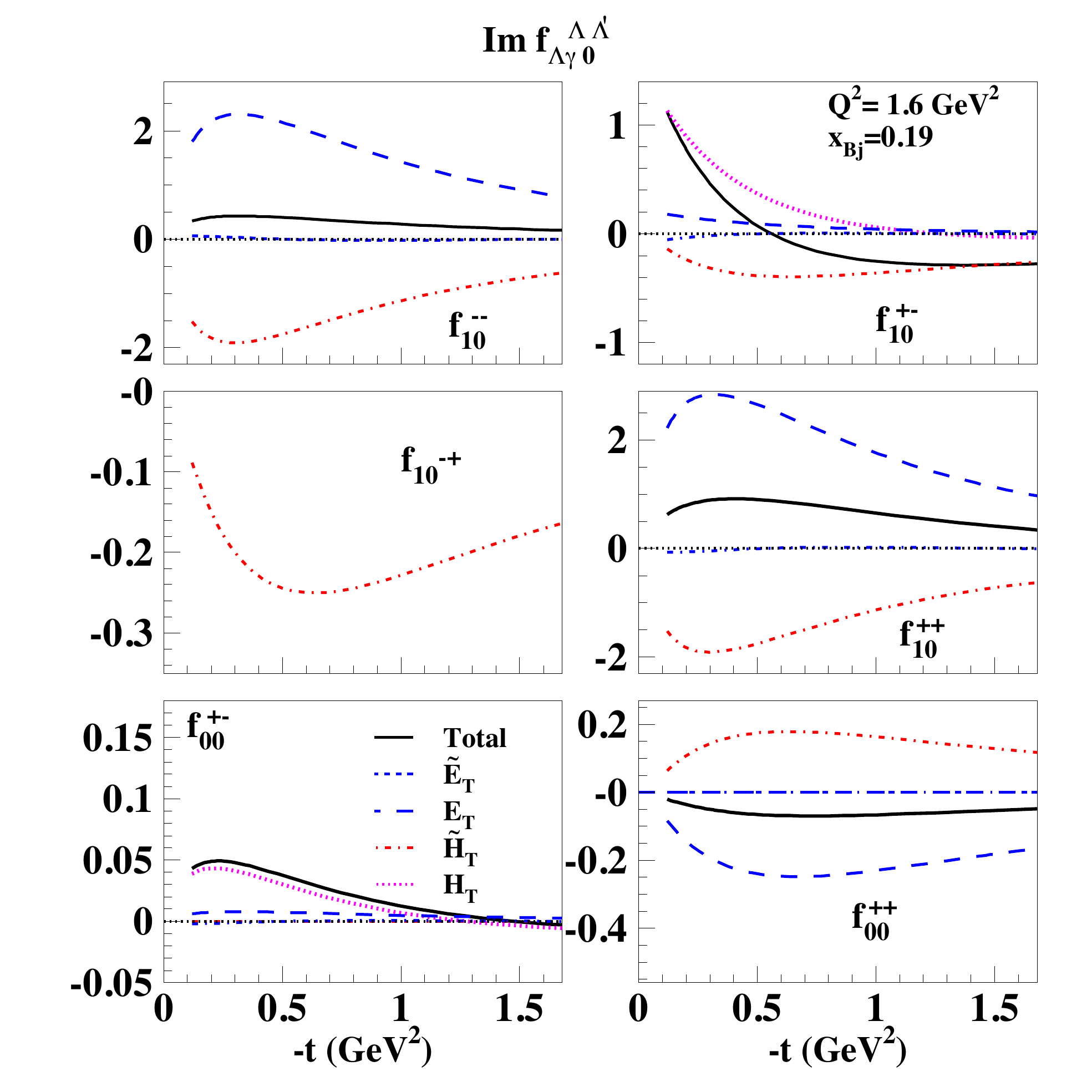}
\hspace{0.3cm}
\includegraphics[width=8.cm]{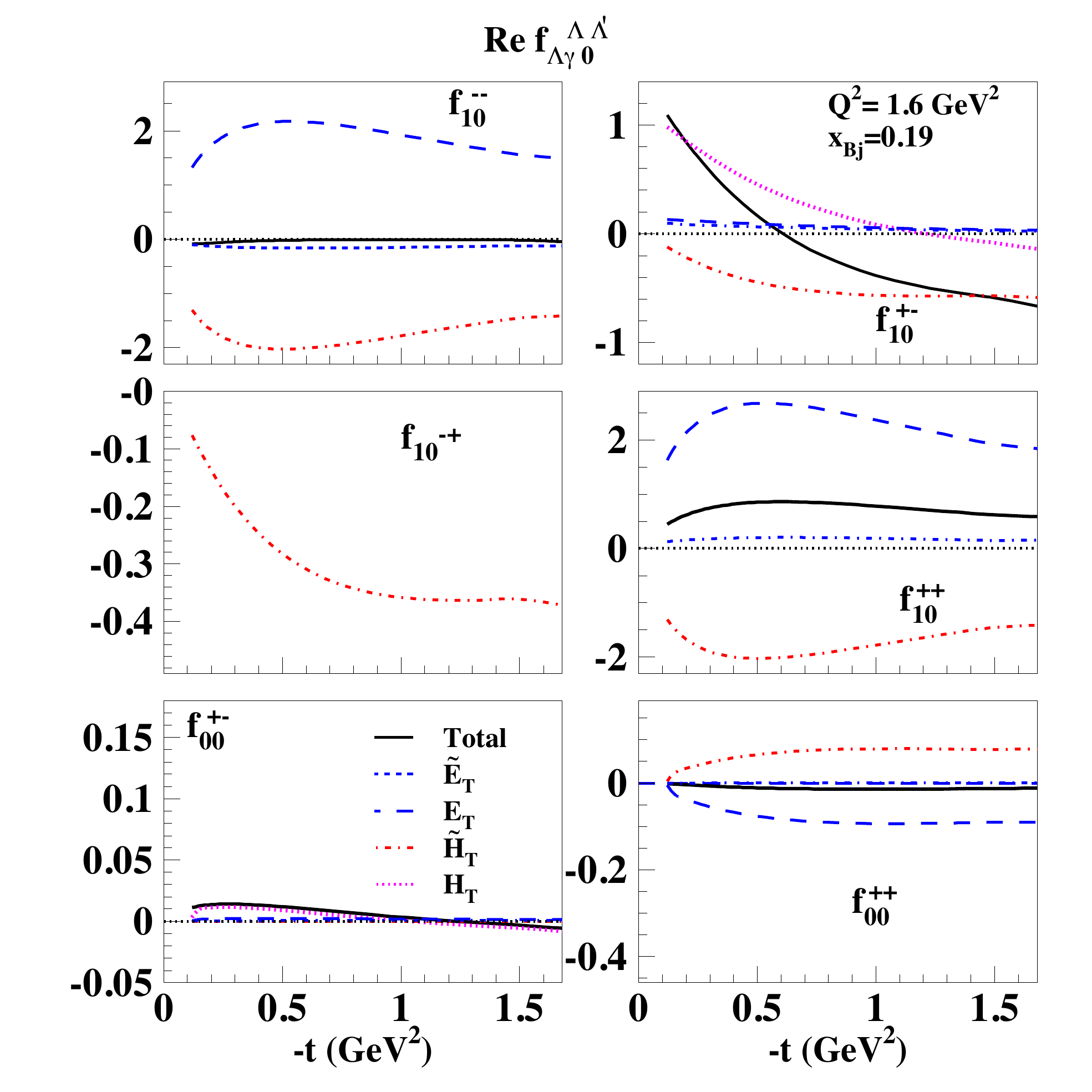}
\caption{(Color online) Helicity amplitudes for both transverse photon polarization, Eqs.(\ref{helamps_gpd2}), and longitudinal photon polarization, Eqs.(\ref{helamps_00}) plotted vs. $-t$ for $x_{Bj} =0.19$, $Q^2= 1.6$ GeV$^2$. 
The imaginary parts are displayed on the left panel, and the real parts on the right panel.}
\label{fig:amps1}
\end{figure}
In order to predict/interpret  experimental results on the various cross section components and the asymmetries constructed through them, 
it is important to devise a scheme that helps us navigate through this elaborate  set of functions. 
For each observable (or set of observables), we will show a decomposition in both the various amplitudes and the various contributing GPDs. Physical 
information will be more easily extracted this way, in cases  where one of the amplitudes (or one particular combination of amplitudes) dominates. 

We discuss in order: {\it i)} the helicity amplitudes in terms of chiral odd GPDs; {\it ii)} the various contributions to Eq.(\ref{xs}) in terms of both helicity amplitudes and GPDs. 

\subsection{Helicity Amplitudes as functions of GPDs}
The  helicity amplitudes are shown in Fig.\ref{fig:amps1} as a function of $-t$, for $x_{Bj}=0.19$, $Q^2 = 1.6$ GeV$^2$ (similar results are obtained for the other kinematical bins in the range of Jefferson Lab data \cite{Kub}). 
The imaginary (real) parts are displayed on the LHS(RHS). The different contributions from the various chiral odd GPDs,  are also shown in the figure.  

We recall the structure of the transverse amplitudes, Eq.(\ref{helamps_gpd2}),
\begin{subequations}
\label{helamps_gpd2_simple}
\begin{eqnarray}
f_{10}^{++} & \propto  & \Delta   \left( 2\widetilde{\cal H}_T  + (1-\xi) {\cal E}_T - (1-\xi)   \widetilde{\cal E}_T  \right)
\\
f_{10}^{+-} & \propto &    {\cal H}_ T + \frac{t_0-t}{4M^2} \widetilde{\cal H}_T    
 +  \frac{\xi^2}{1-\xi^2}  {\cal E}_T  + \frac{\xi}{1-\xi^2} \widetilde{\cal E}_T  \\
 f_{10}^{-+} &  \propto & \Delta^2 \: \widetilde{\cal H}_T  
 \\
f_{10}^{--} & \propto & \Delta  \left(  2\widetilde{\cal H}_ T + (1+\xi)  {\cal E}_T +
(1+\xi) \widetilde{\cal E}_T  \right), 
\end{eqnarray} 
\end{subequations}
Regarding the GPD content of the amplitudes we can deduce the following:

\vspace{0.5cm}
\noindent
(1) All GPDs contributions should be considered separately. In particular, $H_T$, $\widetilde{H}_T$, and $E_T$ are dominating; $\widetilde{E}_T$ is non zero in our model but small. 
Although the combination $2 \widetilde{H}_T + E_T$ might be considered more fundamental in that its spin structure corresponds to the Boer-Mulders function \cite{DieHag_odd,Bur2}, and its first moment yields the proton's transverse anomalous magnetic moment \cite{Bur2}, $\widetilde{H}_T$, and $E_T$ appear separately, and multiplied by different factors in the amplitudes. $2 \widetilde{H}_T + E_T$ should just be viewed as a forward limit. 
\vspace{0.3cm} 

\noindent
(2) The behavior of $f_{10}^{++}$ and $f_{10}^{--}$ is determined by $\widetilde{H}_T$, and $E_T$. As a consequence of what explained in point (1), $f_{10}^{--}$ is sensibly different from $f_{00}^{++}$.  In particular, because of the different multiplicative factors, $f_{10}^{--} <  f_{10}^{++}$.
\vspace{0.3cm} 

\noindent 
(2) $f_{10}^{+-}$ is determined by $H_T$ at small $\mid t \mid$, and by $E_T$ at large $\mid t \mid$.
\vspace{0.3cm} 

\noindent 
(3) $f_{10}^{-+}$ is determined by $\widetilde{H}_T$ only, but it is small due to the $\mid t \mid$ factor suppression. 
\vspace{0.3cm} 

\noindent 
(5) The longitudinal photon contributions, $f_{00}^{+-}$, and $f_{00}^{++}$ are suppressed in the chiral odd case.
\vspace{0.3cm}

\subsection{Unpolarized Target}
The various terms describing scattering from an unpolarized target in Eq.(\ref{xs}) are written in terms of helicity amplitudes, Eqs.(\ref{helamps_gpd2}),  as,
\begin{widetext}
\begin{eqnarray}
\label{dsigT}
F_{UU,T} & = & \frac{1}{2}  (F_{1 1}^{++} + F_{1 1}^{--})  = \frac{1}{2} \sum_{\Lambda'} ( f_{10}^{+ \Lambda' *}  f_{10}^{+ \Lambda'} +  f_{10}^{- \Lambda' *}  f_{10}^{- \Lambda'} )  \nonumber \\
& = &\,  \frac{1}{2} \left( \mid f_{10}^{++} \mid^2 + \mid f_{10}^{+-}  \mid^2 + \mid f_{10}^{-+}  \mid^2 +
\mid f_{10}^{--}  \mid^2 \right) \\ 
\nonumber \\
\label{dsigL}
F_{UU,L} & = &  F_{00}^{++} =  \sum_{\Lambda'}  f_{00}^{+ \Lambda' *}  f_{00}^{+ \Lambda'} =  \mid f_{00}^{++}  \mid^2 + \mid f_{00}^{+-}  \mid^2  \\
\label{dsigTT}
\nonumber \\
F_{UU}^{\cos 2 \phi} & = & -  \Re e  \, F_{1 -1}^{++}  =  -  \Re e  \,  \sum_{\Lambda'}  f_{10}^{+ \Lambda' *}  f_{-10}^{+ \Lambda'}  \nonumber \\
& = &  -  \,    \Re e \left[  (f_{10}^{++})^ * (f_{10}^{--}) - (f_{10}^{+-})^* (f_{10}^{-+}) \right] \\
\nonumber \\
\label{dsigLT}
F_{UU}^{\cos \phi}  & = & \,    \Re e  (F_{10}^{++} + F_{10}^{--}) =     \Re e   \sum_{\Lambda'} ( f_{00}^{+ \Lambda' *}  f_{10}^{+ \Lambda'} +  f_{00}^{- \Lambda' *}  f_{10}^{- \Lambda'} ) \nonumber \\
& = &   \Re e \! \left[   (f_{00}^{+-})^*(f_{10}^{+-} + f_{10}^{-+}) + (f_{00}^{++})^* (f_{10}^{++} - f_{10}^{--}) \right] 
 \\
\nonumber \\
\label{dsigLTp}
F_{LU}^{\sin \phi} & = & \,   - \Im m  (F_{10}^{++} + F_{10}^{--}) =    - \Im m   \sum_{\Lambda'} ( f_{00}^{+ \Lambda' *}  f_{10}^{+ \Lambda'} +  f_{00}^{- \Lambda' *}  f_{10}^{- \Lambda'} ) 
\nonumber \\
& = & - \Im m  \! \left[  (f_{00}^{+-})^*(f_{10}^{+-} + f_{10}^{-+}) +  (f_{00}^{++})^* (f_{10}^{++} - f_{10}^{--}) \right] 
\end{eqnarray}
\end{widetext}
%

\subsubsection{Cross Section Components}
\begin{figure}
\includegraphics[width=8.5cm]{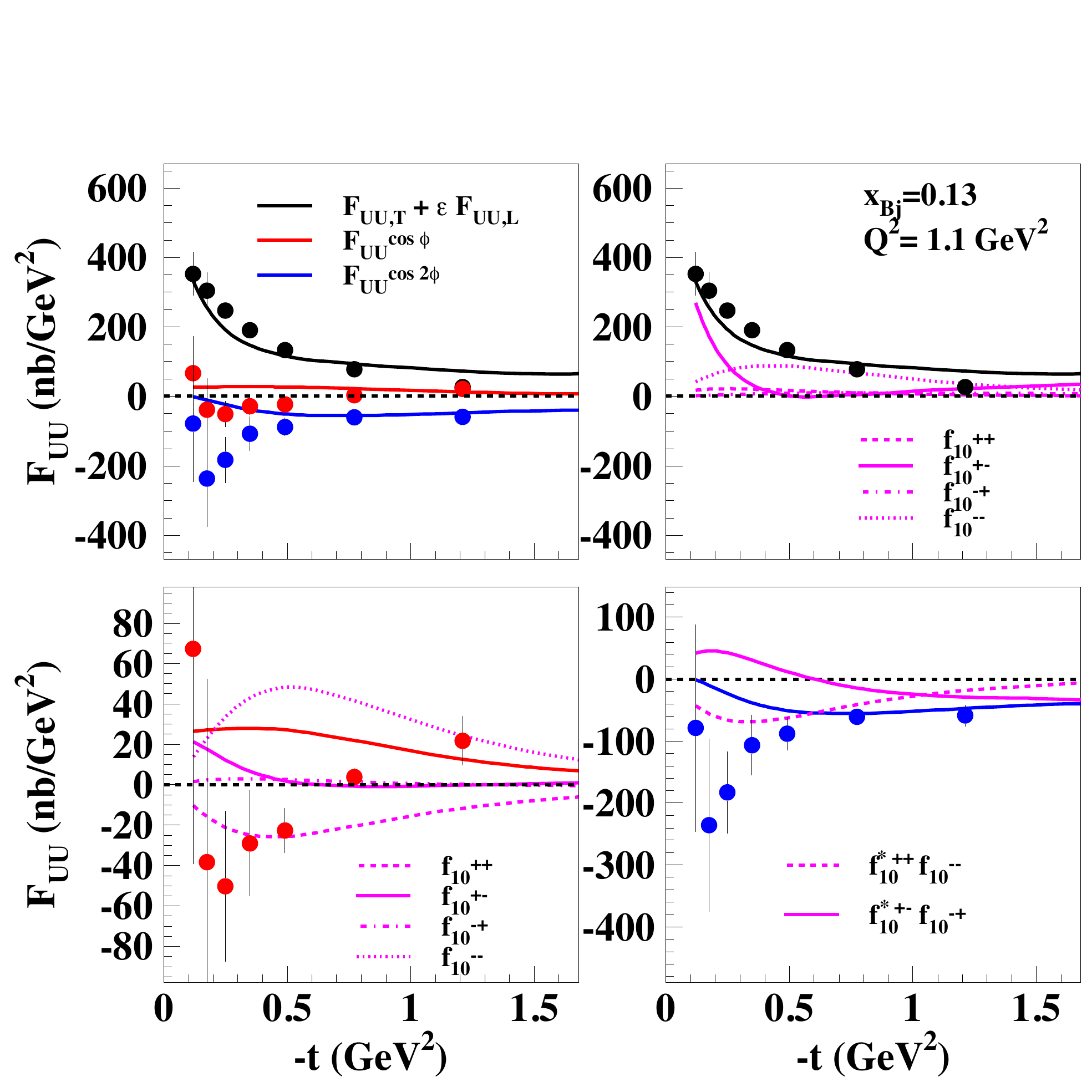}
\hspace{0.3cm}
\includegraphics[width=8.5cm]{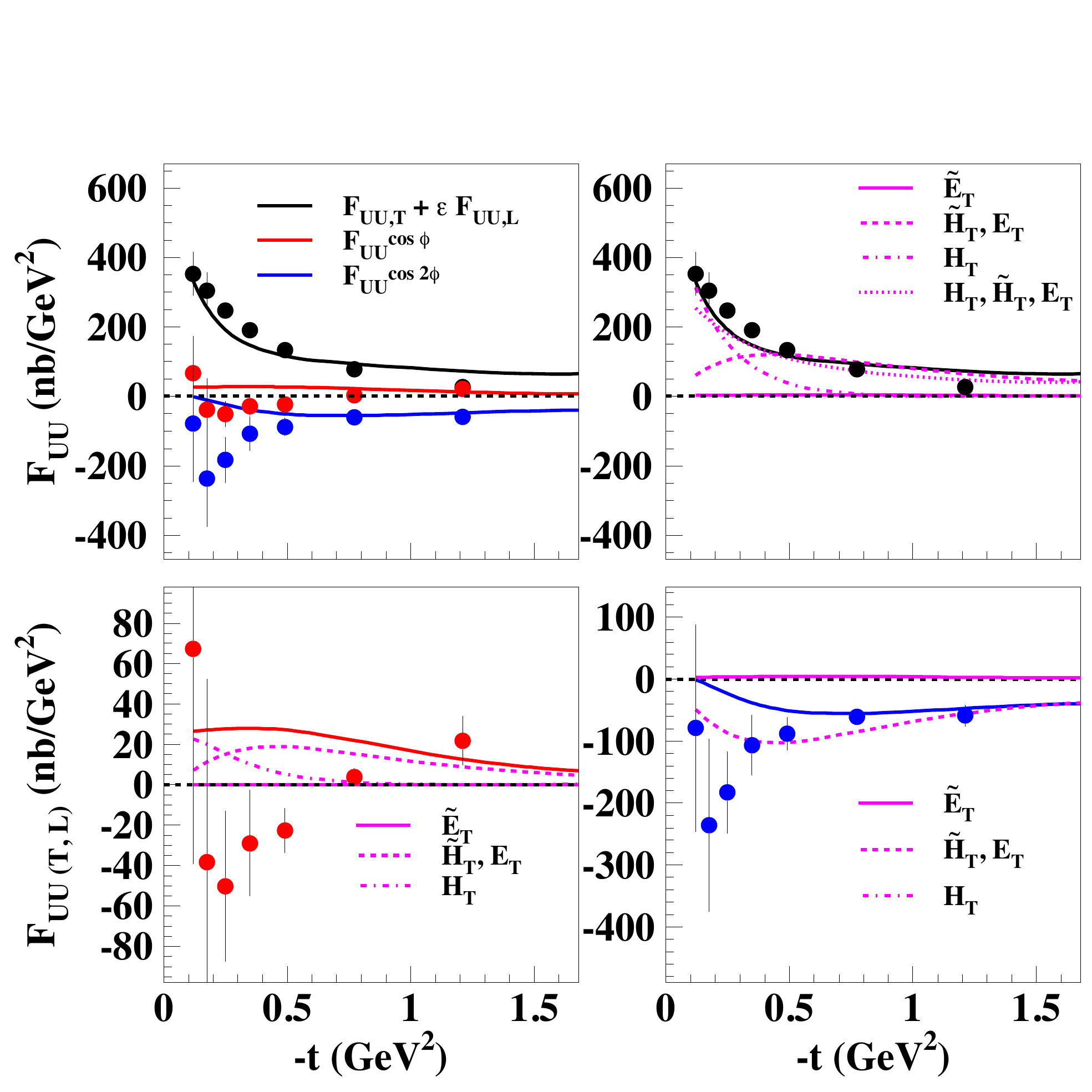}
\caption{(color online) LEFT: Unpolarized cross section components, $F_{UU,T}+\epsilon \, F_{UU,L}$, $F_{UU}^{\cos 2 \phi}$, and $F_{UU}^{\cos \phi}$ in the kinematical bin, $x_{Bj} = 0.13$, $Q^2=1.2$ GeV$^2$. The upper left panel shows all components along with the data from Ref.\cite{Kub}.
The other panels show the contributions from the various helicity amplitudes. 
The right upper panel shows $F_{UU,T}+\epsilon \, F_{UU,L}$, and the contributions from $f_{10}^{++}$,  $f_{10}^{+-}$, $f_{10}^{-+}$  and $f_{10}^{--}$.  
Similarly, the lower left panel and the lower right panel  show the contributions of the various amplitudes to $F_{UU}^{\cos  \phi}$, and $F_{UU}^{\cos 2 \phi}$, respectively; 
RIGHT: Same as LEFT, displaying the GPDs components. The full curve is obtained by using only 
$\widetilde{E}_T$, the dashed curves by including only $2\widetilde{H}_T\pm (1\pm \xi)E_T$, the dot-dashed curve by including only $H_T$, and the dotted curve by including all GPDs, except for  $\widetilde{E}_T $. }
\label{fig_uu1}
\end{figure}

\begin{figure*}
\includegraphics[width=8.cm]{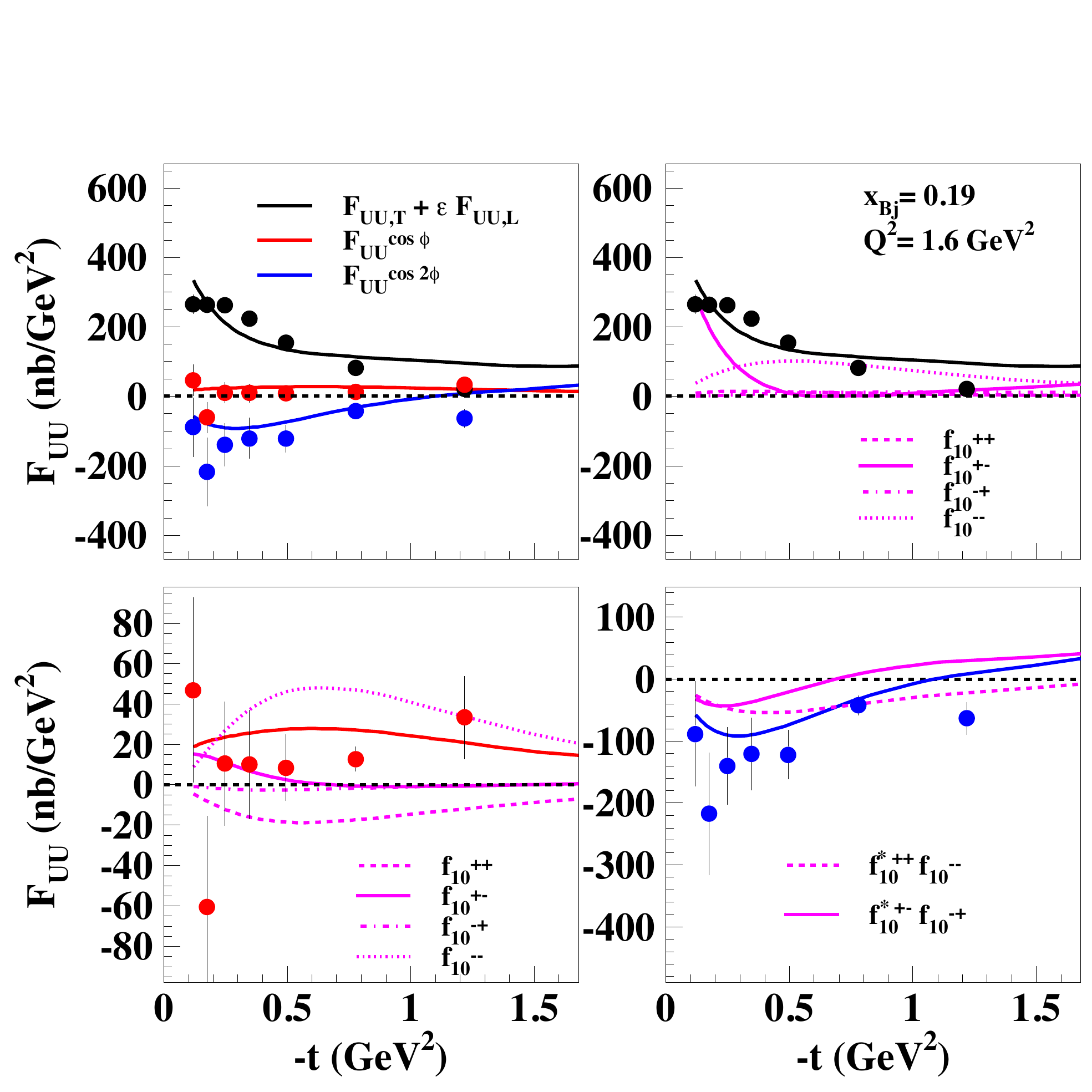}
\hspace{0.3cm}
\includegraphics[width=8.cm]{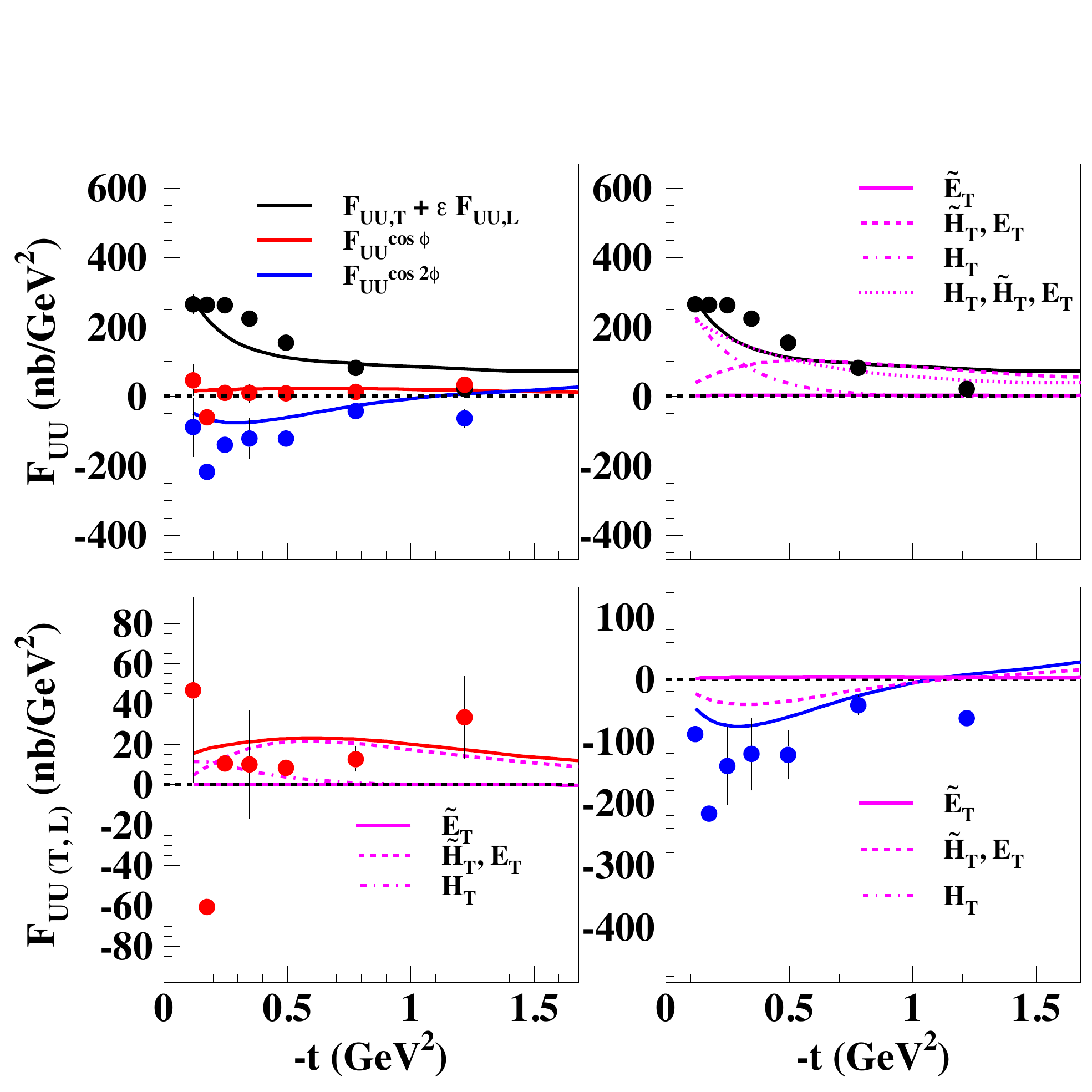}
\caption{(color online) Same as Fig.\ref{fig_uu1} for the kinematics $x_{Bj}=0.19$, $Q^2=1.6$ GeV$^2$.}
\label{fig_uu2}
\end{figure*}

\begin{figure*}
\includegraphics[width=8.cm]{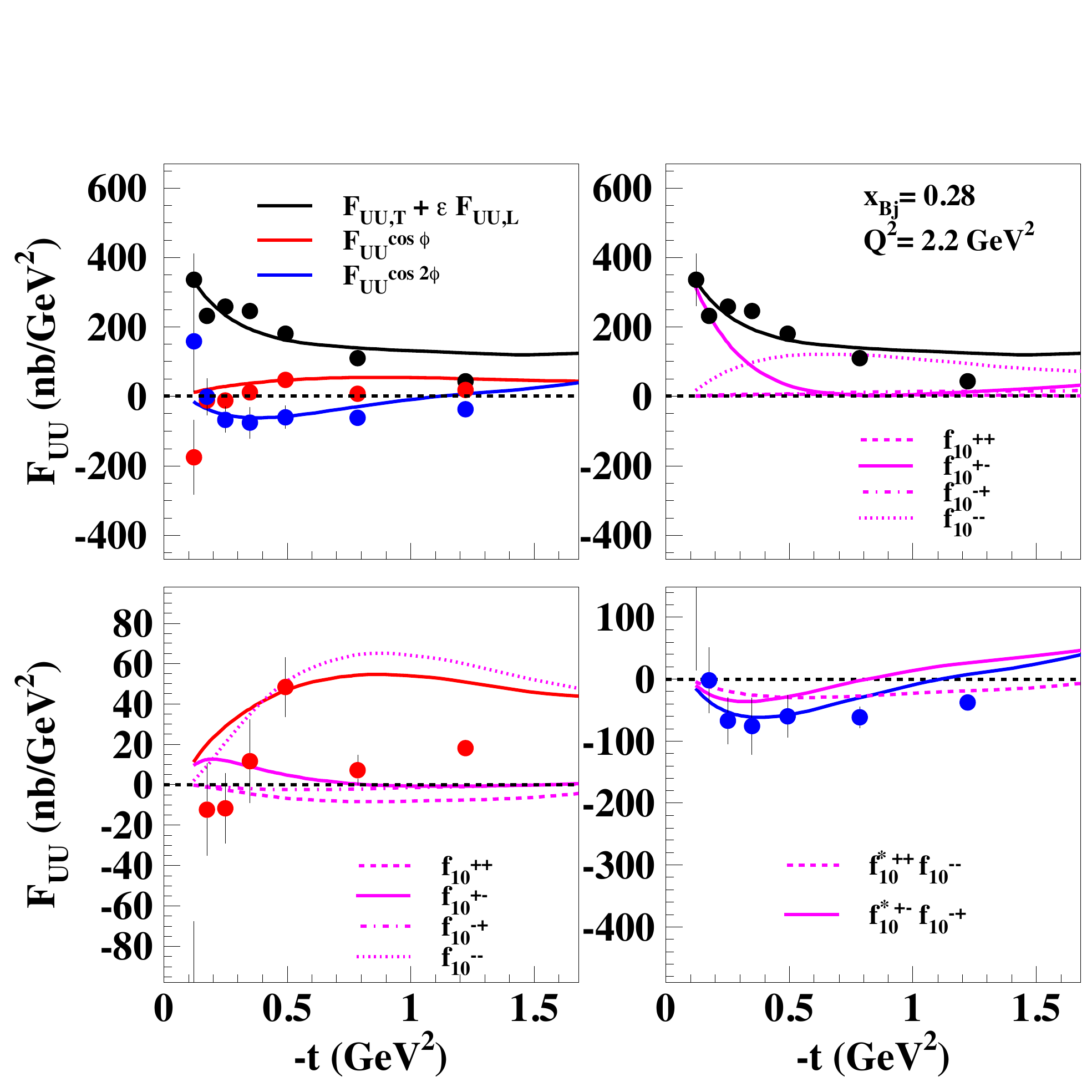}
\hspace{0.3cm}
\includegraphics[width=8.cm]{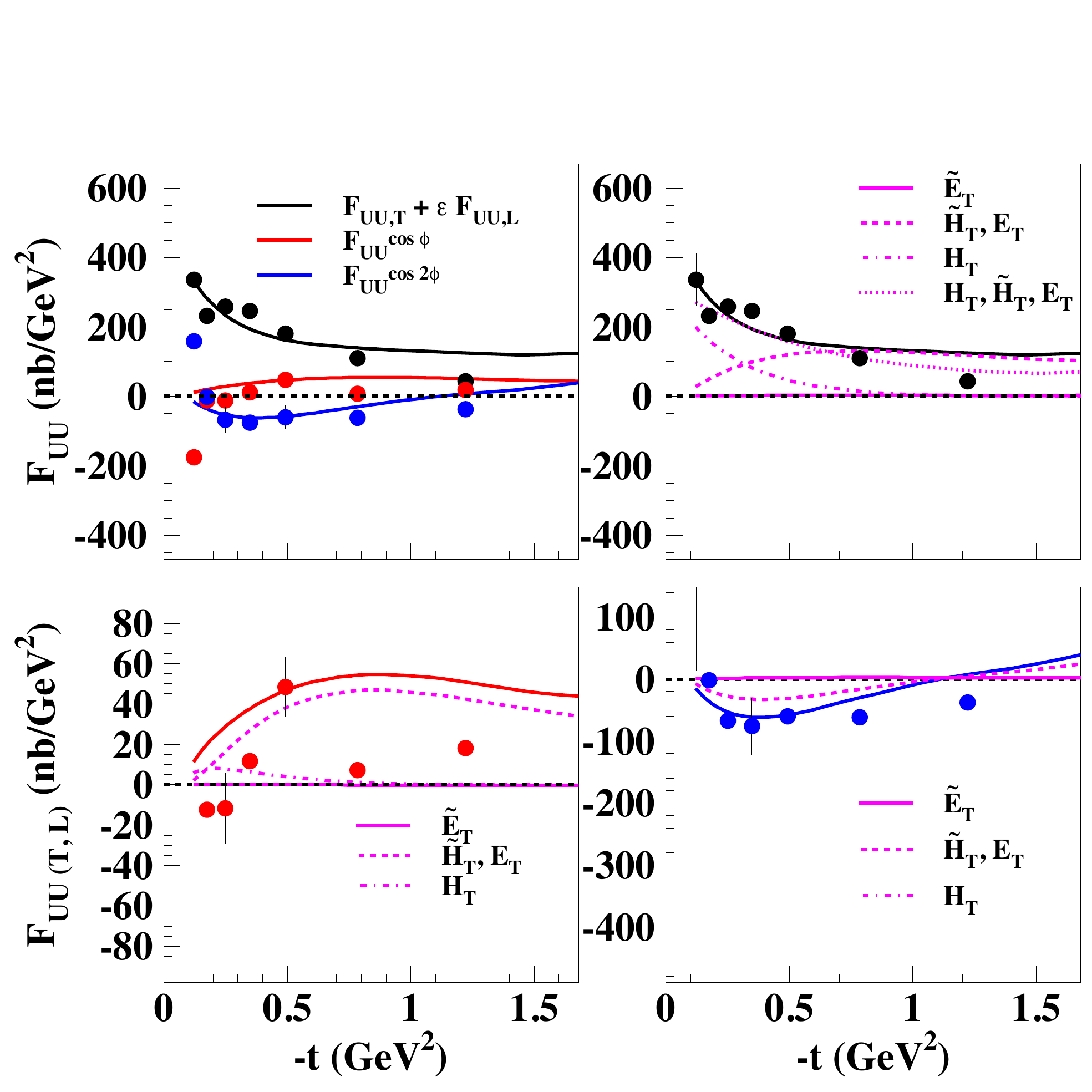}
\caption{(color online) Same as Fig.\ref{fig_uu1} for the kinematics  $x_{Bj}=0.27$, $Q^2=2.2$ GeV$^2$.}
\label{fig_uu3}
\end{figure*}

In Figures  \ref{fig_uu1}, \ref{fig_uu2}, \ref{fig_uu3} we show the  unpolarized cross section components, $F_{UU,T}+\epsilon \, F_{UU,L}$, $F_{UU}^{\cos 2 \phi}$ , and $F_{UU}^{\cos \phi}$ for the kinematics: $x_{Bj} = 0.13$, $Q^2=1.2$ GeV$^2$ (Fig. \ref{fig_uu1}), $x_{Bj} = 0.19$, $Q^2=1.6$ GeV$^2$, (Fig.\ref{fig_uu2}), and $x_{Bj} = 0.28$, $Q^2=2.2$ GeV$^2$ (Fig.\ref{fig_uu3}). 

\noindent In the left panel we show how the various amplitudes contribute to the cross sections components, going clockwise from the upper left corner: 

\vspace{0.3cm}
\noindent \underline{LEFT}

\noindent {\it i)} the unpolarized cross section components, $F_{UU,T}+\epsilon \, F_{UU,L}$, $F_{UU}^{\cos 2 \phi}$, and $F_{UU}^{\cos \phi}$  in the kinematical bin, $x_{Bj} = 0.13$, $Q^2=1.2$ GeV$^2$,  along with the data from Ref.\cite{Kub}

\noindent 
{\it ii)}  the contributions from the various helicity amplitudes $f_{10}^{++}$ (dashes),  $f_{10}^{+-}$, $f_{10}^{-+}$ and $f_{10}^{--}$ 
to $F_{UU,T}+\epsilon \, F_{UU,L}$;

\noindent 
{\it iii)}  the same for $F_{UU}^{\cos  \phi}$; 

\noindent {\it iv)}  the contributions of the combinations $f_{10}^{* ++} f_{10}^{--}$, and $f_{10}^{* +-} f_{10}^{-+}$ to $F_{UU}^{\cos 2 \phi}$.
\vspace{0.3cm}

\noindent In the right panel we show the GPDs content of the various cross sections components independently from which amplitude they enter (clockwise from the upper left corner):  

\vspace{0.3cm}
\noindent\underline{RIGHT}

\noindent {\it i)} the same as in the left panel (upper left); 

\noindent {\it ii)}  the GPDs contributing to $F_{UU,T}+\epsilon \, F_{UU,L}$, ; 

\noindent {\it iii)}   the GPDs contributing to $F_{UU}^{\cos  \phi}$

\noindent {\it  iv)}   the GPDs contributing to $F_{UU}^{\cos 2 \phi}$.
\vspace{0.3cm}

\noindent From {\it ii)} LEFT we see that $F_{UU,T}+\epsilon \, F_{UU,L}$ is dominated by $f_{10}^{+-}$ (low $t$) and $f_{10}^{--}$ (larger $t$). By comparing with {\it ii)} RIGHT we see that $H_T$ dominates at low $t$ but it is taken over by $2\widetilde{H}_T\pm (1\pm \xi)E_T$ at larger $t$. This is consistent with the behavior of the amplitudes displayed in Fig.\ref{fig:amps1}.
From {\it iii)} LEFT we see that the dominant contributions to $F_{UU}^{\cos \phi}$ are given by  $f_{10}^{--}$ and $f_{10}^{++}$, although $f_{10}^{+-}$ contributes at very small $t$. By comparing with {\it ii)} RIGHT in terms of GPDs $H_T$ dominates at low $t$ and $2\widetilde{H}_T\pm (1\pm \xi)E_T$ at larger  $t$.
Finally,  in {\it iv)} LEFT  $F_{UU}^{\cos 2 \phi}$ is given by similar contributions from all the amplitude combinations and it is therefore harder to interpret. From {\it iv)} RIGHT the one can see that the GPDs $2\widetilde{H}_T\pm (1\pm \xi)E_T$ and $H_T$ contribute almost equally in the whole $t$ regime.
%

\begin{figure*}
\includegraphics[width=8.cm]{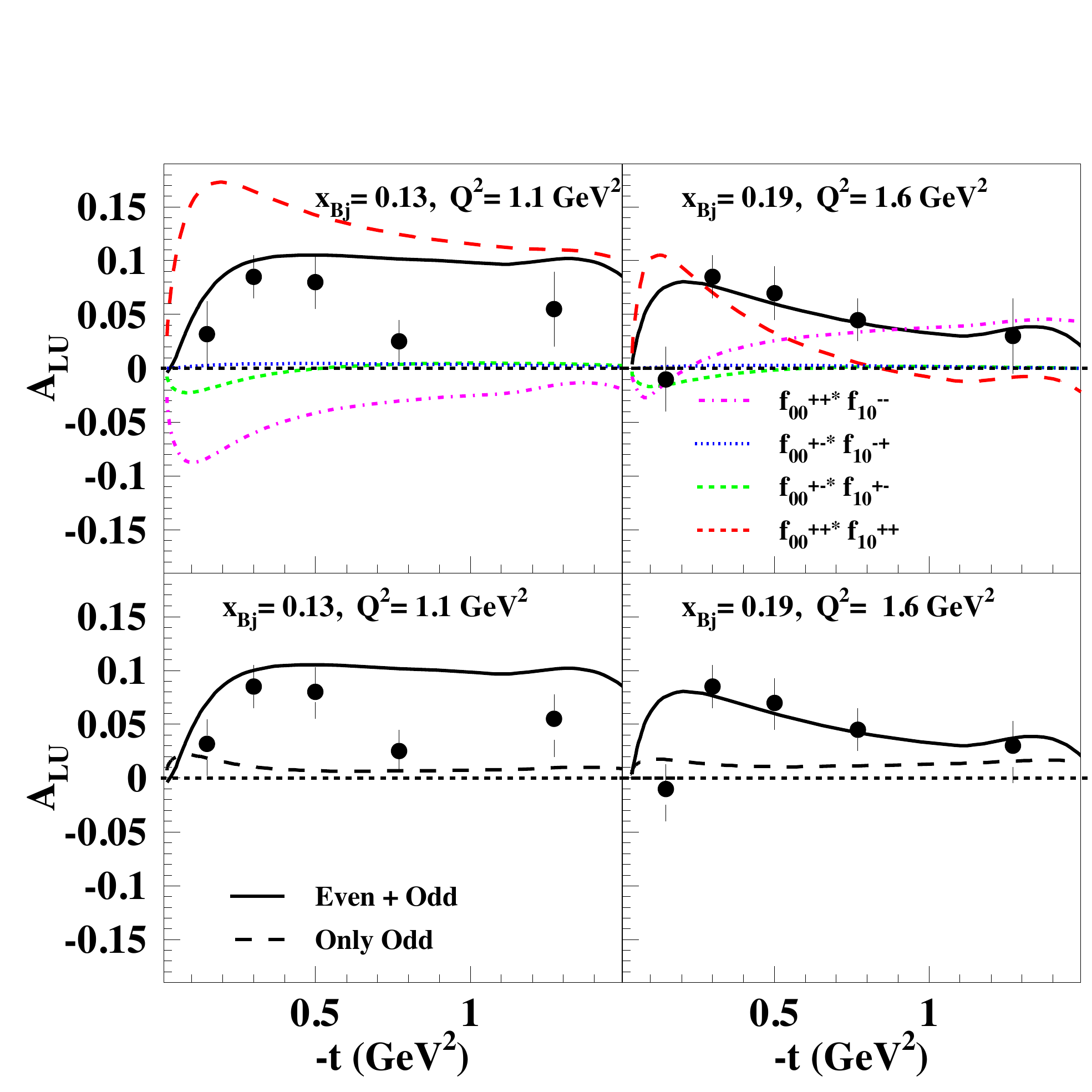}
\caption{(Color online) Beam spin asymmetry, $A_{LU}$, plotted vs. $-t$ for two different kinematics: $Q^2=1.1$ GeV$^2$,  
$x_{Bj}=0.13$ (left), $Q^2=1.6$ GeV$^2$,  
$x_{Bj}=0.19$ (right). 
Experimental data from Ref.\protect\cite{demasi}. In the upper panels the different helicity amplitudes combinations contributing to $A_{LU}$, Eqs.(\ref{dsigLTp},\ref{ALU}), are shown.  The full curve describes the result obtained including all combinations.  In the lower panels we show results obtained including both the chiral even and odd GPDs (full curve) compared to results obtained using the only the chiral odd contribution (dashes). We conclude that the chiral even GPDs dominate this observable.}
\label{fig:ALU}
\end{figure*}

The unpolarized $\sin \phi$ modulation, $F_{LU}^{\sin \phi}$ describes the beam asymmetry, $A_{LU}$,  which is implicit in the term involving $h$ in Eq.(\ref{xs}), 
\begin{equation}
A_{LU}  = \sqrt{\epsilon(1-\epsilon)}  \, \frac{F_{LU}^{\sin \phi} }{F_{UU,T} + \epsilon F_{UU,L}}
\label{ALU}
\end{equation}
$A_{LU}$ is shown in Fig.\ref{fig:ALU} for two of the Jefferson Lab Hall B  kinematical bins along with the different amplitudes contributions, in this case the products: $(f_{10}^{++*} f_{00}^{++})$,  
$(f_{10}^{--*} f_{00}^{++})$, $(f_{10}^{-+*} f_{00}^{+-})$ and $(f_{10}^{+-*} f_{00}^{+-})$, appearing in Eq.(\ref{dsigLTp}). 
Notice that the longitudinally polarized amplitudes receive contributions from both the chiral even and chiral odd GPDs (see Section \ref{even:sec} ). From the graph (lower panels) one can see a definite dominance of the chiral even GPDs. We deduce that  $A_{LU}$ is not favored for the extraction of chiral odd GPDs. 

\subsubsection{$Q^2$ dependence}
\begin{figure*}
\includegraphics[width=8.cm]{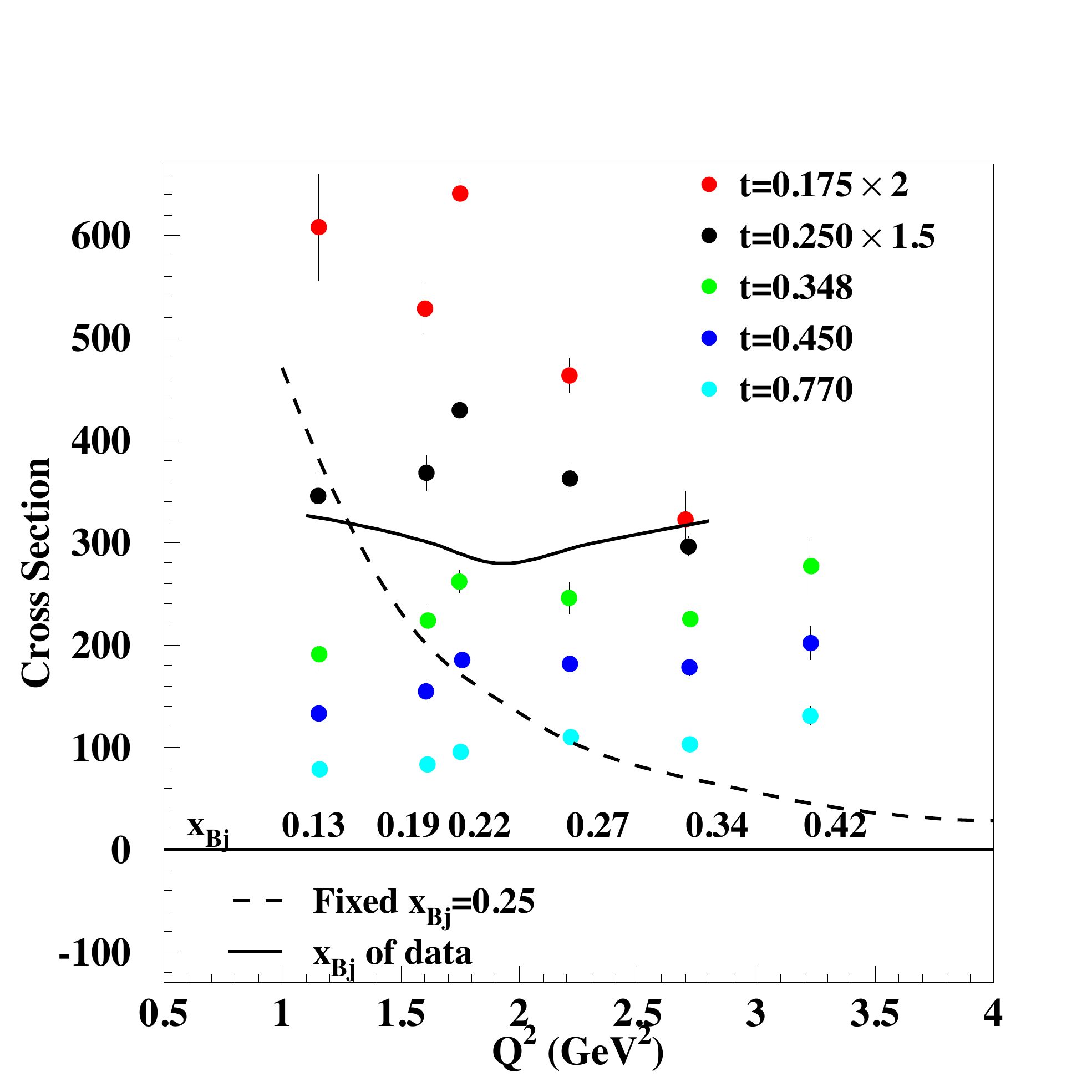}
\caption{(Color online) Cross section, $\sigma_T + \epsilon \sigma_L = F_{UU,T}+\epsilon \, F_{UU,L}$, Eq.(\ref{dsigT},\ref{dsigL}), plotted vs. $Q^2$ at $t$ and $x_{Bj}$ values corresponding to the data from Ref.\cite{Kub}.}
\label{fig:q2dep}
\end{figure*}

In Fig.\ref{fig:q2dep}  we show the $Q^2$ dependence of the cross section term, $F_{UU,T}+\epsilon \, F_{UU,L}$, Eqs.\ref{dsigT},\ref{dsigL}, plotted vs. $Q^2$ at $t$ and $x_{Bj}$ values corresponding to the data from Ref.\cite{Kub}. The full curve was calculated using the same kinematics as the data for $-t = 0.25$ GeV$^2$ (calculations at the other kinematical values give similar results and they are not displayed for simplicity).  The dashed curve was obtained for one of the bins in $t$ at an average value of $x_{Bj} = 0.23$. Our calculation shows that a straightforward comparison with theory can be performed only in this situation, {\it i.e.} at fixed $x_{Bj}$. While in order to unravel the $Q^2$ dependence of the data a more complete coverage of phase space  is needed, our calculation suggests that the trend of the available data does not contradict theoretical expectations based on the $Q^2$-dependent kinematical factors appearing in the expression for the cross section.

\subsection{Longitudinal Target Polarization}
For longitudinal target polarization,
\begin{widetext}
\begin{eqnarray}
\label{ULsinphi:eq}
F_{UL}^{\sin \phi} & = &  -  \Im m  \, (F_{1 0}^{++}  -  F_{1 0}^{--}) =  -  \Im m  \sum_{\Lambda'} [(f_{00}^{+\Lambda'})^*f_{10}^{+\Lambda'} - (f_{00}^{+\Lambda'})^* f_{10}^{-\Lambda'} ] \nonumber \\
& = &   
- \Im m \! \left[   (f_{00}^{+-})^*(f_{10}^{+-} - f_{10}^{-+}) + (f_{00}^{++})^* (f_{10}^{++} + f_{10}^{--})  \right] 
\\
\label{ULsin2phi:eq}
F_{UL}^{\sin 2 \phi} & = &-  \Im m  \, F_{1 -1}^{++}    =  -   \Im m  \,  \sum_{\Lambda'}  f_{10}^{+ \Lambda' *}  f_{-10}^{+ \Lambda'}  \nonumber \\ 
& = & -     \Im m \left[  (f_{10}^{++})^ * (f_{10}^{--}) - (f_{10}^{+-})^* (f_{10}^{-+}) \right] \\
\label{LLcosphi:eq}
F_{LL}^{\cos \phi} & = &  \,    \Re e  (F_{10}^{++} - F_{10}^{--}) =   \Re e   \sum_{\Lambda'} ( f_{00}^{+ \Lambda' *}  f_{10}^{+ \Lambda'} -  f_{00}^{- \Lambda' *}  f_{10}^{- \Lambda'} )  \nonumber \\
& = & \,  \Re e \! \left[   (f_{00}^{+-})^*(f_{10}^{+-} - f_{10}^{-+}) +  (f_{00}^{++})^* (f_{10}^{++} + f_{10}^{--})  \right] 
\\
\label{LL:eq}
F_{LL} & = & \frac{1}{2}   (F_{1 1}^{++} - F_{1 1}^{--})  = \frac{1}{2} \sum_{\Lambda'} ( f_{10}^{+ \Lambda' *}  f_{10}^{+ \Lambda'} -  f_{10}^{- \Lambda' *}  f_{10}^{- \Lambda'} )  \nonumber \\
& = &  \frac{1}{2} \left[  \mid f_{10}^{++} \mid^2 + \mid f_{10}^{+-} \mid^2 - \mid f_{10}^{-+} \mid^2 - \mid f_{10}^{--} \mid^2 
\right] .
\label{UL_LL}
\end{eqnarray}
\end{widetext}
\begin{figure*}
\includegraphics[width=8.cm]{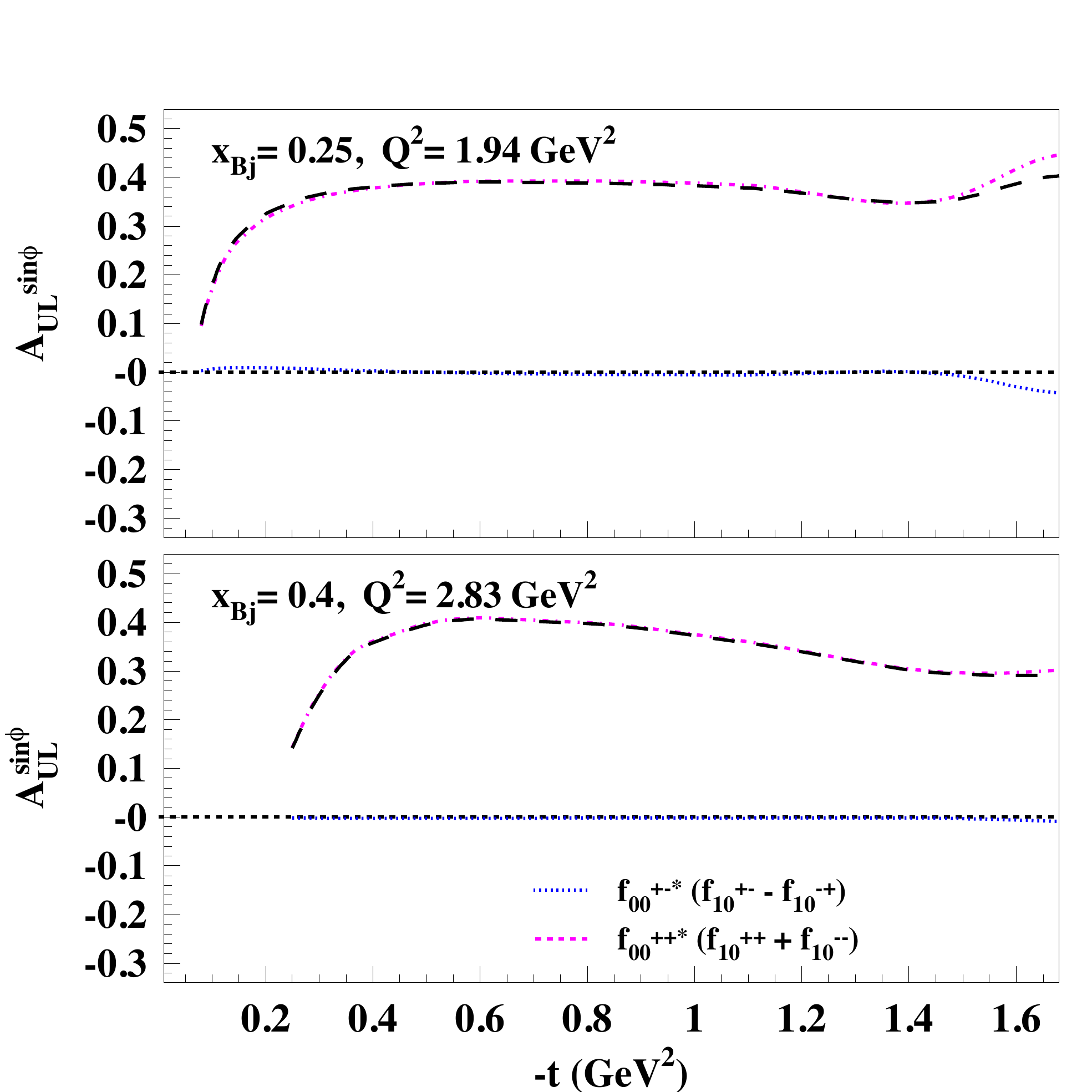}
\includegraphics[width=8.cm]{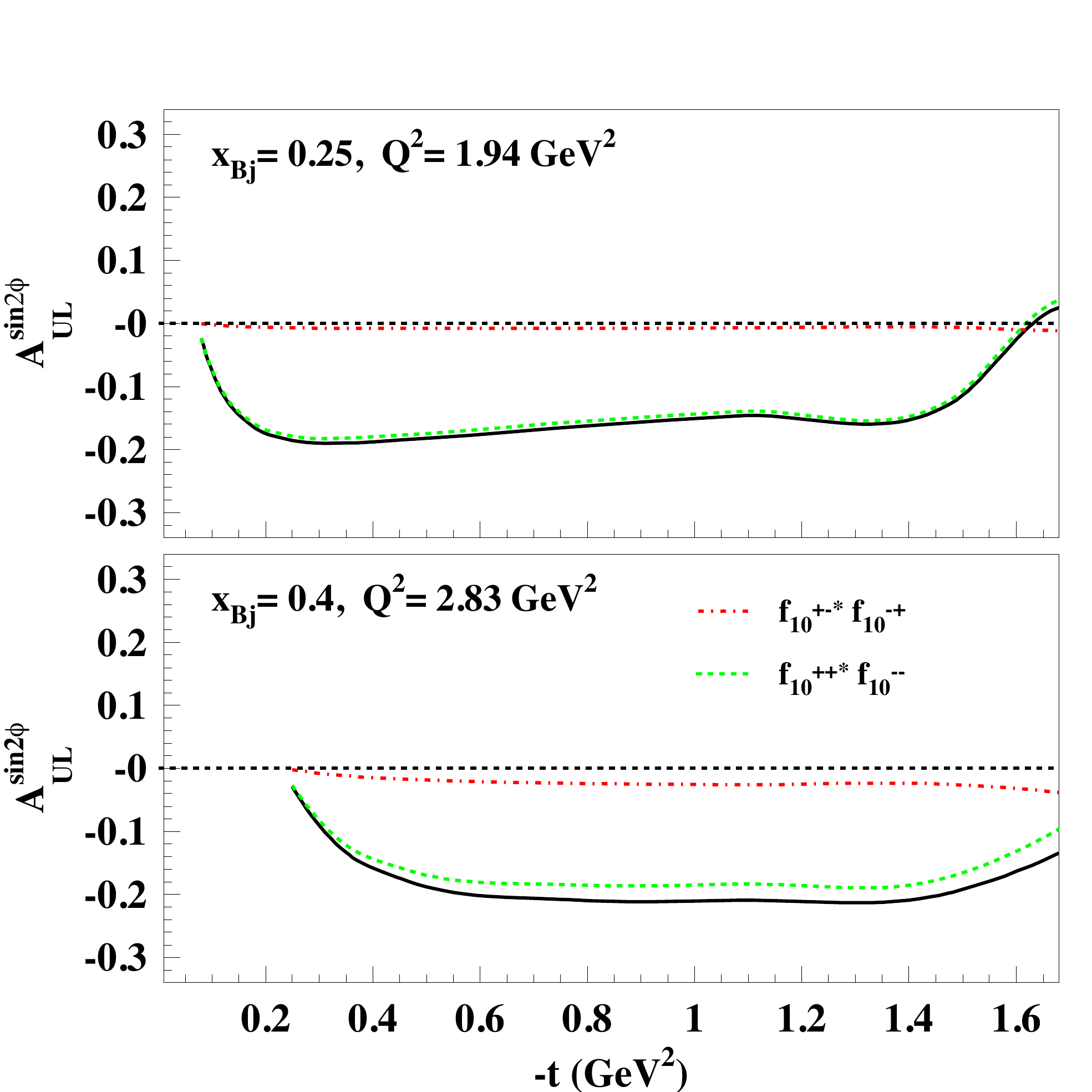}
\caption{(color online) Asymmetry, $A_{UL}$ components, $A_{UL}^{\sin \phi}$ -- first term in Eq.(\ref{AUL:eq}), (left), and $A_{UL}^{\sin 2\phi}$ -- second term in the equation, (right), plotted  vs. $-t$. 
Top panels: $Q^2=1.94$ GeV$^2$,  
Bottom panels: $x_{Bj}=0.25$, and $Q^2=2.85$ GeV$^2$,  
$x_{Bj}=0.4$.  Also shown are the contributions of the different helicity amplitudes combinations described in the text.
}
\label{AUL:fig}
\end{figure*}

\begin{figure*}
\includegraphics[width=7.cm]{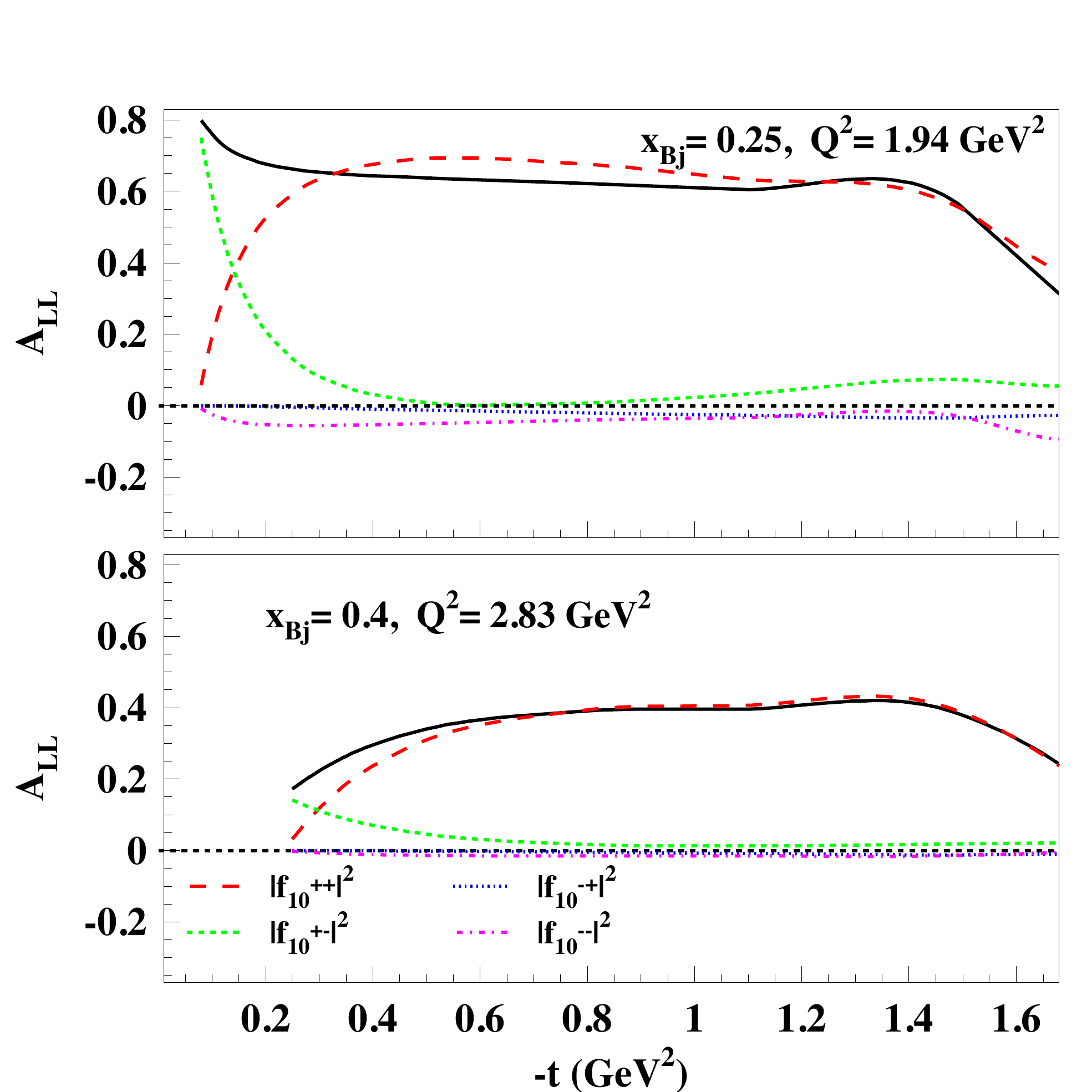}
\hspace{0.3cm}
\includegraphics[width=7.cm]{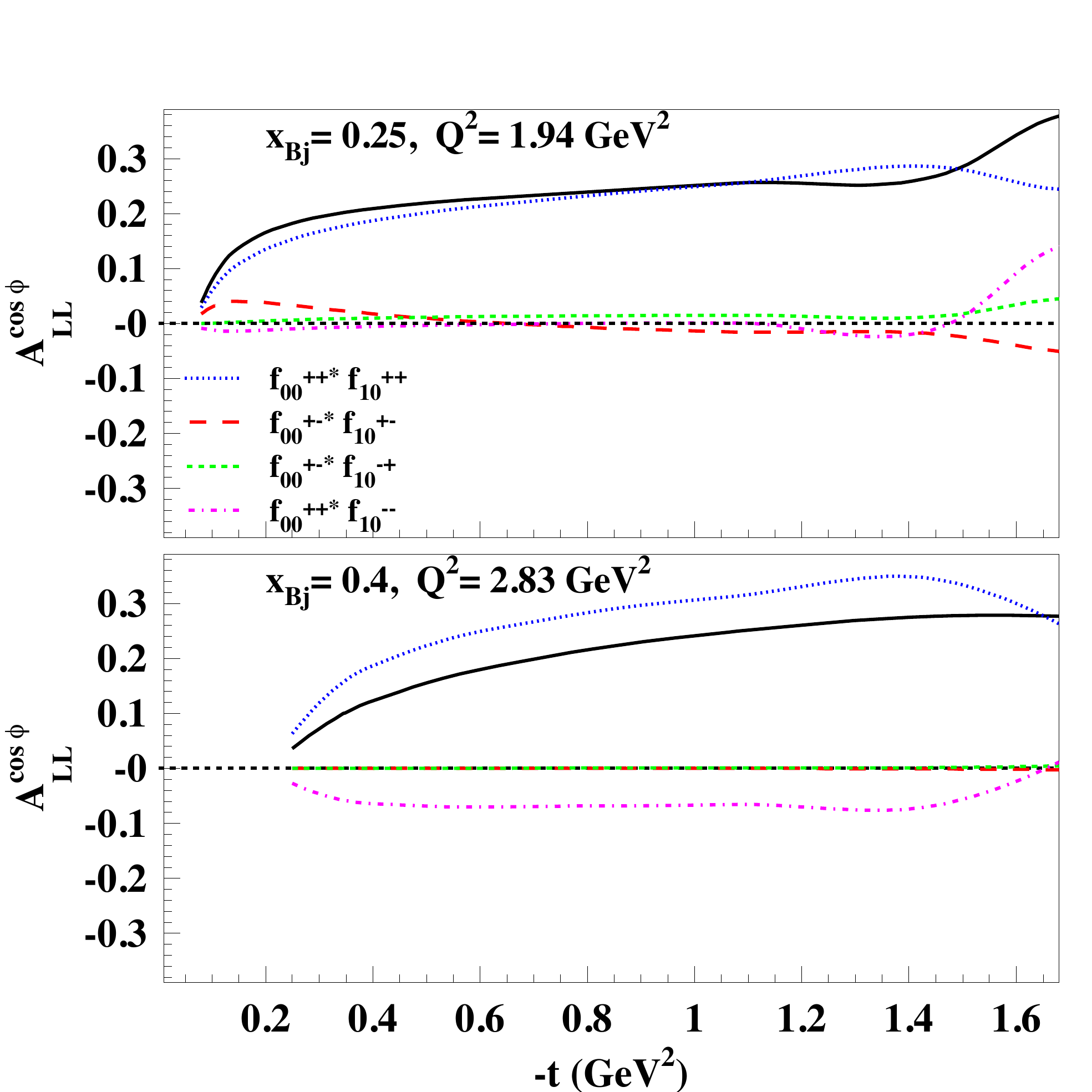}
\caption{(Color online) Components of the asymmetry, $A_{LL}$, Eq.(\ref{ALL:eq}) plotted vs. $-t$ at Jefferson Lab kinematics: $Q^2=1.94$ GeV$^2$,  
$x_{Bj}=0.25$, and $Q^2=2.85$ GeV$^2$,  
$x_{Bj}=0.4$. 
The panel on the left shows the term constant in $\phi$, while the panel on the right shows the $\cos \phi$ modulation of the asymmetry. The solid curve represents the value of the asymmetry, while the other curves labeled in the panels represent the contributions of the various helicity amplitudes to $A_{LL}$ (left) and to  $A_{LL^{\cos \phi}}$ (right). }
\label{ALL:fig}
\end{figure*}

There are several polarization asymmetries that can be constructed. 
For an unpolarized lepton beam on a longitudinally polarized target one has,
\begin{equation}
\label{AUL:eq}
A_{UL} = \frac{N_{s_z=+}-N_{s_z=-}}{N_{s_z=+}+N_{s_z=-}} = \frac{\sqrt{\epsilon(\epsilon+1)} \sin \phi F_{UL}^{\sin \phi}}{{F_{UU,T} +\epsilon \, F_{UU,L} }}  + \frac{\epsilon \sin 2 \phi F_{UL}^{\sin 2 \phi}  }{F_{UU,T} +\epsilon \, F_{UU,L} } = A_{UL}^{\sin \phi}  \sin \phi  + A_{UL}^{\sin 2 \phi}  \sin 2 \phi  
\end{equation}
For a longitudinally polarized lepton beam striking a longitudinally polarized target (relative to the virtual photon direction) the asymmetry  has two components, 
\begin{equation}
\label{ALL:eq}
A_{LL}= \frac{N^\rightarrow_{s_z=+}-N^\rightarrow_{s_z=-}+ N^\leftarrow_{s_z=+}-N^\leftarrow_{s_z=-}}{N_{s_z=+}+N_{s_z=-}}  = \frac{ \sqrt{1 - \epsilon^2} \,  \, F_{LL}  }{F_{UU,T} + \epsilon F_{UU,L}} +    \frac{\sqrt{\epsilon(1-\epsilon)} \, \cos \phi \, F_{LL}^{\cos \phi}}{F_{UU,T} + \epsilon F_{UU,L}} = A_{LL} + A_{LL}^{\cos \phi} \cos \phi
\end{equation}
where $N^{\rightarrow(\leftarrow)}_{s_z=\pm}$ measures a righthanded (lefthanded) lepton scattering on a proton with longitudinal spin, $s_z= \pm 1/2$.  The asymmetries  $A_{UL}$ and $A_{LL}$ are shown in Figures \ref{AUL:fig} and \ref{ALL:fig} at Jefferson Lab kinematics \cite{AvaKim}. 

$A_{UL}^{\sin \phi}$ (Fig.\ref{AUL:fig}, left), is  dominated by the longitudinal components in a similar way as already seen for $A_{LU}$ (Fig.\ref{fig:ALU}). Therefore, for its description one needs to consider simultaneously the chiral even sector. We conclude that this quantity is harder to interpret theoretically since chiral even and odd components cannot be disentangled in a model independent way.  In our model the chiral even component dominates the longitudinal contributions: $f_{00}^{+\pm} = f_{00}^{+\pm, even} + f_{00}^{+\pm, odd} \approx  f_{00}^{+\pm, even}$ (see Section \ref{sec:helamp}) . 
On the other side,  $A_{UL}^{\sin 2 \phi}$ (right) is determined by the chiral odd amplitudes. Specifically, we find that the contribution, $f_{10}^{++ *} f_{10}^{--}$, in Eq.(\ref{ULsin2phi:eq}) dominates, as expected, over the double flip term, $f_{10}^{+- *} f_{10}^{-+}$. As we can see from Eqs.(\ref{helamps_gpd2},\ref{helamps_gpd2_simple}), these amplitudes are almost entirely determined by $\widetilde{H}_T$ and $E_T$, through the combinations: $2 \widetilde{H}_T  + (1 \pm \xi)E_T$. 

\vspace{0.3cm}
We conclude that $A_{UL}$ allows for a clean extraction of the chiral odd GPDs, $\widetilde{H}_T$ and $E_T$. 
\vspace{0.3cm}

$A_{LL}$ and $A_{LL}^{\cos \phi}$  (Fig.\ref{ALL:fig}), can  be interpreted similarly to $A_{LU}$. From Eqs.(\ref{LLcosphi:eq},\ref{LL:eq}) we see that $A_{LL}$  contains only chiral odd GPDs, while $A_{LL}^{\cos \phi}$, because of the longitudinal photon contributions, contains both chiral even and chiral odd terms. In our description, the chiral even terms dominate this quantity through the term $f_{00}^{++})^* (f_{10}^{++}$
(Fig.\ref{ALL:fig}, right).  A straightforward interpretation is instead obtained for $A_{LL}$  where we can see that the term $\mid f_{10}^{+-} \mid^2$, which is dominated by $H_T$, determines the amplitude at low $t$, while at large $t$ because of the form-factor-like fall off of $H_T$ with t, the contribution from $\mid f_{10}^{++} \mid^2$, determined by $\widetilde{H}_T$ and $E_T$, takes over. 
 
\vspace{0.3cm}
We conclude that $A_{LL}$ allows us to extract the chiral odd GPD, $H_T$, and therefore the tensor charge, at small $t$, and $\widetilde{H}_T$ and $E_T$ at larger $t$. 

\subsection{Transverse Target Polarization}
We complete our discussion by listing the six structure functions for the single transversely polarized target in Eq.(\ref{xs}),
\begin{eqnarray}
F_{UT,T}^{\sin(\phi-\phi_S)} &=& - \, \Im m \left[ f_{10}^{++*} f_{10}^{-+} + f_{10}^{+-*} f_{10}^{--} \right] \\
F_{UT,L}^{\sin(\phi-\phi_S)} &=& 2 \, \Im m \left[ f_{00}^{+-*} f_{00}^{++} \right] \\
F_{UT}^{\sin(\phi+\phi_S)} &=&  \, \Im m \left[ f_{10}^{++*} f_{10}^{+-} \right] \\
F_{UT}^{\sin(3\phi+\phi_S)} &=& -  \, \Im m \left[ f_{10}^{-+*} f_{10}^{++} \right]  \\
F_{UT}^{\sin\phi_S} &=& -  \, \Im m \left[ f_{10}^{++*} f_{00}^{+-} - f_{10}^{+-*} f_{00}^{++} \right]   \\
F_{UT}^{\sin(2\phi-\phi_S)} &=&  \, \Im m \left[ f_{10}^{-+*} f_{00}^{++} + f_{10}^{--*} f_{00}^{+-} \right],
\label{UT}
\end{eqnarray}
and three for the longitudinally polarized lepton {\it and} transversely polarized target in Eq.(\ref{xs}),
\begin{eqnarray}
F_{LT}^{\cos(\phi-\phi_S)} &=& \, \Re e \left[ f_{10}^{++*} f_{10}^{-+} + f_{10}^{+-*} f_{10}^{--} \right]
\label{LT} \\
F_{LT}^{\cos\phi_S} &=& - \, \Re e \left[ f_{10}^{++*} f_{00}^{+-} - f_{10}^{+-*} f_{00}^{++} \right] \\
F_{LT}^{\cos(2\phi-\phi_S)} &=& -  \, \Re e \left[ f_{10}^{-+*} f_{00}^{++} + f_{10}^{--*} f_{00}^{+-} \right] .
\end{eqnarray}
For target polarization we distinguish the polarization that is both transverse to the photon direction and to the hadron plane, 
\begin{equation}
A_{UT} = \frac{F_{UT,T}^{\sin(\phi-\phi_S)}}{F_{UU,T} +\epsilon \, F_{UU,L} }
\label{A_UT}
\end{equation}
(for the target polarized along the photon direction there will be no asymmetry because of Parity conservation). For the target at rest, polarized along the incoming lepton direction, there will be a component of nucleon polarization transverse to the photon direction as well as transverse to the nucleon plane. The same $A_{UT}$ will be involved, although modulated by the sine of the photon angle relative to the lepton beam and the $\sin \phi$.

We conclude this Section by noting that transverse asymmetries allow us to best single out the tensor charge \cite{AGL}, namely they are sensitive to the GPD $H_T$, at variance with the quantities reported in detail in this paper which are mostly sensitive to the GPDs $\widetilde{H}_T$ and $E_T$. Both type of measurements are therefore important for interpreting the chiral odd sector. For ease of presentation we will include the detailed discussion of the transverse asymmetries terms in a dedicated paper in preparation.

\section{Conclusions and Outlook}
\label{sec:conclusions}
Once established that the transversity parton distributions in the nucleon
can be accessed through deeply virtual exclusive pseudoscalar meson production which is sensitive to the chiral-odd transversity GPDs, $H_T, E_T, \widetilde{H}_T, \widetilde{E}_T$,
we have now addressed the issue of feasibility of an experimental extraction.

A major goal of this work was to gauge the contributions of the various GPDs to experimental observables, specifically in exclusive $\pi^o$ electroproduction. 
For chiral odd GPDs, contrary to the chiral even case, a big piece of information is missing in that their normalizations are not linked by integral relations to  specific nucleon form factors. This hampers in particular the determination of their $t$ dependence.

Given the structure of the spectator model, parametrized as diquark amplitudes, there are relations between chiral even and chiral odd amplitudes. We applied Parity reflection to one set of the helicity-dependent vertices - the outgoing quark-diquark-nucleon.  A set of linear relations results for the two possible diquark structures (scalar and axial vector) which thereby relates the chiral even helicity amplitudes to the chiral odd amplitudes.  This has led us to parameterizing the chiral odd GPDs, normalized through the chiral even GPDs. By using the more extensive data-driven determination of the chiral even GPDs \cite{GGL}, we are therefore able to provide the full kinematical dependence of all four chiral odd GPDs within a general  class of models: , quark-diquark, two component, or spectator models. 

This represents a consistent quantitative step with respect to our previous work \cite{AGL} where the normalizations in the chiral-odd sector were estimated based on various ans\"atze.  In particular, only $H_T$ and the combination  $2\widetilde{H}_T+E_T$, related \cite{Bur2,Bur3} to the first moment of the Boer-Mulders \cite{BoerMul} TMD, were considered while $\widetilde{E}_T$ was set to zero assuming a straightforward extrapolation of the symmetries in the Regge amplitudes for $\pi^o$ photoproduction.  A similar simplified approach was taken also in Ref.\cite{Kro_new}. 

We see the results in relation to the many measured and measurable observables, in particular in $\pi^0$ electroproduction. What is especially gratifying is that certain asymmetries constrain the GPDs well enough to separately determine $H_T$, and consequently  transversity through the limit $H_T(x,0,0)$, and the combination $2 \widetilde{H}_T + (1\pm \xi)  E_T$, which . 
Data show that transverse virtual photons dominate the process at Jefferson Lab kinematics. 
This is a strong marker for twist-3 contributions to the hard scattering subprocess. 

In upcoming work we have predictions for more pseudoscalar mesons observables including $\eta$ \cite{Girod}, strange and charmed mesons, as well as a refinement of the parametrization to be pursued, including the role of sea quarks. We are considering  the significance of the chiral even/odd duality. The extension of the notions of variable mass spectators to Wigner distributions and Generalized Transverse Momentum Distributions is underway.

We complete our discussion by noting that  the electroproduction of two vector mesons
proposed to extract the transversity GPD, $H_T$, 
is also, in principle, feasible although with a doubling of the technical issues for the method shown here \cite{Pire1,Pire2}.

\acknowledgements
We thank Harut Avakian, Andrey Kim, Valery Kubarovsky and Paul Stoler for many useful discussions and suggestions. This work was supported by the U.S. Department
of Energy grant DE-FG02-01ER4120.

\appendix
\section{Pion transition form factors}
\label{appa}
The hard part of $\gamma^* p \rightarrow \pi^o p^\prime$ involves the $\gamma^* + u(d) \rightarrow \pi^o + u(d)$ amplitudes (Fig.\ref{fig1}). 
The $\pi^o$ vertex  is described in terms of Distribution Amplitudes (DAs) as follows, 
\begin{eqnarray}
{\cal P} & = & K f_\pi \left\{ \gamma_5  \not\!{q}^\prime \phi_\pi(\tau) + \gamma_5 \mu_\pi  \phi_\pi^{(3)} (\tau) \right\}
\label{pi_coupling}
\end{eqnarray}
where $f_\pi$ is the pion coupling,  $\mu_\pi$ is a mass term that can {\it e.g.} be estimated from the gluon condensate, $\phi_\pi(\tau)$ and $\phi_\pi^{(3)} (\tau)$, $\tau$ being the longitudinal momentum fraction, are the twist-2 and twist-3 pion DAs, respectively describing the chiral even and chiral odd processes.

The $\gamma^\mu \gamma^5$ coupling produces the $\pi^0$'s non-flip quark vertex, which corresponds to a twist-2 contribution. This contributes to the longitudinal photon case with no quark helicity flip. 
The non-flip transverse photon contribution is suppressed -- twist-4. 
For transverse $\gamma^*$ the quark can also flip helicity in the near collinear limit. 
This is accomplished through the vertex with $\gamma^5$ coupling giving the same $Q^2$ dependence as in the transverse photon, quark non-flip case. 

Notice that: 
{\it i)} for the chiral-odd coupling the longitudinal term is suppressed relatively to the transverse one, already at tree level; 
{\it ii)} based on collinear factorization, the chiral-even longitudinal term should be dominating. 
In what follows we show, however, that by taking into account both the GPD crossing properties, along with the corresponding $J^{PC}$ quantum numbers in the $t$-channel, the allowed linear combinations of chiral-even GPDs that  contribute to the longitudinal cross section terms are suppressed.

In addition to assessing the impact of the correct GPD combinations to $\pi^o$ electroproduction, we also developed a model for the hard vertex that takes into account the direct impact of spin through different $J^{PC}$ sequencings~\cite{GGL_short}.
According to the modified perturbative approach  (\cite{GolKro} and references therein), one has
\begin{widetext}
\begin{eqnarray}
g_{\Lambda_{\gamma^*},\lambda; 0, \lambda^\prime}  =  \int d \tau \int d^2 b \, \hat{{\cal F}}_{\Lambda_{\gamma^*},\lambda; 0, \lambda^\prime}(Q^2,\tau,b) \alpha_S(\mu_R) \exp[-S] \hat{\phi}_\pi(\tau,b) 
\label{collinear}
\end{eqnarray}
\end{widetext}
where $\hat{{\cal F}}_{\Lambda_{\gamma^*},\lambda; 0, \lambda^\prime}$ is the Fourier transform of the hard (one gluon exchange) kernel, $S$ is the Sudakov form factor,  $\hat{\phi}_\pi$ is the pion distribution amplitude in impact parameter, $b$, space, $\mu_R$ is a renormalization scale. 

Now consider the t-channel perspective. There exist two distinct series  of $J^{PC}$ configurations in the $t$-channel, namely the {\it natural parity} one ($1^{--}, 3^{--} ... $), labeled $V$, and 
the {\it unnatural parity} one ($1^{+-}, 3^{+-} ...$), labeled $A$. We hypothesize that  the two series will generate different contributions to the pion vertex. 
We consider separately the two contributions $\gamma^* (q \bar{q})_V \rightarrow \pi^o$ and 
$\gamma^* (q \bar{q})_A \rightarrow \pi^o$  to the process in Fig.1b.  
What makes the two contributions distinct is that, in the natural parity case (V), L is always the same for the initial and final states, or $\Delta L=0$,
while for unnatural parity (A), $\Delta L =1$. 
We modeled this difference by replacing Eq.(\ref{collinear})  with the following expressions containing a modified kernel 
\begin{widetext}
\begin{eqnarray}
g^V_{\Lambda_{\gamma^*},\lambda; 0, \lambda^\prime}  =  \int dx_1 dy_1 \int  d^2 b  
\, \hat{\psi}_V(y_1,b) \, \hat{{\cal F}}_{\Lambda_{\gamma^*},\lambda; 0, \lambda^\prime}(Q^2,x_1,x_2,b) \alpha_S(\mu_R)
 \exp[-S]     \, \hat{\phi}_{\pi^o}(x_1,b)  && \\
g^A_{\Lambda_{\gamma^*},\lambda; 0, \lambda^\prime}  =  \int dx_1 dy_1 \int d^2  b  
\, \hat{\psi}_A(y_1,b) \, \hat{{\cal F}}_{\Lambda_{\gamma^*},\lambda; 0, \lambda^\prime}(Q^2,x_1,x_2,b) \alpha_S(\mu_R)
\exp[-S]  \, \hat{\phi}_{\pi^o}(x_1,b) &&
\end{eqnarray}
\end{widetext}
where, 
\begin{equation}
\hat{\psi}_{A}(y_1,b) = \int d^2 k_T J_1(y_1 b) \psi_V(y_1,k_T) 
\end{equation}
Notice that we now have an additional longitudinal variable and ``wave function" in order to introduce the effect of 
different $L$ states. The higher order Bessel function describes the situation where $L$ is always larger in the initial state. 
In impact parameter space this corresponds to configurations of larger radius. 
The matching of the $V$ and $A$ contributions to the helicity amplitudes is as follows: $f_{10}^{++}, f_{10}^{--} \propto g^V$, $f_{10}^{+-}  \propto g^V+g^A$, 
$f_{10}^{-+} \propto g^V-g^A$.

\section{Hard scattering process}
\label{appb}
We first give details of the calculation of the hard scattering amplitudes, $g_{10}^{\lambda, \lambda'}$,
\begin{subequations}
\begin{eqnarray}
g_{10}^{++} & = & g_{10}^{- -} = 0  
\\
g_{10}^{+-}  & = &   K N \, N^\prime \, \left[ (g^{\lambda o} g^{\rho +}g^{1 \nu}- g^{\lambda\rho}g^{o+} g^{1 \nu} + g^{\lambda +}g^{o\rho} g^{1 \nu})  \right. \nonumber \\
&- &  \left. i (g^{\lambda o} g^{\rho +}g^{2 \nu}- g^{\lambda\rho}g^{o+} g^{2 \nu} + g^{\lambda +}g^{o\rho} g^{2 \nu} ) \right] k_\lambda k_\rho^\prime \epsilon_\nu^{+1}   \nonumber \\
& \approx & -\frac{K}{\sqrt{k^{\prime \, +} k^+}} \, \left[ k^o p^{\prime \, +} -(k k^\prime) + k^+ k^{\prime \, o} \right] (\epsilon_1^{+1} - i \epsilon_2^{+1})  \\
g_{10}^{-+} & = & -\frac{K}{\sqrt{k^{\prime \, +} k^+}} (k^+k^{\prime \, +}) \nonumber \\
& = &   K N \, N^\prime \, \left[ (g^{\lambda o} g^{\rho +}g^{1 \nu}- g^{\lambda\rho}g^{o+} g^{1 \nu} + g^{\lambda +}g^{o\rho} g^{1 \nu})  \right. \nonumber \\
&+ &  \left. i (g^{\lambda o} g^{\rho +}g^{2 \nu}- g^{\lambda\rho}g^{o+} g^{2 \nu} + g^{\lambda +}g^{o\rho} g^{2 \nu} ) \right] k_\lambda k_\rho^\prime \epsilon_\nu^{+1}   \nonumber \\
& \approx & -\frac{K}{\sqrt{k^{\prime \, +} k^+}} \, \left[ k^o k^{\prime \, +} -(k k^\prime) + k^+ k^{\prime \, o} \right] (\epsilon_1^{+1} + i \epsilon_2^{+1})  \nonumber \\
& = & 0 .
\end{eqnarray}
\label{ampg2_app}
\end{subequations}

We now turn to the longitudinal amplitude, $g^{+-}_{00}$,
\begin{subequations}
\begin{eqnarray}
g_{00}^{+-}   & = &  
K N \, N^\prime \, \left[ (g^{\lambda o} g^{1 \rho}g^{+ \nu}- g^{\lambda 1}g^{o \rho} g^{+ \nu}  + g^{\lambda 1}g^{o \nu} g^{\rho +}  \right.
\nonumber \\ 
& &  \left. + g^{\lambda +}g^{o \nu} g^{1 \rho}) - i  ( 1 \leftrightarrow 2) \right]  k_\lambda k_\rho^\prime \epsilon_\nu^0   \nonumber \\
& \approx & \frac{K}{\sqrt{k^{\prime \, o} k^o}} \, \left[ (k^o k^{\prime \, 1} -  k^1 k^{\prime \, o} )\epsilon_+^0 + 
(k^1 k^{\prime \, +} + k^1 k^{\prime \, +} ) \epsilon_o^0  \right. \nonumber \\
& - & \left.  i ( 1 \leftrightarrow 2) \right]
 \\
g_{00}^{++}  & = & 0 
\end{eqnarray}
\label{ampg5_app}
\end{subequations}

Notice that in Eqs.(\ref{g5final1}), (\ref{g5final2}) the following term appears,
\begin{equation}
\frac{(k^\prime_1 - i k^\prime_2)}{\sqrt{k_o k_o^\prime}} = \frac{(k^\prime_1 - i k^\prime_2)}{k_o^\prime} \, \sqrt{\frac{k_o^\prime}{k_o}} \equiv \frac{(k^\prime_1 - i k^\prime_2) }{P^+ \left[ \displaystyle k^{\prime \, 2} _\perp \, \left(\frac{2M\zeta^2}{Q^2}\right)^2  + (X-\zeta)^2 \right]^{1/2} } \sqrt{\frac{k_o^\prime}{k_o}}.
\label{a1}
\end{equation}
The first factor corresponds to $\sin \theta' = \hat{ {\bf k}' {\bf P}}$, {\it i.e.} of the angle between the returning quark's momentum, ${\bf k}^\prime = {\bf k} - {\bf \Delta}$, and the initial proton's momentum, ${\bf P}$, which lies along the $z$ axis.
With the choice of kinematical variables in this paper, $P^+ = Q^2/2M \zeta^2$. 
The four-vector components, using $v \equiv(v_o; v_\perp, v_3)$, are
\[ k^\prime \equiv ((X-\zeta) P^+; {\bf k}^\prime_\perp, (X-\zeta) P^+), \;\;\;\;  
P \equiv (P^+; 0,P^+) \]
In terms of these we define
\begin{equation}
\sin \theta^\prime = \frac{\mid {\bf k^\prime}_\perp \mid }{\sqrt{k_\perp^{\prime 2} + k _3^{\prime 2}}}
\end{equation}
\begin{figure}
\includegraphics[width=9.cm]{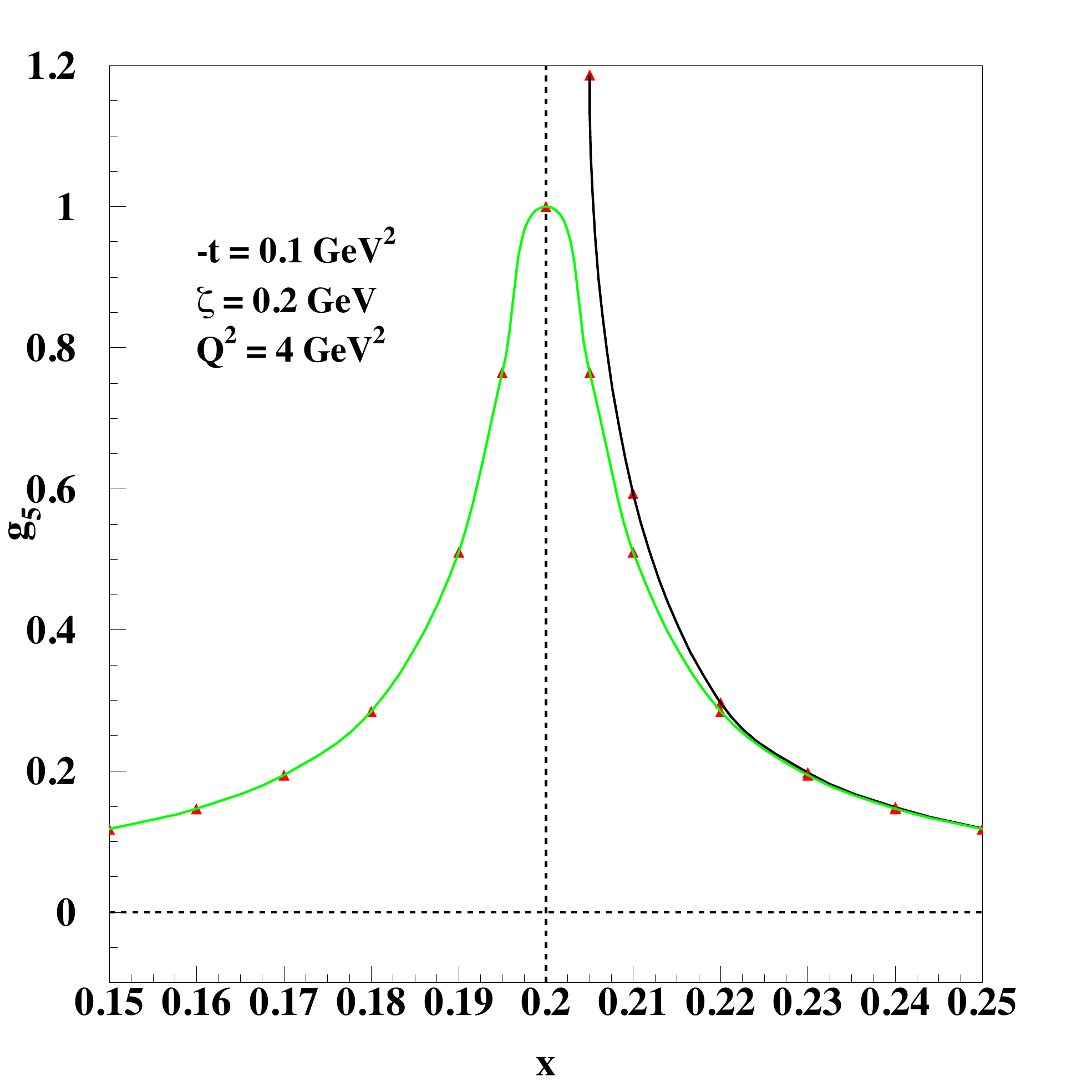}
\caption{(color online) The function $\sin \theta'$ (in green), where $\theta'$, Eq.(\ref{a1}) is the angle between the returning quark's momentum, $k^\prime$, and the initial proton momentum,
$P$, which is along the $z$ axis. The black curves shows the approximated form obtained disregarding $k_\perp'$, which is valid for $X > \zeta$ (symbols shown to guide the eye, are in the the points where the numerical calculation was performed).}
\label{figappA}
\end{figure}
By inserting these expressions in  $g_{00}^{+-}$, Eq.(\ref{g5final2}), we have,
\begin{equation}
g_{00}^{+-} = \displaystyle\frac{\mid {\bf k^\prime}_\perp \mid}{\sqrt{Q^2}}\zeta \sqrt{X(X-\zeta)}  \frac{1}{P^+\left[(X-\zeta)^2 + \langle k_\perp^2 \rangle/(Q^2/2 M \zeta^2)\right]^{1/2} } \; C^+
\end{equation}
Two kinematical limits are relevant: {\it i)}  $X=\zeta$, where the quark is perpendicular to the $z$ axis, and $\sin \theta^\prime =1$; {\it ii)} 
$X \neq \zeta$, $\mid k^\prime_\perp \mid << P^+$, where the denominator in Eq.(\ref{a1}) becomes $\approx P^+ (X-\zeta)$.   
The two distinct limits are shown in Fig.\ref{figappA}, as a function of $X$, at fixed $k_\perp = 0.3 $ GeV. 

\newpage
\section{Chiral Even GPDs new set of parameters}
\label{appc}
We give the complete set of chiral even parameters. We first performed a fit  for $t=\zeta=0$, of the PDF global parameterizations in the valence quark sector, and obtained the parameters $m_q$, $M_X^q$, $M_\Lambda^q$, and $\alpha_q$ for all four GPDs, as well as the normalization
factors,  ${\cal N}_q$, for $H_q$ and $\widetilde{H}_q$ \protect\cite{GGL}. Because we did not use the actual data at this stage, these parameters assume the fixed vaues in the table, with no error bar. 
In the next step, we took  $t \neq 0$ and, by keeping the first set of parameters fixed, we performed a fit  of the nucleon form factors. We obtained: {\it i)} the parameters $\alpha^\prime_q$, $p_q$, and the normalization, ${\cal N}_q$, for $E_q$,  \cite{newFF} by fitting the proton and neutron flavor separated form factors data from Ref.\cite{Cates}; {\it ii)} the parameters $\alpha^\prime_q$, $p_q$, and the normalizations for $\widetilde{H}_q$ $\widetilde{E}_q$ by fitting available data on the axial  (\cite{Schindler} and references therein), and pseudoscalar  \cite{Fearing} form factors, respectively. Also shown are the $\chi^2$ values for the separate contributions to the fit.
\begin{table}[h]
\begin{tabular}{|c|c|c|c|c|}
\hline
\hline
Parameters             &  $H$                &  $E$                & $\widetilde{H}$       & $\widetilde{E} $  \\ 
\hline
\hline
$m_u $ (GeV)                & 0.420                                 &  0.420                              &  2.624                &  2.624            \\
$M_X^u$ (GeV)            & 0.604                                 &  0.604                              &  0.474                &  0.474            \\
$M_\Lambda^u$ (GeV)    & 1.018                            &  1.018                              &  0.971                &  0.971            \\
$\alpha_u$                    & 0.210                                  &  0.210                              &  0.219                &  0.219            \\
$\alpha^\prime_u$      & 1.814  $\pm$ 0.022       &  2.835 $\pm$ 0.051     &  1.543  $\pm$ 0.296     &  5.130  $\pm$ 0.101\\
$p_u$                            & 0.449 $\pm$ 0.017       &  0.969 $\pm$ 0.031      &  0.346  $\pm$ 0.248     &  3.507 $\pm$ 0.054\\
${\cal N}_u$                  & 2.043                                 &  1.803                               &  0.0504                          &  1.074 \\ 
$ \chi^2$ &  0.5   &  3.2 &   0.12    &     2.0  \\   
\hline

$m_d $ (GeV)                      & 0.275                              & 0.275                                &  2.603                               &  2.603                 \\
$M_X^d$ (GeV)                  & 0.913                              & 0.913                                &  0.704                               &  0.704                 \\
$M_\Lambda^d$ (GeV)    & 0.860                               & 0.860                                &  0.878                               &  0.878                 \\
$\alpha_d$                          & 0.0317                            & 0.0317                              &  0.0348                            &  0.0348                \\
$\alpha^\prime_d$             & 1.139 $\pm$ 0.056      & 1.281  $\pm$ 0.031       &  1.298  $\pm$ 0.245      &  3.385  $\pm$ 0.145    \\
$p_d$                                   & -0.113 $\pm$ 0.104      & 0.726 $\pm$ 0.0631        &  0.974  $\pm$ 0.358      &  2.326  $\pm$ 0.137    \\
${\cal N}_d$                        & 1.570                              & -2.800                               & -0.0262                            &  -0.966  \\ 
$ \chi^2$ &   0.9  & 4.8   &  0.11     &    1.0   \\ \hline


\end{tabular}
\caption{%
\label{table_1} 
Parameters obtained from our recursive fitting procedure applied to $H_q$, $E_q$, $\widetilde{H}_q$, and $\widetilde{E}_q$, $q=u,d$.  }
\end{table}
The form factor fit performed in Ref.\cite{newFF} taking into account the new flavor separated data on the nucleon's Dirac and Pauli form factors from Ref.\cite{Cates}, while keeping the parameters from the PDFs fit fixed to their previously determined values \cite{GGL}, allowed us to sensibly reduce the error on the GPDs.

\newpage
\section{Kinematic rotation for light cone wave functions with $S=1$}
\label{appd}
For $S=1$ one must calculate the following structures for spin non flip,
\begin{eqnarray}
\label{nonflip}
&& \bar{u}(k,\pm) \gamma_5 \gamma^\mu U(P,\pm)  \epsilon_\mu^{\lambda^{\prime\prime}}(P_X)  = \nonumber \\
&&  \frac{1}{4} {\rm Tr} \left\{ (\not\!{P} +m )(1+\gamma^o) (1 \mp \gamma_5 \gamma_3) (\not\!{k} +m )\gamma^\mu \gamma_5 \right\} = \nonumber  \\
&&  \frac{1}{4} \left[ {\rm Tr} \left\{ \gamma^\lambda \gamma^o \gamma^\nu \gamma^\mu \gamma_5 \right\} +  \right.  
 \left. {\rm Tr} \left\{ \gamma^\lambda \gamma^3 \gamma^\nu \gamma^\mu  \right\} \right] P_\lambda k_\nu \epsilon_\mu  =  \nonumber  \\ 
&& \left( i \epsilon^{\lambda o \nu \mu}  + g^{\lambda 3}g^{\nu \mu} - g^{\lambda \nu}g^{\mu 3} + g^{\lambda \mu}g^{\nu 3} \right) P_\lambda k_\nu \epsilon_\mu 
\nonumber  \\
\end{eqnarray}
and spin flip,
\begin{eqnarray}
\label{flip}
&& \bar{u}(k,\pm) \gamma_5 \gamma^\mu U(P,\mp)  \epsilon_\mu^{\lambda^{\prime\prime}}(P_X)  = \nonumber \\ 
&=& \frac{1}{4}   {\rm Tr} \left\{ (\not\!{P} +m ) (1+\gamma^o) \gamma_5(\gamma_1 \pm i\gamma_2)  (\not\!{k} +m )\gamma_\mu \gamma_5 \right\}  \nonumber  \\
&=&  \frac{M}{4} {\rm Tr} \left\{ \gamma^o (\gamma_1 \pm i\gamma_2)   \gamma^\nu \gamma^\mu  \right\}   k_\nu \epsilon_\mu  +    \nonumber \\
&& \frac{m}{4} {\rm Tr} \left\{ \gamma^\lambda   \gamma^o (\gamma_1 \pm i\gamma_2) \gamma^\mu  \right\} 
P_\lambda \epsilon_\mu  .
\end{eqnarray}
We performed the calculation in a frame which is rotated with respect to the original one where $P$ was lying along the $z$-axis, so that $P_X$ is along the $z$-axis, and the polarization vectors have the usual form
\begin{eqnarray} 
\epsilon^{(\pm1)}_\mu = \frac{1}{\sqrt{2}} \left( 0; \mp1, i, 0 \right). 
\end{eqnarray}
The coordinates of the relevant four-vectors expressed as $v \equiv (v_3,v_\perp)$, in the rotated frame are,
\begin{subequations}
\begin{eqnarray}
P_X & \equiv & \left( (1-X)P^+,0 \right) \\
P & \equiv & \left( P^+ \cos \theta_X, - P^+ \sin \theta_X \right) \\
k & \equiv & \left( XP^+ \cos (\theta + \theta_X), XP^+ \sin (\theta+\theta_X) \right) 
\end{eqnarray}
\end{subequations}
where,
\begin{eqnarray} 
\sin \theta & = & \frac{k_\perp}{P^+\sqrt{X^2+k_\perp^2/P^{+ \, 2}} } \\
\sin \theta_X & = & \frac{k_\perp}{P^+ \sqrt{(1-X)^2 + k_\perp^2/P^{+ \, 2}} } 
\end{eqnarray}
define the angle between the struck quark and the initial proton, 
and the angle between  the diquark and the initial proton, respectively. 
The components in this frame, expressed as $v \equiv (v^+,v^-,v_\perp)$ are,
\begin{subequations}
\label{LHScomp}
\begin{eqnarray}
P_X & \equiv & \left( (1-X)P^+, \frac{M_X^2}{2(1-X)P^+}, 0 \right) \\
P & \equiv & \left( P^+,  \frac{M^2+ P_\perp^2}{2P^+}, - \frac{k_\perp}{\sqrt{(1-X)^2 + k_\perp^2/P^{+ \, 2} } } \right) \\
k & \equiv & \left( XP^+,  \frac{m^2+ k_\perp^2}{2X P^+}, \frac{k_\perp}{\sqrt{(1-X)^2 + k_\perp^2/P^{+ \, 2}}}  \right)
\end{eqnarray}
\end{subequations}
For the RHS vertex one finds, similarly,
\begin{widetext} 
\begin{subequations}
\label{RHScomp}
\begin{eqnarray}
P^\prime & \equiv & \left( (1-\zeta)P^+,  \frac{M^2+ P^{\prime \, 2}_\perp}{2(1-\zeta)P^+}, - \frac{\tilde{k}_\perp}{\sqrt{[(1-X)/(1-\zeta)]^2 + \tilde{k}_\perp^2/P^{+ \, 2} } }  \right) \\
k^\prime & \equiv & \left( (X-\zeta)P^+,  \frac{m^2+ k^{\prime \, 2}_\perp}{2(X-\zeta) P^+}, \frac{\tilde{k}_\perp}{\sqrt{[(1-X)/(1-\zeta)]^2 + \tilde{k}_\perp^2/P^{+ \, 2} } }  \right) 
\end{eqnarray}
\end{subequations}
\end{widetext}
where $\tilde{{\bf k}} = {\bf k} - \frac{1-X}{1-\zeta}{\bf \Delta}$, and $P^+ = (Pq)/q^- = Q^2/2M\zeta^2$, as in Eq.(\ref{g5final1}).



\begin{thebibliography}{90}
\bibitem{Mulders} P. J. Mulders and R. D. Tangerman, Nucl. Phys. B461, 197 (1996).

\bibitem{BogMul} M.~Boglione and P.~J.~Mulders,
Phys.\ Rev.\  D {\bf 60}, 054007 (1999).

\bibitem{BarBradMar} V.~Barone, F.~Bradamante and A.~Martin,
  Prog.\ Part.\ Nucl.\ Phys.\  {\bf 65}, 267 (2010)
    
\bibitem{AGL} S.~Ahmad, G.~R.~Goldstein and S.~Liuti, 
Phys. Rev. D{\bf 79}, 054014 (2009).

\bibitem{GL_charm} S.~Liuti and G.~R.~Goldstein,
  arXiv:1009.1334 [hep-ph].

\bibitem{Ji_even} X.~D.~Ji, Phys.\ Rev.\ D {\bf 55}, 7114 (1997)

\bibitem{Ji_odd} P. Hoodbhoy and X. Ji, Phys. Rev. D 58, 054006 (1998).

\bibitem{Diehl_odd}  M. Diehl, Eur. Phys. Jour. C {\bf 19}, 485 (2001).
 
\bibitem{GGL} G.~R.~Goldstein, J.~O.~Hernandez and S.~Liuti,
  Phys.\ Rev.\ D {\bf 84}, 034007 (2011)

\bibitem{newFF} J.~O.~Gonzalez-Hernandez, S.~Liuti, G.~R.~Goldstein and K.~Kathuria,
  Phys.\ Rev.\ C {\bf 88}, 065206 (2013).

\bibitem{Bur2} M. Burkardt, Phys. Rev. D {\bf 72}, 094020 (2005); {\it ibid}
Phys. Lett. {\bf B} 639 (2006) 462.

\bibitem{DieHag_odd}  M.~Diehl and P. Hagler,
  Eur.\ Phys.\ J.\ C {\bf 44}, 87 (2005)
  
\bibitem{GolKro} S.~V.~Goloskokov, P.~Kroll,
  Eur.\ Phys.\ J.\  {\bf A47}, 112 (2011).

\bibitem{Kro_new} S.~V.~Goloskokov and P.~Kroll,
  arXiv:1310.1472 [hep-ph];
 S.~V.~Goloskokov,
  arXiv:1211.5416 [hep-ph].

\bibitem{ColFraStr} J.C. Collins, L. Frankfurt and M. Strikman, 
 Phys.\ Rev.\  D {\bf 56}, 2982 (1997)

\bibitem{ManPil} L.~Mankiewicz, G.~Piller and A.~Radyushkin,
  Eur.\ Phys.\ J.\  C {\bf 10}, 307 (1999)
  
\bibitem{MankWeigl} L.~Mankiewicz, G.~Piller and T.~Weigl,
  Eur.\ Phys.\ J.\ C {\bf 5}, 119 (1998)
  
\bibitem{VdH} M.~Vanderhaeghen, P.~A.~M.~Guichon and M.~Guidal, Phys. Rev. Lett. {\bf 80}, 5064 (1998); 
{\it ibid}  Phys.\ Rev.\  D {\bf 60}, 094017 (1999)

\bibitem{ChenJi} Z.~Chen and X. D. Ji,
  Phys.\ Rev.\ D {\bf 71}, 016003 (2005)
 
\bibitem{Hagler} P. Hagler,
  Phys.\ Lett.\ B {\bf 594}, 164 (2004)

\bibitem{BelJiYuan_F2} A. Belitsky, X. D. Ji, and F. Yuan, Phys. Rev. Lett. {\bf 91}, 092003 (2003). 

\bibitem{Rad_new} A.~V.~Radyushkin,
  Phys.\ Rev.\ D {\bf 80}, 094009 (2009)

\bibitem{Metz} S.~Meissner, A.~Metz and M.~Schlegel,
  JHEP {\bf 0908}, 056 (2009)

\bibitem{Ji1} X.~D.~Ji, Phys.\ Rev.\ D {\bf 55}, 7114 (1997)

\bibitem{Diehl_hab}  M.~Diehl,
Phys.\ Rept.\  {\bf 388}, 41 (2003).

\bibitem{GolOwe} G.~R.~Goldstein and J.~F.~Owens,
  Phys.\ Rev.\  D {\bf 7} (1973) 865.

\bibitem{VGG} M . Vanderhaeghen, P. A. M. Guichon and M. Guidal,
  Phys.\ Rev.\  D {\bf 60}, 094017 (1999); P.~A.~M.~Guichon and M.~Vanderhaeghen,
  Prog.\ Part.\ Nucl.\ Phys.\  {\bf 41}, 125 (1998).
  
  
\bibitem{Radyush_pos} A.~V.~Radyushkin,
  Phys.\ Lett.\ B {\bf 380}, 417 (1996); {\it ibid} 
  Phys.\ Rev.\ D {\bf 56}, 5524 (1997).
  
\bibitem{HERMES1} A.~Airapetian {\it et al.} [ HERMES Collaboration ],
  JHEP {\bf 0806}, 066 (2008).

\bibitem{HERMES2} A.~Airapetian {\it et al.} [ HERMES Collaboration ],
  JHEP {\bf 0911}, 083 (2009).

\bibitem{HallB} F.~X.~Girod {\it et al.} [ CLAS Collaboration ],
  Phys.\ Rev.\ Lett.\  {\bf 100}, 162002 (2008).

\bibitem{Cates} G.~D.~Cates, C.~W.~de Jager, S.~Riordan and B.~Wojtsekhowski,
  Phys.\ Rev.\ Lett.\  {\bf 106}, 252003 (2011). 
  
\bibitem{BroCloGun} S.~J.~Brodsky, F.~E.~Close, J.~F.~Gunion,
   Phys.\ Rev.\  {\bf D8}, 3678 (1973).
   
 \bibitem{Roberts2} C.~D.~Roberts,
  arXiv:1203.5341 [nucl-th].
   
\bibitem{BoerMul} D. Boer and P. J. Mulders, Phys. Rev. D {\bf 57}, 5780 (1998).
   
\bibitem{Forshaw} J.A. Forshaw and D.A. Ross, {\em ``Quantum ChromoDynamics and the Pomeron"}, Cambridge University Press, 1997. 
  
\bibitem{AHLT} S.~Ahmad, H.~Honkanen, S.~Liuti {\it et al.}; {\it ibid} 
  Eur.\ Phys.\ J.\  {\bf C63}, 407-421 (2009); {\it ibid}   Phys.\ Rev.\  {\bf D75}, 094003 (2007).

\bibitem{Rad11} A.~V.~Radyushkin,
  Phys.\ Rev.\ D {\bf 83}, 076006 (2011).
  
\bibitem{Szcz11} A.~P.~Szczepaniak, J.~T.~Londergan and F.~J.~Llanes-Estrada,
  Acta Phys.\ Polon.\ B {\bf 40}, 2193 (2009). 
  
\bibitem{BroDieHwa}   S.~J.~Brodsky, M.~Diehl and D.~S.~Hwang,
  Nucl.\ Phys.\ B {\bf 596}, 99 (2001).

\bibitem{BroHwa} S.~J.~Brodsky and D.~S.~Hwang,
  Nucl.\ Phys.\ B {\bf 543}, 239 (1999).

\bibitem{GolLiu_z_diag} G.~R.~Goldstein and S.~Liuti,
  Int.\ J.\ Mod.\ Phys.\ Conf.\ Ser.\  {\bf 04}, 179 (2011).
  
  \bibitem{BroEst}   S.~J.~Brodsky, F.~J.~Llanes-Estrada,
  Eur.\ Phys.\ J.\  {\bf C46}, 751 (2006);  S.~J.~Brodsky, F.~J.~Llanes-Estrada, A.~P.~Szczepaniak,
  Phys.\ Rev.\  {\bf D79}, 033012 (2009).
  
  \bibitem{GolLiu_disp} G.~R.~Goldstein and S.~Liuti,
  Phys.\ Rev.\ D {\bf 80}, 071501 (2009).
  
\bibitem{GGL_param}  J.~O.~Gonzalez-Hernandez, S.~Liuti, G.~R.~Goldstein and K.~Kathuria,
  arXiv:1206.1876 [hep-ph], and J.~O.~Gonzalez-Hernandez and S.~Liuti, {\it in preparation}.

\bibitem{BacConRad} 
A.~Bacchetta, F.~Conti and M.~Radici,
  Phys.\ Rev.\ D {\bf 78}, 074010 (2008).
 
\bibitem{MusRad} I.~V.~Musatov and A.~V.~Radyushkin,
  Phys.\ Rev.\ D {\bf 61}, 074027 (2000).
 
\bibitem{GolMar} K.~J.~Golec-Biernat and A.~D.~Martin,
  Phys.\ Rev.\ D {\bf 59}, 014029 (1999).
  
\bibitem{Bur3} M.~Burkardt and G.~Schnell,
  Phys.\ Rev.\  D {\bf 74}, 013002 (2006).

\bibitem{Soffer}  J.~Soffer,
  Phys.\ Rev.\ Lett.\  {\bf 74}, 1292 (1995).

\bibitem{Losini} S.~Liuti, G.~R.~Goldstein and J.~O.~Gonzalez Hernandez,
  Nuovo Cim.\ C {\bf 035N2}, 321 (2012).
  
\bibitem{Goeke} K.~Goeke, M.~V.~Polyakov and M.~Vanderhaeghen,
  Prog.\ Part.\ Nucl.\ Phys.\  {\bf 47}, 401 (2001).

\bibitem{GGL_short} G.~R.~Goldstein, J.~.O.~Gonzalez~Hernandez and S.~Liuti, J. Phys. G {\bf 39}, 115001 (2012).
  

\bibitem{Bacchetta_review} A.~Bacchetta, M.~Diehl, K.~Goeke, A.~Metz, P.~J.~Mulders and M.~Schlegel,
  JHEP {\bf 0702}, 093 (2007).

\bibitem{DieSap} M.~Diehl and S.~Sapeta,
  Eur.\ Phys.\ J.\ C {\bf 41}, 515 (2005).


\bibitem{Kub} I. Bedlinskiy,  {\it et al.}, Phys.Rev.Lett. 109, 112001(2012).

\bibitem{AvaKim} H. Avakian and A. Kim, {\em private communication}.

\bibitem{demasi} R.~De Masi {\it et al.},
  Phys.\ Rev.\  C {\bf 77}, 042201 (2008).
  
\bibitem{Girod} F.X. Girod, V. Kubarovsky and P. Stoler, {\em private communication}. 

\bibitem{Pire1} D.~Y.~.Ivanov, B.~Pire, L.~Szymanowski and O.~V.~Teryaev,
  Phys.\ Lett.\ B {\bf 550}, 65 (2002).

\bibitem{Pire2} M.~El Beiyad, B.~Pire, M.~Segond, L.~Szymanowski and S.~Wallon,
  Phys.\ Lett.\ B {\bf 688}, 154 (2010).
  
\bibitem{Schindler}   M.~R.~Schindler, S.~Scherer,
  Eur.\ Phys.\ J.\  {\bf A32}, 429-433 (2007).

\bibitem{Fearing} T.~Gorringe, H.~W.~Fearing,
  Rev.\ Mod.\ Phys.\  {\bf 76}, 31-91 (2004).

%
  


\end{thebibliography}
\end{document}